\documentclass[12pt,a4paper]{article}
\usepackage[onehalfspacing]{setspace}
\usepackage{amssymb,amsmath,amsthm,amsfonts,geometry,graphicx,caption,natbib,booktabs,xcolor,bm,float,array}
\usepackage{lmodern}
\usepackage[colorlinks=true,linkcolor=cyan,urlcolor=cyan,citecolor=cyan,hypertexnames=false]{hyperref}
\makeatletter
\def\input@path{{exhibits/}}
\makeatother
\graphicspath{{exhibits/}}
\newtheorem{thm}{Theorem}
\newtheorem{prop}{Proposition}
\newtheorem{lem}{Lemma}
\theoremstyle{definition}
\newtheorem{assumption}{Assumption}

\newtheorem{assumptionb}{Assumption}
\newtheorem{assumptionc}{Assumption}
\theoremstyle{definition}\newtheorem{example}{Example}
\theoremstyle{definition}\newtheorem{definition}{Definition}

\newcommand{\todo}[1]{{\color{red}[\textbf{TODO:} #1]}}
\newcommand{\safegraphic}[2][]{\IfFileExists{#2}{\includegraphics[#1]{#2}}{\fbox{\todo{figure file \detokenize{#2} missing}}}}
\newcommand{\safeinput}[1]{\IfFileExists{#1}{\input{#1}}{\fbox{\todo{table file \detokenize{#1} missing}}}}
\newcommand{\E}{\mathbb{E}}

\title{Control VAR: a counterfactual based approach to inference in macroeconomics}
\author{Raimondo Pala\thanks{University of Rome Tor Vergata. Email: raimondopala@gmail.com. The code and data that reproduce every table and figure of the paper and the supplement are available at \url{https://github.com/Raimondo-P/Control-VAR}.}}
\date{\today}

\begin{document}
\maketitle

\begin{abstract}
\noindent A recursive VAR is read as causal whenever the policy is ordered first. I show that this ordering is not sufficient when the policy reacts to the economy it is meant to affect: the impulse response then combines the causal effect with a bias from the common trends that the policy reacts to and a bias specific to the treated unit. I develop the Control VAR (CVAR), which compares the treated unit with untreated controls that share its stochastic trends, and removes the first bias and, under exogeneity with respect to the counterfactual cycle, the second. The method allows repeated, continuous and multiple policies, separates the effect of a single policy change from that of its later values, measures the bias of a VAR without controls through the response of the controls, and bounds the remaining bias. I estimate the weights, derive bootstrap inference, show that the CVAR in changes remains valid when the treated unit has a stochastic trend of its own, test whether the more efficient CVAR in levels is supported, and characterise the bias of weights estimated in changes. I apply the CVAR to the Alaska Permanent Fund Dividend, extended unemployment benefits and California's Low Carbon Fuel Standard.

\end{abstract}

\newpage
\section{Introduction\label{sec:intro}}

What would have happened to inflation had central banks around the world not introduced unconventional monetary policy at the onset of the Great Financial Crisis? Is monetary policy effective in controlling inflation? What is the impact of monetary policy on the economy?

While some of these questions may appear too tough to tackle, economists have them frequently in mind, and often find themselves at a loss as to which methods to use to answer them. Long-run, short-run and sign restrictions in VARs, local projections and linear regressions are just some of the many tools developed to address them, yet economists still face the difficult reality of being unable to give a properly causal answer. Many of these questions naturally involve the formulation of a counterfactual. When we think about the effectiveness of unconventional monetary policy, we have in mind a world in which central banks preferred inaction in the face of a looming disinflationary pressure, and when we think about the effects of a natural disaster, we have in mind a world in which no disaster happened. We rarely make the case, however, that we are truly identifying the effect of the policy over inaction, or the effect of the disaster, and for a long time causality has been almost a forbidden topic in macroeconomics.

The difficulty is specific to the social sciences, where the lack of experiments leaves economists to make informed guesses. \citet{Nakamura2018a} compare economists to weather forecasters, painting a gloomy picture of economic science, but they also express optimism about the direction the field has taken. Much of the recent literature has developed tools that formulate a counterfactual in the potential outcomes framework of \citet{Rubin1980} for VARs and local projections \citep{bojinovshephard2019,RambachanSheppard2021,dube2023local}, and another stream defines counterfactuals through optimal policies \citep{mckay2023can}. Both define the effect of a macroeconomic policy as the difference between two alternative assignments of the policy.

The tool that macroeconomists most often use to measure such effects is the recursive vector autoregression, in which the policy is ordered before the outcomes and its unexpected movements are treated as shocks whose impulse responses give the effect of the policy \citep{Sims1980,ChristianoEichenbaumEvans1999,Bloom2009}. The causal literature has given this reading a foundation, since the impulse response is a causal effect when the policy does not react, within the period, to the shocks that move the outcome \citep{RambachanSheppard2021}. Many policies fail this requirement by design, because they are meant to react to the economy: central banks set rates in response to current inflation and activity, fiscal stabilisers expand when output falls, and unemployment benefits are extended when a state's unemployment rises. A countercyclical policy is therefore applied when the economy is at its weakest, so the recovery that would have followed anyway is attributed to the policy, which appears to work even when it has no effect. I show that ordering such a policy first is not sufficient for a causal reading of its impulse response, and I give the exact form of the bias.

Applied microeconomics faces treatments that depend on the outcome in many settings, and it has well-established answers when units that do not receive the treatment are available, such as difference-in-differences \citep{Card1990,CardKrueger,Abadie2005} and synthetic control \citep{AbadieGardeazabal2003,Abadie2010}. Two features of macroeconomic data limit their use. Comparable units are few, and macroeconomic series wander without returning to a fixed mean, so that regressions among them can produce spurious relations \citep{GrangerNewbold1974}. Cointegration showed that such series can share a common stochastic trend, so that a combination of them is stable \citep{EngleGranger1987,Johansen1991}, and synthetic control rests on the same idea when it reconstructs the path a treated unit would have followed from a weighted combination of untreated units that share its trends \citep{Abadie2015,HsiaoChingWan2012,Arkhangelsky2021}.

This paper brings the two traditions together in the control VAR (CVAR), which uses control units to remove the common trends of the treated unit and estimates a VAR in the policy and the remaining cycle. Since the common trends carry the shocks the policy reacts to, removing them removes the confounding they create. Theorem \ref{thm:var} makes the confounding precise for a recursive VAR without controls, whose impulse response combines a causal component with two biases. The causal component is a weighted average of effects whose form depends on the distribution of the policy, an average treatment effect on the treated for a dummy and an average causal response for a continuous policy. The first bias comes from the common trends the policy reacts to, so that an intervention that only partly offset a recession suffered also by economies that did not intervene can appear harmful, and the second comes from the shocks that only the treated unit experiences. Replacing the outcome with its cycle removes the first bias (Theorem \ref{thm:cycle}), and when the policy is exogenous with respect to the cycle the second vanishes, so that the response of the cycle is the effect of a policy surprise (Theorem \ref{thm:ident}). That response includes the later reaction of the policy to its own consequences, which is the object macroeconomists usually report, and Theorem \ref{thm:direct} separates it from the effect of a single change in the policy with its later values held fixed, the impulse causal effect of \citet{RambachanSheppard2021}.

The controls also give users of VARs a direct measure of the bias they remove. When the weighted controls are added to a VAR, their response to a policy that never reached them is the bias from the common trends (Proposition \ref{prop:placebo}), and the same response bounds the bias specific to the treated unit when exogeneity is in doubt, giving the multiple of it at which a conclusion breaks down (Proposition \ref{prop:bounds}). Because the weights on the controls are estimated, Theorem \ref{thm:est} and Proposition \ref{prop:J} show how their estimation enters the distribution of the response and when it can be ignored.

The controls remove the common trends only if they share every stochastic trend of the treated unit. Over several decades a region can also follow a path of its own, through the rise or decline of an industry or a population that grows faster than that of its neighbours, and comparisons in levels, synthetic control among them, assume such a path away. The CVAR does not need this assumption, because estimated on the changes of the cycle it identifies the same effects when the treated unit carries a trend of its own (Theorem \ref{thm:diff}), and I use it as the default. The CVAR in levels is more efficient when the treated unit shares all its trends with the controls, and Theorem \ref{thm:po} gives a test of the null hypothesis that it does not, so that a rejection allows the estimator in levels while a non-rejection only costs efficiency. Applied to the employment, income, house prices and consumer prices of US states and regions, each compared with its neighbours, the test rejects about as often as its nominal size, which shows that the assumption behind comparisons in levels cannot be supported for most units even over several decades.

The results rest on five assumptions: the consistency condition of the potential outcomes framework for time series \citep{RambachanSheppard2021} together with a stable unit treatment value assumption \citep{Rubin1980}, under which the policy does not reach the controls; the factor structure of synthetic control \citep{Abadie2010}; an effect linear and additive in the policy, as in the literature on VARs \citep{KilianLutkepohl2017}; a conditional exogeneity condition of the kind used for policy evaluation in time series \citep{AngristJordaKuersteiner2018}, imposed on the cycle instead of the outcome; and stationarity. The Supplementary Appendix relaxes the first and the third, recovering the effect when the policy spills over to the controls through a known exposure mapping and when the effect is not linear in the surprise. Its simulations, calibrated to the application to unemployment benefits, show that a VAR without controls overstates the effect of a policy that reacts to common shocks while the CVAR is unbiased, and that with long enough lags a bootstrap that re-estimates the weights in every draw delivers bands with close to nominal coverage.

I apply the CVAR to three policies that react to the economy they affect. The dividend that Alaska pays every year to its residents follows the returns of a sovereign wealth fund, so its largest payments came after the stock market booms that preceded the recessions of 2001 and 2008. A VAR on Alaska alone consequently finds that a larger dividend lowers employment, a response that the states that never receive the dividend also show, while the CVAR finds no fall in employment and a rise in leisure and hospitality, the sector where the spending of the dividend lands \citep{Kueng2018}. Where synthetic control has evaluated the introduction of the dividend \citep{JonesMarinescu2022}, the CVAR uses the variation in its size over more than thirty payments. Extended unemployment benefits switch on in recessions, so a VAR without controls finds that they sharply lower unemployment, and the states that never extended benefits show almost the same response; the CVAR finds effects close to zero, in line with the evidence for periods in which benefit durations were already long \citep{AcostaEtAl2025}. Finally, the cost of the Low Carbon Fuel Standard of California changes every month with the price of its credits, and the CVAR finds a pass-through to the retail price of gasoline consistent with full pass-through, with narrower bands than a VAR without controls.

The CVAR belongs to the methods that build counterfactuals for a treated time series from untreated units. The artificial counterfactual \citep{CarvalhoMasiniMedeiros2018,MasiniMedeiros2021} and CausalImpact \citep{Brodersen2015} target the effect of a single intervention, and synthetic control \citep{Abadie2010} and synthetic difference-in-differences \citep{Arkhangelsky2021} estimate the weights from the periods before a single adoption. Difference-in-differences with staggered adoption \citep{callawaysantannamultipletime2021,sunabraham2021} and with non-binary and non-absorbing treatments \citep{callawaysantannabacon2021,deChaisemartinDHaultfoeuille2020} identifies the effect when the timing of the treatment is unrelated to the outcomes once unit and time effects are removed. Local-projection difference-in-differences \citep{dube2023local} and panel VARs with a causal interpretation \citep{pala2024pvarcontrol} allow a recurring policy that reacts to the past, although they are restricted to dummy policies and do not account for cointegration among the variables. The CVAR covers the remaining case of a policy that recurs, takes several values and reacts to shocks it shares with the controls, and because it estimates the weights from the stochastic trends over the whole sample it needs no period free of the policy. It thereby relaxes the central condition under which the causal interpretation of VARs and local projections holds \citep{RambachanSheppard2021,GoncalvesHerreraKilianPesavento2021,kolesar2024dynamic}. Table \ref{tab:methods} summarises what each approach allows.

\begin{table}[!t]
\centering
\caption{What each approach allows\label{tab:methods}}
\footnotesize
\setlength{\tabcolsep}{4pt}
\begin{tabular}{>{\raggedright\arraybackslash}p{4.6cm}ccccc}
\toprule
 & Recurring & Continuous & Reaction to & Heterogeneous & Stochastic \\
 & policy & intensity & the past & exposure to & trends \\
 & & & & common shocks & \\
\midrule
DiD, staggered adoption & $\times$ & $\times$ & $\times$ & $\times$ & $\times$ \\
DiD, non-binary and non-absorbing & \checkmark & \checkmark & $\times$ & $\times$ & $\times$ \\
Local-projection DiD & \checkmark & $\times$ & \checkmark & $\times$ & $\times$ \\
Synthetic control, synthetic DiD & $\times$ & $\times$ & $\times$ & \checkmark & $\times$ \\
Artificial counterfactual, CausalImpact & $\times$ & $\times$ & $\times$ & \checkmark & \checkmark \\
Recursive VARs, local projections & \checkmark & \checkmark & \checkmark & $\times$ & -- \\
Panel VARs & \checkmark & $\times$ & \checkmark & $\times$ & $\times$ \\
\midrule
CVAR & \checkmark & \checkmark & \checkmark & \checkmark & \checkmark \\
\bottomrule
\end{tabular}
\par\smallskip\parbox{0.95\textwidth}{\footnotesize DiD: difference-in-differences. Recurring policy: the policy switches on and off or changes repeatedly. Continuous intensity: the policy takes more than two values. Reaction to the past: the policy may respond to the past of the outcome it affects. Heterogeneous exposure to common shocks: the treated and untreated units may respond differently to the shocks they share, and the policy may react to those shocks. Stochastic trends: the outcomes may share stochastic trends, with cointegration among the units. Recursive VARs and local projections use no untreated units, so the last column does not apply to them. Difference-in-differences, staggered adoption: \citet{callawaysantannamultipletime2021,sunabraham2021}; non-binary and non-absorbing: \citet{callawaysantannabacon2021,deChaisemartinDHaultfoeuille2020}; local projections: \citet{dube2023local}; synthetic control: \citet{Abadie2010,Arkhangelsky2021}; artificial counterfactual: \citet{CarvalhoMasiniMedeiros2018,MasiniMedeiros2021,Brodersen2015}; VARs and local projections: \citet{RambachanSheppard2021,kolesar2024dynamic}; panel VARs: \citet{pala2024pvarcontrol}.}
\end{table}

Section \ref{sec:causal} presents the causal framework and Section \ref{sec:assumptions} the assumptions. Section \ref{sec:theorems} contains the theorems, estimation and inference, Section \ref{sec:test} the test of the design, Section \ref{sec:ext} the case of a treated unit with a trend of its own, Sections \ref{sec:alaska} to \ref{sec:lcfssum} the applications, and Section \ref{sec:conclusions} concludes.

\section{Causal framework\label{sec:causal}}

\citet{RambachanSheppard2021} define the dynamic causal effects of a policy in a time series through potential outcomes indexed by the whole path of the policy. I follow their example and adapt the framework to a setting with one treated unit and a set of control units, building on the panel formulation of \citet{bojinovshephard2019} and on the potential outcomes of \citet{Rubin1980}.

There are $K+1$ units: a treated unit, indexed by $1$, and $K$ control units, indexed by $k=1,\dots,K$. For $t=1,\dots,T$, each unit $i$ receives a policy $W_{it}$, and a vector $R_t$ of other policies may also be in place. In the application, the units are states, $W_{it}$ is the dividend that state $i$ pays to each of its residents in month $t$, in thousands of dollars, which is positive only in Alaska and only in the month of the yearly payment, and $R_t$ is the monthly change in the price of oil. The policy may also be a dummy, such as an indicator for whether a programme is in place.

Let $\mathbf{w}_{1:T}=(w_{i,1:T})_{i}$ collect a path of the policy for every unit over the whole sample, and let $r_{1:T}$ be a path of the other policies. The potential outcome of unit $i$ in period $t$ is
\begin{equation}
Y_{it}(\mathbf{w}_{1:T},r_{1:T}),
\label{eq:po}
\end{equation}
the value its outcome would take if the policies of all units had followed those paths. In the application, $Y_{1t}(\mathbf{w}_{1:T},r_{1:T})$ is employment in Alaska in month $t$ under a given history of dividends. In this general form, employment in Alaska today may depend on the whole path of the policies: on past dividends, on the expectation of future ones, and on the other policies, such as the current price of oil. The observed outcomes are the potential outcomes at the realised paths, $Y_{it}=Y_{it}(\mathbf{W}_{1:T},R_{1:T})$.

Causal effects are comparisons of potential outcomes across paths. The most general is the effect of a whole path, $Y_{1,t+h}(\mathbf{w}_{1:T},r_{1:T})-Y_{1,t+h}(\mathbf{w}'_{1:T},r_{1:T})$. Its realised counterpart compares the realised path with a path in which the treated unit never receives the policy,
\begin{equation}
\Delta_{t+h}=Y_{1,t+h}(\mathbf{W}_{1:T},R_{1:T})-Y_{1,t+h}(\mathbf{W}^{0}_{1:T},R_{1:T}),
\label{eq:effectpath}
\end{equation}
where $\mathbf{W}^0_{1:T}$ replaces the path of the treated unit with zeros and leaves the paths of the other units unchanged. It is the effect on employment in Alaska of all the dividends that Alaska actually paid up to $t+h$.

A policy that reacts to the economy is itself an outcome of its history, and a change in the policy today changes the policy tomorrow. I therefore also treat future policies as potential outcomes. For a period $t$, let $W_{1,t+k}(w)$, $k\geq0$, be the policy of the treated unit in period $t+k$ if its policy in period $t$ is set to $w$, the history before $t$ is the realised one, and the policy then follows its usual determinants, so that $W_{1t}(w)=w$ and the realised policy is $W_{1,t+k}=W_{1,t+k}(W_{1t})$. This is the formulation of dynamic treatment regimes \citep{Robins1986,Murphy2003}, in which the treatment in each period responds to the history that earlier treatments helped to create.

Two causal effects of the policy in period $t$ follow from this formulation, and they differ in what happens to the policy after period $t$. The first, introduced by \citet{RambachanSheppard2021}, changes the policy in period $t$ and holds the rest of the path fixed.

\begin{definition}[Impulse causal effect\label{def:ice}]
The impulse causal effect at horizon $h$ of setting the policy of the treated unit in period $t$ to $w$ instead of $w'$ is
\begin{equation}
Y_{1,t+h}(\mathbf{W}_{1:t-1},w,\mathbf{W}_{t+1:T},R_{1:T})-Y_{1,t+h}(\mathbf{W}_{1:t-1},w',\mathbf{W}_{t+1:T},R_{1:T}),
\label{eq:ice}
\end{equation}
with the policies of every unit in every other period at their realised values.
\end{definition}

It measures the direct effect of changing the policy in period $t$ on the outcome $h$ periods later. The second lets the policy follow its usual determinants after period $t$.

\begin{definition}[Regime effect\label{def:regime}]
The regime effect at horizon $h$ of setting the policy of the treated unit in period $t$ to $w$ instead of $w'$ is
\begin{equation}
\begin{split}
&Y_{1,t+h}\langle w\rangle-Y_{1,t+h}\langle w'\rangle,\quad\text{where}\\
&Y_{1,t+h}\langle w\rangle=Y_{1,t+h}\big(\mathbf{W}_{1:t-1},w,W_{1,t+1}(w),\dots,W_{1,t+h}(w),\mathbf{W}_{-1,t:T},R_{1:T}\big),
\end{split}
\label{eq:regime}
\end{equation}
where $\mathbf{W}_{-1,t:T}$ are the realised paths of the other units.
\end{definition}

It measures the effect of changing the policy in period $t$ together with the changes in its later values that the usual determinants of the policy then bring about. A policymaker who considers a different policy today, knowing that the policy will then evolve as usual, is interested in the regime effect.

A recursive VAR reports a third object, which connects $\Delta_{t+h}$ to the timing of the policy through its unexpected part. Let $Z_t=(R_t',W_{1t},G_t)'$, where $G_t$ is the cycle of the treated unit, let $\mathcal{F}_{t-1}$ be the information in $Z_{t-1},Z_{t-2},\dots$, and let
\begin{equation}
u_t=W_{1t}-\E^\ast[W_{1t}\mid R_t,\mathcal{F}_{t-1}]
\label{eq:innovation}
\end{equation}
be the innovation of the policy, where $\E^\ast$ denotes the linear projection. In the application, $u_t$ is the part of this month's dividend that could not be predicted from the past dividends, the price of oil and the past of the economy of Alaska. The third causal effect weighs the effect of the realised path by this innovation.

\begin{definition}[Effect of a policy surprise\label{def:surprise}]
The effect of a policy surprise at horizon $h$ is
\begin{equation}
\theta_h=\frac{\E[\Delta_{t+h}\,u_t]}{\E[u_t^2]}.
\label{eq:estimand}
\end{equation}
\end{definition}

In essence, $\theta_h$ is the part of the effect of the dividends on employment $h$ months later that is associated with an unexpected dividend of one unit in month $t$. Like the regime effect, it includes the effect of the dividends that follow the surprise. 

\section{Assumptions\label{sec:assumptions}}

The first assumption restricts how the policies enter the potential outcomes. Its first two parts are needed by any time-series estimate of the effect of a policy, with or without controls, and its third part is specific to the use of controls.

\begin{assumption}[No anticipation, no interference and untreated controls\label{assu:A1}]
\begin{enumerate}
\item[(i)] No anticipation: $Y_{it}(\mathbf{w}_{1:T},r_{1:T})=Y_{it}(\mathbf{w}_{1:t},r_{1:t})$ for every unit $i$ and period $t$.
\item[(ii)] No interference: $Y_{1t}(\mathbf{w}_{1:T},r_{1:T})=Y_{1t}(w_{1,1:T},r_{1:T})$, and $Y_{kt}(\mathbf{w}_{1:T},r_{1:T})=Y_{kt}(\mathbf{w}'_{1:T},r_{1:T})$ for every control $k$ and every pair of paths $\mathbf{w}_{1:T}$, $\mathbf{w}'_{1:T}$ that differ only in the path of the treated unit.
\item[(iii)] Untreated controls: $W_{kt}=0$ for every control $k$ and period $t$.
\end{enumerate}
\end{assumption}

Part (i) is the no anticipation condition of \citet{RambachanSheppard2021}: a policy affects the outcome only once it is in place, so that, in the example of the dividend, payments that have not yet been made do not affect employment in Alaska. Part (ii) is the stable unit treatment value assumption of \citet{Rubin1980}, adapted to time series. As in any comparison of treated and untreated units, it makes the units comparable by ruling out spillovers: the policy of the treated unit does not change the potential outcomes of the controls. In the example, it rules out an effect of the dividend on employment in other states, through the spending of Alaskans elsewhere or through movements of people into or out of Alaska.\footnote{Supplementary Appendix \ref{sec:spillnl} relaxes this part when the policy spills over to the controls through a known exposure.} Part (iii) requires that the controls never receive the policy, which holds by construction when they are states that pay no dividend.

Under Assumption \ref{assu:A1} the notation simplifies. I write $W_t$ for the policy of the treated unit, $Y^1_t(w_{-\infty:t})$ for the potential outcome of the treated unit under its own path, holding the other policies at their realised values, $Y^1_t(0)$ for its value had it never received the policy, and $Y^0_{kt}$ for the observed outcomes of the controls, which are their potential outcomes without the policy. The effect of the policy actually applied in \eqref{eq:effectpath} becomes $\Delta_{t+h}=Y^1_{t+h}-Y^1_{t+h}(0)$.

\begin{assumption}[Common trends\label{assu:A2}]
There is an $r$-dimensional process $F_t$, integrated of order one with stationary first difference, and a $K\times r$ matrix $\Lambda$ of rank $r\leq K$, such that
\begin{equation}
Y^0_t=\Lambda F_t+v_t,\qquad Y^1_t(0)=\lambda'F_t+e_t,
\label{eq:factors}
\end{equation}
where $v_t$ and $e_t$ are stationary and $\lambda=\Lambda'b$ for some $b\in\mathbb{R}^K$.
\end{assumption}

Assumption \ref{assu:A2} is a factor model: the outcome of the treated unit, such as employment in Alaska, follows stochastic trends that it shares with the outcomes of the controls, such as employment in other states. The vector $F_t$ collects the forces behind these trends, such as the national business cycle or the price of oil, and each unit loads on them differently. The condition $\lambda=\Lambda'b$ says that the treated unit loads on these forces as a weighted combination of the controls does. The model resembles the factor model of synthetic control \citep{Abadie2010}, from which it differs in the nature of the trends, which are integrated as in the cointegration literature \citep{EngleGranger1987}, and in the role of the weights. Under it, the counterfactual cycle of the treated unit,
\begin{equation}
G^0_t=Y^1_t(0)-b'Y^0_t=e_t-b'v_t,
\label{eq:cycle0}
\end{equation}
is stationary, with mean $a$. The weights $b$ play the role of the synthetic control weights, and I call $C_t=b'Y^0_t$ the weighted controls, so that the observed cycle is $G_t=Y^1_t-C_t$.

Two features of $b$ differ from the usual practice of synthetic control, and both follow from the role of the weights here. The weights are not required to sum to one. Employment in Alaska may be more or less cyclical than in the controls, and the weights must reproduce the amplitude of its exposure to the trends, in addition to its direction. The weights are also not required to be unique. When there are more controls than common trends, $r<K$, the controls are cointegrated among themselves, the combinations $S_t=H'Y^0_t=H'v_t$ are stationary for any $K\times(K-r)$ matrix $H$ whose columns span the null space of $\Lambda'$, and every vector in $\mathcal{B}=\{b+H\phi:\phi\in\mathbb{R}^{K-r}\}$ removes the trends. This is the situation of synthetic control with more donors than factors \citep{AbadieLHour2021}. The economist should therefore pick the controls that can best approximate movements in the treated economy.

\begin{assumption}[Additive dynamic effects\label{assu:A3}]
$Y^1_t(w_{-\infty:t})=Y^1_t(0)+\sum_{j\geq0}\tau_jw_{t-j}$, with $\sum_{j\geq0}|\tau_j|<\infty$.
\end{assumption}

Assumption \ref{assu:A3} is the linearity that the literature on VARs imposes when it interprets impulse responses \citep{KilianLutkepohl2017}. Supplementary Appendix \ref{sec:identification} gives the weights with which the response averages heterogeneous effects when the effect is not linear, and Supplementary Appendix \ref{sec:spillnl} shows how to recover the response to a surprise of a given size and sign.

Under Assumption \ref{assu:A3}, the impulse causal effect \eqref{eq:ice} of a unit change in the policy in period $t$ on the outcome $h$ periods later is $\tau_h$, and the effect of a policy surprise in \eqref{eq:estimand} decomposes as
\begin{equation}
\theta_h=\sum_{j=0}^h\tau_j\,\psi_{h-j},\qquad \psi_k=\frac{\E[W_{t+k}u_t]}{\E[u_t^2]},
\label{eq:decomp}
\end{equation}
where $\psi_k$ is the path that the policy is expected to follow $k$ periods after a surprise.\footnote{The decomposition uses the orthogonality of $u_t$ to the past of the policy, so that only the policy from period $t$ onward contributes.} The regime effect requires an assumption on the potential policies.

\begin{assumptionb}[Linear policy response\label{assu:A3b}]
$W_{t+k}(w)=W_{t+k}(0)+\phi_kw$ for $k\geq0$, with $\phi_0=1$ and $\sum_{k\geq0}|\phi_k|<\infty$.
\end{assumptionb}

Assumption \ref{assu:A3b} says that a change of $w$ in the policy today changes its value $k$ periods later by $\phi_kw$, whatever path it would otherwise have followed, so that $\phi_k$ is the propagation of the policy through its own determinants. Under Assumptions \ref{assu:A3} and \ref{assu:A3b}, the regime effect \eqref{eq:regime} of a unit change in the policy in period $t$ is $\theta^\ast_h=\sum_{j=0}^h\tau_j\phi_{h-j}$. It combines the direct effects of the policy with its own propagation, and it is the object that macroeconomists usually report as the response of the outcome to a policy shock. The effect of a surprise satisfies
\begin{equation}
\theta_h=\theta^\ast_h+\sum_{j=0}^{h-1}\tau_j\,\kappa_{h-j},\qquad \kappa_k=\frac{\E[W_{t+k}(0)\,u_t]}{\E[u_t^2]}.
\label{eq:pathbias}
\end{equation}
Since $\psi_k=\phi_k+\kappa_k$, the coefficient $\kappa_k$ is the part of the path of the policy $k$ periods ahead that a surprise predicts without causing it. It is non-zero when the surprise carries news about later values of the policy that would have occurred anyway, and the effect of a surprise then attributes to the policy the effect of those later values. When $\kappa_k=0$ for every $k$, the surprise moves the later policy only through its propagation, and the effect of a surprise equals the regime effect.

Under Assumptions \ref{assu:A1} to \ref{assu:A3}, the observed cycle $G_t=Y^1_t-b'Y^0_t$ satisfies
\begin{equation}
G_t=G^0_t+\sum_{j\geq0}\tau_jW_{t-j}.
\label{eq:cycle}
\end{equation}

\begin{assumption}[Exogeneity with respect to the counterfactual cycle\label{assu:A4}]
$\E[G^0_{t+h}u_t]=0$ for all $h\geq0$.
\end{assumption}

Assumption \ref{assu:A4} is a conditional exogeneity condition of the kind used to evaluate policies in time series \citep{AngristJordaKuersteiner2018}, imposed on the counterfactual cycle in place of the outcome. It allows the policy to react to the past of the treated unit and of the controls, to the other policies, and to the contemporaneous innovations of the common trends. It requires that, given the past, the policy surprise does not predict the part of the outcome of the treated unit that the controls do not share and that would have occurred without the policy; in the example, a larger dividend must not predict a movement of employment specific to Alaska.

\begin{assumption}[Stationarity and moments\label{assu:A5}]
$(R_t',W_t,G^0_t,\Delta F_t',v_t')'$ is covariance stationary with finite fourth moments, and $\E[u_t^2]>0$.
\end{assumption}

Assumption \ref{assu:A5} requires that the policies, the counterfactual cycle and the growth of the common trends have stable means and variances.

\section{Theorems\label{sec:theorems}}

\subsection{What a VAR without controls estimates\label{sec:var}}

Before turning to the CVAR, I show what a standard recursive VAR without controls estimates. Let $\tilde u_t$ be the innovation of the policy in a recursive VAR in $(R_t',W_t,Y^1_t)'$, with the past of $Y^1_t$ among the regressors. The impulse response of the outcome of the treated unit $h$ periods after a surprise is then $\gamma^{V}_h=\E[(Y^1_{t+h}-Y^1_{t-1})\tilde u_t]/\E[\tilde u_t^2]$, and $\tilde\theta_h$ denotes the effect of a policy surprise \eqref{eq:estimand} computed with $\tilde u_t$.\footnote{By Lemma \ref{lem:outcome} in Supplementary Appendix \ref{sec:identification}, $\gamma^V_h$ is the response of a VAR in levels, or of a VAR in differences cumulated to levels, when the lags of the system span the predictable part of the policy.}

\begin{thm}[What a VAR without controls identifies\label{thm:var}]
Under Assumptions \ref{assu:A1}, \ref{assu:A2}, \ref{assu:A3} and \ref{assu:A5},
\begin{equation}
\gamma^{V}_h=\tilde\theta_h+\lambda'\frac{\E[(F_{t+h}-F_{t-1})\tilde u_t]}{\E[\tilde u_t^2]}+\frac{\E[(e_{t+h}-e_{t-1})\tilde u_t]}{\E[\tilde u_t^2]}.
\label{eq:bias}
\end{equation}
\end{thm}

The response of the VAR has three parts. The first, $\tilde\theta_h$, is the causal component: the effect on the treated unit of a surprise in the policy, which is the object the researcher wants to estimate.\footnote{When the effect of the policy is not linear, the causal component is a weighted average of effects, with weights that depend on the distribution of the policy. Supplementary Appendix \ref{sec:identification} characterises it for dummy, continuous and non-negative policies, with results close to those of \citet{kolesar2024dynamic} for local projections and VARs.} The second is the bias from the common trends: the average movement of the common trends from the period before a surprise to $h$ periods after it, per unit of surprise, which $\lambda$ translates into the outcome of the treated unit. In the example of the dividend, the common trends include the national business cycle and the financial markets whose returns set the size of the dividend, and the VAR attributes to the dividend the movements of these trends that follow it. The common trends are not the only source of bias. The third term depends on $e_t$, the part of the outcome of the treated unit that neither the common trends nor the policy explain. It captures the reverse causality that arises when the policy reacts to developments specific to the treated unit, and it is zero when the policy surprise is unrelated to the later movements of $e_t$, the exogeneity under which \citet{RambachanSheppard2021} give the VAR its causal interpretation.

\subsection{The CVAR\label{sec:cvar}}

The CVAR subtracts the weighted controls from the outcome of the treated unit. Because the controls share the common trends, the subtraction removes the first source of bias, the common developments that the policy reacts to, and leaves the bias specific to the treated unit, now measured on the counterfactual cycle, as the next theorem shows.

\begin{thm}[What the CVAR identifies\label{thm:cycle}]
Under Assumptions \ref{assu:A1}, \ref{assu:A2}, \ref{assu:A3} and \ref{assu:A5}, the response of the cycle to the policy innovation is
\begin{equation}
\gamma_h\equiv\frac{\E[G_{t+h}u_t]}{\E[u_t^2]}=\theta_h+\frac{\E[G^0_{t+h}u_t]}{\E[u_t^2]},\qquad h=0,1,2,\dots
\label{eq:cyclebias}
\end{equation}
If $Z_t$ follows a VAR, possibly of infinite order, $\gamma_h$ is the response of $G_t$ at horizon $h$ to a unit innovation in $W_t$ in the recursive identification with $R_t$ first, $W_t$ second and $G_t$ last.
\end{thm}

Comparing Theorem \ref{thm:cycle} with Theorem \ref{thm:var}, the common-trend term has disappeared, because the weighted controls load on the common trends as the treated unit does. The remaining term concerns the counterfactual cycle $G^0_t=e_t-b'v_t$ of \eqref{eq:cycle0}, the deviation of the outcome the treated unit would have had without the policy from the counterfactual that its controls provide, and it measures how much of that deviation a policy surprise predicts. Assumption \ref{assu:A4} sets it to zero. The assumption does not require the deviation to vanish: the treated unit may still move away from its controls for reasons of its own, as long as the policy surprise does not predict those movements. In the example, the control states must provide a credible counterfactual for employment in Alaska, in the sense that the movements of Alaska away from that counterfactual which are not due to the dividend are unrelated to the dividend surprise.

\begin{thm}[Identification\label{thm:ident}]
Under Assumptions \ref{assu:A1} to \ref{assu:A5}, $\gamma_h=\theta_h=\sum_{j=0}^h\tau_j\psi_{h-j}$ for all $h\geq0$.
\end{thm}

The response of the cycle is therefore the effect of the policy on the treated unit: the impulse causal effects of the policy, combined with the path the policy follows after a surprise. When the effect of the policy varies over time, $\theta_h$ averages it with weights proportional to the surprise, and the average is an average treatment effect on the treated for a dummy policy and an average causal response for a continuous one (Supplementary Appendix \ref{sec:identification}).

Under the conditional exogeneity of Assumption \ref{assu:A4}, the CVAR therefore recovers the causal effect without bias, which a VAR without controls does not, as the following example shows.

\begin{example}
Consider a system with one control and one common trend, in which the control is $Y^0_t=F_t+v_t$, the treated unit without the policy is $Y^1_t(0)=F_t+e_t$, and the common trend follows $\Delta F_t=\eta_t$. The policy reacts to the common shock, $W_t=c\,\eta_t+\varepsilon_t$, and has only an impact effect, so that $Y^1_t=Y^1_t(0)+\tau_0W_t$, with all shocks independent, serially uncorrelated and of unit variance. The impact response of the VAR without controls is
\begin{equation*}
\gamma^V_0=\tau_0+\frac{c}{1+c^2}\qquad\text{(VAR)}.
\end{equation*}
If the policy has no effect, $\tau_0=0$, and reacts one for one to the common shock, $c=1$, the VAR reports an impact effect of one half. The cycle is $G_t=e_t-v_t+\tau_0W_t$, which does not contain the common shock, and the CVAR returns
\begin{equation*}
\gamma_0=\tau_0=0\qquad\text{(CVAR)}.
\end{equation*}
\end{example}

By ignoring the common shock, the VAR reports an effect where there is none, and the mistake has consequences for policy. A local government that responds to a downturn may conclude that its intervention worked, or that it failed, when the downturn and the recovery that followed came from a broader disruption in the sectors on which its economy depends. Comparing the treated unit with units of similar economic composition that went through the same disruption without the policy separates the two. The difference between the response of a VAR without controls and that of the CVAR then measures the gap between the conclusion that the policymaker reaches by observing the treated unit alone and the one reached by comparing it with units that did not adopt the policy.

The response of the cycle combines the effect of a policy surprise with that of the later values of the policy that follow it, and the same VAR separates the two, because it also estimates the path $\psi_h$ that the policy follows after a surprise.

\begin{thm}[Impulse causal effects and regime effects\label{thm:direct}]
Let Assumptions \ref{assu:A1} to \ref{assu:A5} hold.
\begin{enumerate}
\item[(a)] The impulse causal effects are identified recursively from the responses of the cycle and of the policy, $\tau_0=\gamma_0$ and $\tau_h=\gamma_h-\sum_{j=0}^{h-1}\tau_j\psi_{h-j}$ for $h\geq1$.
\item[(b)] If, in addition, Assumption \ref{assu:A3b} holds and $\E[W_{t+k}(0)\,u_t]=0$ for all $k\geq1$, then $\gamma_h=\theta^\ast_h$, the regime effect.
\end{enumerate}
\end{thm}

Part (a) of Theorem \ref{thm:direct} requires nothing about how the policy evolves after the surprise: whatever path the policy follows, the CVAR measures it, and removing it from the response of the cycle leaves the direct effects. Part (b) gives the condition under which no correction is needed: when the surprise does not predict the path the policy would have followed without it, the impulse response is already the regime effect.

\subsection{Bounds\label{sec:bounds}}

The controls never receive the policy, so under Assumption \ref{assu:A1} any response they show to its innovation comes from the shocks the policy reacted to, and that response is the part of the VAR estimate that a control removes. The controls are therefore useful beyond the construction of the cycle: they measure the bias of a VAR without controls and, as this subsection shows, they give a scale for the bias that the CVAR leaves when Assumption \ref{assu:A4} is in doubt. The gap between the two responses can be measured in a VAR that contains both the treated unit and the weighted controls. Let $\breve u_t$ be the innovation of the policy in a recursive VAR in $(R_t',W_t,Y^1_t,C_t)'$ with $R_t$ first and $W_t$ second, and for any variable $X_t$ of the system let $\gamma^X_h=\E[(X_{t+h}-X_{t-1})\breve u_t]/\E[\breve u_t^2]$ be its response.

\begin{prop}[Response of the controls\label{prop:placebo}]
\begin{enumerate}
\item[(a)] For every $h\geq0$, $\gamma^{Y^1}_h=\gamma^G_h+\gamma^C_h$.
\item[(b)] Under Assumptions \ref{assu:A1}, \ref{assu:A2} and \ref{assu:A5}, $\gamma^C_h=\lambda'\E[(F_{t+h}-F_{t-1})\breve u_t]/\E[\breve u_t^2]+b'\E[(v_{t+h}-v_{t-1})\breve u_t]/\E[\breve u_t^2]$. If the idiosyncratic components of the controls are uncorrelated with $\breve u_t$ at all leads and lags, $\gamma^C_h$ is the common-trend term of Theorem \ref{thm:var} computed with $\breve u_t$.
\item[(c)] Under Assumptions \ref{assu:A1} to \ref{assu:A5}, with $\breve u_t$ in place of $u_t$, $\gamma^G_h$ is the causal effect of a policy surprise.
\end{enumerate}
\end{prop}

A researcher who has estimated a VAR without controls can add the weighted controls to the system and read the size of the bias from their response, and a test that this response is zero at all horizons is a test of the absence of the common-trend bias. The response of the treated outcome splits exactly into the causal response of the cycle and the response of the controls.

The CVAR still rests on Assumption \ref{assu:A4} for the bias specific to the treated unit, and a reader may doubt it in a given application. The response of the controls also helps here, because it gives a natural scale for how large a violation would have to be to change a conclusion: instead of setting the bias specific to the treated unit to zero, one can bound it by a multiple of the bias from the common trends, which the controls measure. In the VAR of Proposition \ref{prop:placebo}, write the response of the cycle as $\gamma^G_h=\breve\theta_h+\breve B_h$, where $\breve\theta_h$ is the effect of a policy surprise and $\breve B_h=\E[G^0_{t+h}\breve u_t]/\E[\breve u_t^2]$ the bias specific to the treated unit, and let $\gamma^C_h$ be the response of the weighted controls, which measures the bias from the common trends.

\begin{assumptionc}[Bounded bias specific to the treated unit\label{assu:A4b}]
For a given $\lambda\geq0$, $|\breve B_h|\leq\lambda\,|\gamma^C_h|$ for every horizon $h$ of interest.
\end{assumptionc}

Assumption \ref{assu:A4b} replaces exogeneity with a limit on how much the policy surprise can be confounded by developments specific to the treated unit. The limit is set relative to a quantity that the data reveal, the response of the weighted controls, which measures how strongly the surprise is confounded by the common trends. The assumption allows the bias specific to the treated unit to reach at most a multiple $\lambda$ of that measured bias, in either direction. With $\lambda=1/2$, for example, suppose that employment in the control states falls after a dividend surprise, because of the national developments that come with the surprise. The developments specific to Alaska that come with the same surprise may then move employment in Alaska by at most half as much, up or down. The multiple concerns the confounding that comes with the surprise, and it says nothing about an effect of the policy on the controls, which Assumption \ref{assu:A1} excludes. Assumption \ref{assu:A4} is the case $\lambda=0$, and $\lambda=1$ allows the confounding specific to the treated unit to be as strong as the confounding from the common trends. The comparison follows the logic of \citet{AltonjiElderTaber2005} and \citet{Oster2019}, who bound the selection on unobserved variables by the selection on observed ones: here the observed confounding is the one that the common trends carry, which the controls reveal, and the unobserved confounding is the one specific to the treated unit.

\begin{prop}[Bounds\label{prop:bounds}]
Under Assumptions \ref{assu:A1}, \ref{assu:A2}, \ref{assu:A3}, \ref{assu:A5} and \ref{assu:A4b}, and if the idiosyncratic components of the controls are uncorrelated with $\breve u_t$ at all leads and lags:
\begin{enumerate}
\item[(a)] $\breve\theta_h\in\big[\gamma^G_h-\lambda|\gamma^C_h|,\ \gamma^G_h+\lambda|\gamma^C_h|\big]$, an interval that reduces to $\gamma^G_h$ when $\lambda=0$.
\item[(b)] The response of the VAR without controls, $\gamma^{Y^1}_h=\gamma^G_h+\gamma^C_h$, belongs to the interval only if $\lambda\geq1$: attributing to the policy the response of a VAR without controls requires a bias specific to the treated unit as large as the bias from the common trends.
\end{enumerate}
\end{prop}

The bounds are estimated by the corresponding responses of the VAR of Proposition \ref{prop:placebo}, and bands for their end points follow from bootstrapping that VAR. The breakdown value of $\lambda$, at which a conclusion stops holding, summarises how strong the violation of Assumption \ref{assu:A4} would have to be to overturn it, as in the sensitivity analysis of \citet{RambachanRoth2023} for parallel trends.

\subsection{Estimation and inference\label{sec:estimation}}

The theorems take the weights on the controls as given. In practice they are unknown, so the response of the CVAR is computed from a cycle built with estimated weights, and inference has to account for the fact that a variable of the VAR is itself estimated. The problem is familiar in econometrics, where it arises with generated regressors \citep{Pagan1984} and with two-step estimators in general \citep{MurphyTopel1985,NeweyMcFadden1994}: the sampling error of the first step enters the distribution of the second unless it vanishes faster than the error of the second step. In the CVAR the first step has a feature of its own, because the weights come from a cointegrating regression, in which the part that multiplies the common trends is estimated faster than the rest. I estimate the weights and the response in four steps: the first two build the counterfactual from the controls, and the last two estimate the VAR on the resulting cycle.

\textit{Step 1.} I estimate how many common trends the controls share. When there are more controls than trends, the controls are cointegrated among themselves, and the number of their cointegrating relations, $K-\hat r$, has to be estimated, as in a cointegrated VAR \citep{Johansen1991}. The difference lies in the use of the estimate. In a cointegrated VAR the cointegrating relations are the object of interest, while in the CVAR they tell the next step which combinations of the controls carry the trends and which are stationary and serve only to absorb idiosyncratic variation. I estimate $K-\hat r$ with an information criterion on the eigenvalues of the reduced-rank regression of \citet{Johansen1991}, with a penalty that grows with the logarithm of the sample size \citep{AznarSalvador2002,ChengPhillips2009}, collect the estimated cointegrating vectors in $\hat H$, and let $\hat D_1$ span the orthogonal complement of $\hat H$, with $\hat D_1=I_K$ when $\hat r=K$. The combinations $\hat X_{1t}=\hat D_1'Y^0_t$ carry the common trends of the controls. The usual estimation of a cointegrating relation treats the regressors as having independent trends and has no such step.

\textit{Step 2.} I estimate the weights $(a,b)$ by dynamic least squares \citep{Saikkonen1991,StockWatson1993} in the regression
\begin{equation}
Y^1_t=a+b'Y^0_t+\sum_{j=0}^{q}\kappa_j'\begin{pmatrix}R_{t-j}\\W_{t-j}\end{pmatrix}+\sum_{j=-L}^{L}\delta_j'\Delta\hat X_{1,t+j}+\varepsilon_t,
\label{eq:dols}
\end{equation}
optionally with the constraint $b\geq0$. Two features of \eqref{eq:dols} depart from the standard regression. The policy terms are included because, by \eqref{eq:cycle}, the error of a regression of $Y^1_t$ on $Y^0_t$ alone would contain the effect of the policy, which is correlated with the innovations of the trends whenever the policy reacts to them. The leads and lags are taken only in the directions of the trends, because leads and lags of the stationary combinations of the controls would span those combinations as $L$ grows and leave their weights unidentified. The levels enter as $b'Y^0_t$, so the fit does not depend on the choice of $\hat D_1$.

\textit{Step 3.} With the weights in hand, I compute the cycle $\hat G_t=Y^1_t-\hat a-\hat b'Y^0_t$. The estimated policy terms are not subtracted, since the cycle must contain the effect of the policy.

\textit{Step 4.} I estimate a VAR($p$) with a constant in $\hat Z_t=(R_t',W_t,\hat G_t)'$, and compute the response of $\hat G_t$ to an innovation in $W_t$ in the recursive identification with $R_t$ first, normalised to a unit impact on the policy.

I call the weights that the estimator selects in the limit the pseudo-true weights $b^\ast$, and $G^\ast_t=Y^1_t-b^{\ast\prime}Y^0_t$ the cycle they produce, for which Assumption \ref{assu:A4} is required.\footnote{Formally, write the controls in the rotated coordinates $D'Y^0_t=(X_{1t}',S_t')'$ with $D=(D_1,H)$, so that $b'Y^0_t=c_1'X_{1t}+c_2'S_t$ with $c=D^{-1}b$; $c_1$ multiplies the trends and $c_2$ the stationary combinations. The pair $(a^\ast,b^\ast)$ is the limit, as $L\to\infty$, of the population minimisers of the squared error of \eqref{eq:dols} over $b\in\mathcal{B}$, with the constraint when it is imposed. It exists and is unique when the spectral density of $(\Delta X_{1t}',S_t',R_t',W_t)'$ is bounded away from singularity.}

\begin{thm}[Estimated weights\label{thm:est}]
Let Assumptions \ref{assu:A1} to \ref{assu:A5} hold with $b=b^\ast$, let the spectral density of $(\Delta X_{1t}',S_t',R_t',W_t)'$ be bounded away from singularity, let $L\to\infty$ with $L^3/T\to0$, let the regularity conditions of \citet{Saikkonen1991} hold for \eqref{eq:dols},\footnote{The vector of the changes of the trends and of the error of \eqref{eq:dols} with all leads and lags, $(\Delta X_{1t}',\varepsilon^\infty_t)'$, is a stationary linear process with absolutely summable coefficients and independent innovations with finite fourth moments, and its spectral density at frequency zero is positive definite. The coefficients $\delta_j$ of the projection of $\varepsilon^\infty_t$ on all leads and lags of $\Delta X_{1t}$ satisfy $\sqrt{T}\sum_{|j|>L}\|\delta_j\|\to0$, which makes the omitted leads and lags negligible at the rate $T^{-1/2}$ that the coefficients on the stationary regressors require.} let $(R_t',W_t,G^\ast_t,\Delta X_{1t}',S_t')'$ be strictly stationary and strong mixing with coefficients $\alpha(m)$ such that $\sum_{m\geq1}\alpha(m)^{\epsilon/(2+\epsilon)}<\infty$ and have finite moments of order $4+2\epsilon$ for some $\epsilon>0$, let $\mathbb{P}(\hat r=r)\to1$,\footnote{The information criterion of the first step satisfies this condition \citep{AznarSalvador2002,ChengPhillips2009}. The trace test of \citet{Johansen1991} with a fixed level selects the true rank with probability approaching one minus the level, and with it the conclusions of the theorem hold on an event of that probability.} and let $\hat c=D^{-1}\hat b$ and $c^\ast=D^{-1}b^\ast$.
\begin{enumerate}
\item[(a)] $\hat c_1-c_1^\ast=O_p(T^{-1})$, $\hat c_2-c_2^\ast=O_p(T^{-1/2})$ and $\hat a-a^\ast=O_p(T^{-1/2})$. Without the constraint, or when $b^\ast$ is in the interior of the constraint set, $\sqrt{T}(\hat c_2-c_2^\ast)$ is asymptotically normal.
\item[(b)] If $Z_t$ follows a stationary VAR($p$) with $G^\ast_t$ as the cycle, let $\tilde\gamma_h$ be the estimator of the response with $(a^\ast,b^\ast)$ known. Then, for every fixed $h$,
\begin{equation}
\sqrt{T}(\hat\gamma_h-\gamma_h)=\sqrt{T}(\tilde\gamma_h-\gamma_h)+J_h'\sqrt{T}(\hat c_2-c_2^\ast)+o_p(1),
\label{eq:firststep}
\end{equation}
where $J_h$ is the derivative of the population response with respect to the weights on $S_t$. Without the constraint, or when $c^\ast$ is in the interior of the constraint set, the two terms on the right are jointly asymptotically normal.
\item[(c)] If $r=K$, the second term in \eqref{eq:firststep} is absent, with or without the constraint.
\end{enumerate}
\end{thm}

The error in the weights on the trends is of order $T^{-1}$ and disappears from the distribution of the response, whereas the weights on the stationary combinations of the controls, which absorb idiosyncratic variation of the treated unit as synthetic control weights do, are estimated at the same rate as the coefficients of the VAR, so their error enters the distribution of the response.

The error of these synthetic control weights enters the response only indirectly, through the cycle on which the VAR is estimated and the policy innovation it implies. Proposition \ref{prop:J} in Supplementary Appendix \ref{sec:supp-inf} gives its coefficient, which is zero when the policy surprise $u_t$ is orthogonal to the stationary movements of the controls $S_{t+j}$ at every lead and lag, as it is when the controls never receive the policy and their idiosyncratic shocks are unrelated to the policy of the treated unit, and when the policy does not react to the past of the cycle, so that a change in the weights leaves the policy innovation unchanged.

When the conditions fail, inference must account for the first step. I compute bands with a residual bootstrap that regenerates the controls together with the policies and the cycle, and re-estimates the weights in every draw. Let $\hat H$ span the estimated cointegrating vectors of the controls and $\hat D_1$ its orthogonal complement, as in the first estimation step, and let $\hat X_{1t}=\hat D_1'Y^0_t$ and $\hat S_t=\hat H'Y^0_t$. The bootstrap proceeds as follows.
\begin{enumerate}
\item[(i)] Estimate a VAR($p$) with a constant in $\hat Q_t=(\Delta\hat X_{1t}',\hat S_t',R_t',W_t,\hat G_t)'$ and collect the centred residual vectors.
\item[(ii)] In each draw, resample the residual vectors with replacement and rebuild $(\Delta X^\ast_{1t},S^\ast_t,R^\ast_t,W^\ast_t,G^\ast_t)$ recursively from the observed initial values.
\item[(iii)] Rebuild the controls, cumulating $\Delta X^\ast_{1t}$ from the observed initial value and rotating back, $Y^{0\ast}_t=(\hat D_1,\hat H)'^{-1}(X^{\ast\prime}_{1t},S^{\ast\prime}_t)'$, and the treated outcome, $Y^{1\ast}_t=\hat a+\hat b'Y^{0\ast}_t+G^\ast_t$.
\item[(iv)] Re-estimate the weights by \eqref{eq:dols} on the bootstrap sample, with the rank and $\hat D_1$ held at their estimates, compute the bootstrap cycle, estimate the VAR($p$) in $(R^\ast_t,W^\ast_t,\hat G^\ast_t)$ and compute the response.
\item[(v)] Simulate one long path of the VAR of step (i), many times longer than the sample, apply steps (iii) and (iv) to it, and call the resulting response $\gamma^\ast_h$; shift every draw by $\hat\gamma_h-\gamma^\ast_h$.
\end{enumerate}
The bands are percentile bands of the shifted draws.

The bootstrap of the CVAR differs from that of a standard or a cointegrated VAR in two respects. In a VAR estimated by least squares, or a cointegrated VAR estimated by maximum likelihood, the parameters of the bootstrap world are the estimates themselves, so the bootstrap distribution is centred at the estimate by construction. The CVAR adds a first step: the weights that remove the common trends are a projection of the treated unit on the controls, and they determine the cycle on which the VAR is estimated. The error in the weights therefore enters the residuals of the cycle, and through them the response. A bootstrap that holds the controls at their observed values cannot reproduce this error, because it regenerates a cycle that is independent of the observed controls.\footnote{By the argument of Proposition \ref{prop:J}, the first-step term of such a bootstrap vanishes asymptotically, so the bootstrap is valid only when $J_h=0$.} Steps (ii) and (iii) regenerate the controls together with the policies and the cycle, which reproduces the dependence between them that Proposition \ref{prop:J} makes explicit. The second respect is the centring. In the bootstrap world generated by the joint VAR, the weights to which the projection converges, and the response they produce, are functions of the autocovariances of the fitted joint VAR, which differ from those of the sample, so the bootstrap distribution is centred at a value that differs from the estimate. The problem is the one that \citet{HallHorowitz1996} solve for overidentified GMM estimators, whose moment conditions do not hold in the bootstrap world \citep{Horowitz2001}, and it is close to the one that \citet{Kilian1998} addresses when he estimates the bias of impulse responses in the bootstrap world. Step (v) applies their recentring principle: it computes the response of the bootstrap world by simulation, as indirect inference computes its binding function \citep{GourierouxMonfortRenault1993}, and centres the draws at it. When the weights are known, a bootstrap of the VAR in $Z_t$ alone suffices.

With the weights known, the residual bootstrap of the VAR in $Z_t$ gives percentile bands with asymptotic coverage $1-\alpha$ by the standard argument for autoregressions (Theorem \ref{thm:resid} in Supplementary Appendix \ref{sec:supp-inf}).

The bootstrap, like the estimator, depends on the lag length of the VAR in the fourth step. The counterfactual cycle sums the idiosyncratic components of the treated unit and of the controls, so $Z_t$ is a VARMA whose autoregressive representation has infinite order and slowly declining coefficients, and the relevant theory lets the lag length grow with the sample, $p_T\to\infty$ with $p_T^3/T\to0$ \citep{LewisReinsel1985,InoueKilian2002}. Information criteria select too few lags in designs of this kind (Supplementary Appendix \ref{sec:montecarlo}), so I use one year of monthly lags, and when the sample is too short for long lags, local projections of the cycle on the policy innovation estimate the same response \citep{PlagborgMollerWolf2021,LiPlagborgMollerWolf2024}.

With the weights estimated and a lag length that grows with the sample, the following result covers both the estimator and the bootstrap of steps (i) to (iv).

\begin{thm}[Estimated weights, long autoregressions and the bootstrap\label{thm:joint}]
Let the conditions of Theorem \ref{thm:est}, except the finite order of the VAR in part (b), hold, with $c^\ast$ in the interior of the constraint set when the constraint is imposed.\footnote{When a weight is on the boundary, the limit in part (a) is the projection of a normal vector on a cone, and the bootstrap is inconsistent \citep{Andrews2000}.} Let $Q_t=(\Delta X_{1t}',S_t',R_t',W_t,G^\ast_t)'$ have the representation $Q_t=\mu+\sum_{l\geq1}\Phi_lQ_{t-l}+\epsilon_t$, with $\det(I-\sum_{l\geq1}\Phi_lz^l)\neq0$ for $|z|\leq1$, $\sum_{l\geq1}l\|\Phi_l\|<\infty$, and i.i.d. innovations $\epsilon_t$ with finite fourth moments and positive definite covariance, and let $Z_t$ have a representation with the same properties. Let the lag length of the VARs in $\hat Z_t$ and in $\hat Q_t$ be $p=p_T$, with $p_T\to\infty$, $p_T^3/T\to0$ and $\sqrt{T}\sum_{l>p_T}\|\Phi_l\|\to0$ and the same for the coefficients of $Z_t$, and let the number of leads and lags $L$ and of lags of the policies $q$ in \eqref{eq:dols} be at most $p_T$. Then, for every fixed $h$:
\begin{enumerate}
\item[(a)] the expansion \eqref{eq:firststep} of Theorem \ref{thm:est}(b) holds, with $\gamma_h$ the response of the VAR($\infty$) representation of $Z_t$ and $J_h$ its derivative with respect to the weights on $S_t$, and the two terms on its right are jointly asymptotically normal;
\item[(b)] the distribution of $\sqrt{T}(\hat\gamma^\ast_h-\gamma^\ast_h)$ under the bootstrap of steps (i) to (iv), where $\gamma^\ast_h$ is the response in the bootstrap world, converges in probability to the limit of the distribution of $\sqrt{T}(\hat\gamma_h-\gamma_h)$, and the percentile band of the draws shifted in step (v) has asymptotic coverage $1-\alpha$ as the length of the long path grows with $T$.
\end{enumerate}
\end{thm}

Part (b) holds whether or not $J_h$ is zero,\footnote{Theorems \ref{thm:resid} and \ref{thm:joint} assume i.i.d. innovations. Under conditional heteroskedasticity of unknown form, resampling residual vectors one at a time does not reproduce the fourth moments that enter the variance of the innovation covariance, and the residual-based moving-block bootstrap of \citet{bruggemann2016inference} gives valid inference for the responses of a VAR at every horizon, including impact; it can replace the resampling of step (ii). The wild bootstrap reproduces the variance of the slope coefficients \citep{goncalveskilian2004}, and Supplementary Appendix \ref{sec:supp-inf} shows that it does not reproduce that of the innovation covariance.} so the bands account for the estimation of the weights in every case covered by Proposition \ref{prop:J}. When $r=K$ there are no stationary combinations of the controls, $S_t$ drops out of $Q_t$, and the first-step term is absent by Theorem \ref{thm:est}(c). In the simulations of Supplementary Appendix \ref{sec:montecarlo}, the joint bootstrap gives bands with coverage close to that obtained with the weights known at every horizon up to a year. The simulations also show that a wild bootstrap with Rademacher weights is not a valid alternative at impact (Supplementary Appendix \ref{sec:supp-inf}).\footnote{Theorem \ref{thm:wild} in Supplementary Appendix \ref{sec:supp-inf} shows that, with Rademacher weights, the bootstrap covariance of the residuals differs from the estimate by a term of smaller order than its sampling error, so the bands at impact are too narrow.}

\section{Testing the design\label{sec:test}}

The CVAR rests on Assumption \ref{assu:A2}: the treated unit shares the stochastic trends of the controls, so that the counterfactual cycle is stationary. When the assumption fails, the treated unit carries a trend of its own, the cycle has a unit root, and the responses of Section \ref{sec:theorems} are spurious. Testing the assumption is a test of the null hypothesis of no cointegration between the treated unit and its controls. The standard tests for this null estimate the cointegrating vector first and test the residual for a unit root \citep{EngleGranger1987,PhillipsOuliaris1990}, possibly after GLS detrending \citep{PerronRodriguez2016}, and the stationarity tests reverse the null \citep{Shin1994,HarveyThiele2021}. Three features of the CVAR make their tables inapplicable: the weights are estimated together with the policy terms, they may be constrained, and the controls may be cointegrated among themselves. The simulations of Supplementary Appendix \ref{sec:montecarlo} show that the residual-based tests then reject a true null far more often than their nominal level. The power of tests of this null has been studied by \citet{Pesavento2004}, \citet{ElliottJanssonPesavento2005} and \citet{ElliottPesavento2009}, who show that no uniformly most powerful test exists and derive point-optimal tests and power bounds. I adapt the point-optimal approach to the CVAR.

Write the treated unit as
\begin{equation}
Y^1_t=\mu+b'Y^0_t+\delta(L)'P_t+u_t,\qquad u_t=\rho_Tu_{t-1}+v_t,\qquad \rho_T=1+c/T,
\label{eq:testmodel}
\end{equation}
where $P_t$ collects the current value and $q$ lags of the policies and $v_t$ is a stationary linear process with long-run variance $\omega^2$. The null hypothesis $c=0$ is that the treated unit has a stochastic trend of its own, and the CVAR is not valid; the alternatives $c<0$ are that the cycle is stationary. For $\alpha\in(0,1]$ let $S(\alpha)$ be the sum of squared residuals of the regression of $Y^1_t-\alpha Y^1_{t-1}$ on the quasi-differenced constant, controls and policy terms, with the first observation in levels, and let $\hat\omega^2$ be the autoregressive spectral estimator of $\omega^2$ from the augmented Dickey-Fuller regression on the residuals of the regression with $\alpha=\bar\alpha$, with the lag order chosen by the modified information criterion of \citet{NgPerron2001}. For a fixed $\bar c<0$ and $\bar\alpha=1+\bar c/T$, the test rejects for small values of
\begin{equation}
P_T=\frac{S(\bar\alpha)-\bar\alpha\,S(1)}{\hat\omega^2}.
\label{eq:PT}
\end{equation}
The regression is unconstrained: the restriction $b\geq0$ belongs to the estimator and plays no role in the test. The weights are concentrated out of the quasi-differenced regression under both hypotheses, which is the flat-weight limit of the family of tests of \citet{ElliottPesavento2009}. The difference with the residual-based tests is that the controls enter the regression quasi-differenced at the same $\alpha$ as the treated unit, so that their coefficient is identified mainly from the co-movement of the changes, under the null and the alternative alike.

The properties of the test rest on Assumption \ref{assu:T} in Supplementary Appendix \ref{sec:supp-testboot}. It requires the policies to be stationary under both hypotheses, the shocks specific to the treated unit to be unrelated to the changes of the controls and of the policies at all leads and lags, which excludes a reaction of the policy to the past of the component specific to the treated unit, and a small initial condition.

\begin{thm}[Asymptotic equivalence with the known-weights test\label{thm:po}]
Let Assumption \ref{assu:T} hold and let $P^u_T$ be the statistic \eqref{eq:PT} computed from $u_t$ itself, with only the constant as regressor. Under the null and under local alternatives $c$ fixed, $P_T=P^u_T+o_p(1)$. Consequently:
\begin{enumerate}
\item[(a)] $P_T\Rightarrow\bar c^2\int_0^1W_c(s)^2ds-\bar c\,W_c(1)^2$, where $W_c$ is an Ornstein-Uhlenbeck process with parameter $c$ and $W_0$ a standard Brownian motion; under the null the limit does not depend on the number of controls $K$, on the number of their trends $r$, or on the policy terms, and the critical values are those of \citet{ERS1996} for the case with a constant.\footnote{With $\bar c=-7$, the choice for which the envelope has power one half at the 5 percent level \citep[p.~821]{ERS1996}, the asymptotic critical values at 1, 2.5, 5 and 10 percent are 1.99, 2.55, 3.26 and 4.48 \citep[Table I]{ERS1996}.}
\item[(b)] The asymptotic local power of the test equals that of the point-optimal test computed from the counterfactual cycle with known weights and policy coefficients. At $c=\bar c$ it equals the asymptotic Gaussian power envelope, which bounds the power of any test of a unit root in the cycle \citep{ERS1996}; since the test does not use the weights, it attains at $\bar c$ the envelope of the tests that do not know them.
\end{enumerate}
\end{thm}

Part (a) says that estimating the weights inside the test leaves its null distribution unchanged in large samples, which is what distinguishes it from the residual-based tests of the null of no cointegration. Those tests estimate the weights by a regression in levels, and under the null the regression is spurious: least squares chooses the weights that make the residual as small as possible, which also makes it look closer to stationarity than the true deviation of the treated unit from its controls, and the more so the more controls the regression uses. The critical values of the residual-based tests therefore depend on the number of controls in the regression, and a separate table is needed for each number of controls \citep{PhillipsOuliaris1990,PerronRodriguez2016}. In the point-optimal test the controls enter quasi-differenced at $\bar\alpha$, close to one, so their coefficients are identified mainly from the co-movement of the changes, where there is no spurious relation to fit, and the estimated weights cannot make a random walk look stationary. The statistic then behaves as if it were computed from the true deviation $u_t$, and its critical values are those of the univariate test of \citet{ERS1996} with a constant, whatever the number of controls, of their trends or of the policy terms. Part (b) means that, under Assumption \ref{assu:T}, not knowing the weights costs no power in large samples: the test is as powerful as if the counterfactual cycle were observed. In finite samples the estimation of the weights and of the policy coefficients shifts the null distribution slightly, and the long-run variance of a persistent cycle is estimated with error. I therefore compute the p-value with a sieve bootstrap under the null, which rebuilds the treated unit by cumulating its fitted changes and resampled autoregressive errors, holds the controls and the policy terms at their observed values, and has asymptotic size equal to its nominal level (Proposition \ref{prop:sieve} in Supplementary Appendix \ref{sec:supp-testboot}).

The test decides whether the more efficient estimator in levels can be used. Estimated on the changes of the cycle, the CVAR identifies the effect whether or not the treated unit has a trend of its own, so a non-rejection costs efficiency and leaves the identification unchanged. A rejection supports Assumption \ref{assu:A2}, and the CVAR can then be estimated in levels with the weights of \eqref{eq:dols}. The test is consistent, so when the cycle is stationary it selects the estimator in levels with probability approaching one.\footnote{When the cycle has a trend of its own, the test selects the estimator in levels with probability equal to its size, so the size of the test also sets the protection of the estimator in levels against a stochastic trend in the cycle.} The power of any test of this null is low against a cycle that returns to its mean slowly relative to the length of the sample, because such a cycle is hard to tell apart from a random walk. Supplementary Appendix \ref{sec:screen} applies the test to the states of the United States, each compared with its neighbouring states, and reports its power in panels built like those data.

\section{A treated unit with a trend of its own\label{sec:ext}}

Assumption \ref{assu:A2} asks the weighted controls to reproduce every stochastic trend of the treated unit. When the test of Section \ref{sec:test} does not reject, the treated unit may carry, alongside the trends it shares with the controls, a slow-moving component of its own that no combination of the controls reproduces. In the application, the national cycle and the price of oil move employment in every state, and Alaska can also follow a path of its own over four decades, through the rise and decline of oil production on the North Slope. The controls still remove the shared trends, and the CVAR can be estimated on the changes of the cycle, which is the version I use in the applications.

\begin{assumption}[Common trends and a trend of the treated unit\label{assu:A2d}]
There is an $r$-dimensional process $F_t$, integrated of order one with stationary first difference, a $K\times r$ matrix $\Lambda$ of rank $r\leq K$ and a known vector $b\in\mathbb{R}^K$ such that
\begin{equation*}
Y^0_t=\Lambda F_t+v_t,\qquad Y^1_t(0)=\lambda'F_t+e_t,\qquad \lambda=\Lambda'b,
\end{equation*}
where $v_t$ is stationary and $e_t$ is integrated of order at most one, with stationary first difference.
\end{assumption}

Assumption \ref{assu:A2d} keeps the condition that the weighted controls share the trends of the treated unit and drops the stationarity of $e_t$. The counterfactual cycle $G^0_t=e_t-b'v_t$ can then be integrated of order one, and its change $\Delta G^0_t$ is stationary. The weights are set by the researcher or estimated. When $e_t$ has a stochastic trend, a regression of the treated outcome on the controls in levels has no well-defined limit, and a regression in changes estimates the projection of $\Delta Y^1_t$ on $\Delta Y^0_t$, in which the changes of $v_t$ act as measurement error in the regressors and move the weights away from those that remove the trends. In the application to Alaska I estimate the weights from the changes. Under Assumptions \ref{assu:A1}, \ref{assu:A2d} and \ref{assu:A3}, the decomposition \eqref{eq:cycle} holds unchanged, because it uses only $\lambda=\Lambda'b$.

Let $Z^\Delta_t=(R_t',W_t,\Delta G_t)'$, let $u_t$ be the innovation of the policy in the recursive VAR in $Z^\Delta_t$, defined as in \eqref{eq:innovation} with the past of $\Delta G_t$ in place of that of $G_t$, let $\theta_h$ and $\psi_k$ be defined as in \eqref{eq:estimand} and \eqref{eq:decomp} with this innovation, and let
\begin{equation}
\Gamma_h=\frac{\E[(G_{t+h}-G_{t-1})u_t]}{\E[u_t^2]}
\label{eq:cumcycle}
\end{equation}
be the response of the cycle, cumulated from the month before the surprise.

\begin{assumption}[Exogeneity with respect to the changes of the counterfactual cycle\label{assu:A4d}]
(i) $(R_t',W_t,\Delta G^0_t,\Delta F_t',\Delta v_t')'$ is covariance stationary with finite fourth moments, and $\E[u_t^2]>0$. (ii) $\E[(G^0_{t+h}-G^0_{t-1})u_t]=0$ for all $h\geq0$.
\end{assumption}

Part (i) plays the role of Assumption \ref{assu:A5} for the changes of the cycle. Part (ii) plays the role of Assumption \ref{assu:A4}: it allows the policy to react to the past changes of the cycle, which the VAR conditions on, and requires that the surprise does not predict how the component specific to the treated unit moves from the month before the surprise onward. In the application, it requires that a surprise in the dividend carries no news about the path that employment specific to Alaska would have followed without it.

\begin{thm}[Response of the cumulated cycle\label{thm:diff}]
Let Assumptions \ref{assu:A1}, \ref{assu:A2d}, \ref{assu:A3} and \ref{assu:A4d}(i) hold.
\begin{enumerate}
\item[(a)] For every $h\geq0$,
\begin{equation*}
\Gamma_h=\theta_h+\frac{\E[(G^0_{t+h}-G^0_{t-1})u_t]}{\E[u_t^2]}.
\end{equation*}
If $Z^\Delta_t$ follows a VAR, possibly of infinite order, $\Gamma_h=\sum_{j=0}^{h}\gamma^\Delta_j$, where $\gamma^\Delta_j$ is the response of $\Delta G_t$ at horizon $j$ to a unit innovation in $W_t$ in the recursive identification with $R_t$ first, $W_t$ second and $\Delta G_t$ last.
\item[(b)] If Assumption \ref{assu:A4d}(ii) also holds, $\Gamma_h=\theta_h$ for all $h\geq0$, and Theorem \ref{thm:direct} holds with $\Gamma_h$ in place of $\gamma_h$.
\item[(c)] If the imposed weights $b$ do not satisfy $\lambda=\Lambda'b$, then for every $h\geq0$
\begin{equation*}
\Gamma_h=\theta_h+(\lambda-\Lambda'b)'\frac{\E[(F_{t+h}-F_{t-1})u_t]}{\E[u_t^2]}+\frac{\E[(G^0_{t+h}-G^0_{t-1})u_t]}{\E[u_t^2]},
\end{equation*}
with $G^0_t=e_t-b'v_t$.
\end{enumerate}
\end{thm}

Theorem \ref{thm:diff} is the counterpart of Theorems \ref{thm:cycle} and \ref{thm:ident} for a cycle with a stochastic trend. The controls remove the common trends exactly as before when the imposed weights are the right ones, and part (c) gives the bias when they are not: the common-trend term of \eqref{eq:bias}, with the difference between the loadings of the treated unit and those of the weighted controls in place of the loadings of the treated unit. The VAR is estimated on the changes of the cycle, and the assumption that closes the identification concerns the changes of the counterfactual cycle. The response of a VAR without controls in Theorem \ref{thm:var} is written in cumulated changes, so the comparison between the two estimators carries over: the response of the VAR without controls in $(R_t',W_t,\Delta Y^1_t)'$ equals $\Gamma_h$ plus the common-trend term of \eqref{eq:bias}, up to the difference between the two innovations. Proposition \ref{prop:placebo} and the bounds of Proposition \ref{prop:bounds} are also stated in cumulated changes, and they hold with the VAR in $(R_t',W_t,\Delta Y^1_t,\Delta C_t)'$.

With known weights, the cycle $G_t=Y^1_t-b'Y^0_t$ is observed and there is no first step. The vector $Z^\Delta_t$ is stationary, so Theorem \ref{thm:resid} applies to the responses $\gamma^\Delta_j$ and, because $\Gamma_h$ is a finite sum of them, to $\Gamma_h$ at every fixed horizon: the residual bootstrap of the VAR in $Z^\Delta_t$, with the responses cumulated in each draw, gives percentile bands with asymptotic coverage $1-\alpha$. The change of the cycle is far less persistent than the cycle in levels, so shorter orders suffice: in the applications I use twelve lags or the smallest order at which the residuals show no autocorrelation, with longer orders as a robustness check.\footnote{The bands treat the selected order as fixed, as is standard in applied work with VARs.}

When the weights are estimated, the regression runs on the changes of the series, where the idiosyncratic movements of the controls are as large as their common movements, and the estimated weights inherit them.

\begin{prop}[Weights estimated in changes\label{prop:wchanges}]
Let Assumptions \ref{assu:A2d} and \ref{assu:A4d}(i) hold, let the policies affect $\Delta Y^1_t$ only through the current and lagged values that the regression projects out, and, after that projection, let $\Delta F_t$, $\Delta v_t$ and $\Delta e_t$ be mutually uncorrelated, with $\operatorname{Var}(\Delta F_t)=\Sigma_F$ and $\operatorname{Var}(\Delta v_t)=D$ positive definite. The least-squares weights of the regression of $\Delta Y^1_t$ on $\Delta Y^0_t$ converge in probability to
\begin{equation*}
w^\ast=(\Lambda\Sigma_F\Lambda'+D)^{-1}\Lambda\Sigma_F\lambda,
\end{equation*}
and the response of the cumulated cycle built with $w^\ast$ is given by part (c) of Theorem \ref{thm:diff} with $b=w^\ast$ and
\begin{equation*}
\lambda-\Lambda'w^\ast=\big(I_r+\Lambda'D^{-1}\Lambda\,\Sigma_F\big)^{-1}\lambda .
\end{equation*}
\end{prop}

The weighted controls reproduce the part of the common movements of the treated unit that their changes reveal above their own noise, and the rest stays in the cycle. With a single common trend, the weights are proportional to $D^{-1}\Lambda$, so their composition favours the controls that follow the trend most closely relative to their noise, and their scale is shrunk by the factor $s/(1+s)$, with $s=\Sigma_F\Lambda'D^{-1}\Lambda$; the loading left in the cycle is $\lambda/(1+s)$, and the term it adds to the response equals $1/s$ times the response of the weighted controls in Proposition \ref{prop:placebo}, with the same sign. The term vanishes as $\Lambda'D^{-1}\Lambda$ grows, when the controls are many and their idiosyncratic movements independent or when those movements are small. It is the counterpart in changes of the bias of synthetic control with an imperfect fit before the treatment \citep{FermanPinto2021}, which disappears as the number of controls grows \citep{Ferman2021}. Consistent estimation of the weights in changes with a small set of controls is an open problem, and I leave it to future work. The size of the term can be measured. With a large panel of untreated units, the principal components of their changes estimate the common movements \citep{Bai2003}, and the ratio of the loading of the weighted controls to that of the treated unit on each component gives the share of that movement the weights reproduce. Equal weights avoid the estimation noise, and the same ratio checks whether they reproduce the loadings of the treated unit.

\section{The Alaska Permanent Fund Dividend\label{sec:alaska}}

I apply the CVAR to three policies. The first is the dividend of the Alaska Permanent Fund, whose size changes from year to year with the income of the fund. The second, extended unemployment benefits, switches on in recessions, and it is the standard example of a policy that reacts to the cycle it is meant to affect. The third, the Low Carbon Fuel Standard of California, is a cost that changes every month and that fuel suppliers can pass on to consumers. In all three I estimate the CVAR in changes; the test of Section \ref{sec:test} does not reject, so the data do not support the estimator in levels. In the first I estimate the weights on the controls, and in the other two I give them equal weights.

Since 1982 the State of Alaska has paid every resident a yearly dividend from the Alaska Permanent Fund, a sovereign wealth fund built from the royalties of oil production. A public formula sets the total distribution at about a tenth of the income that the fund earned over the last five fiscal years, the total is divided equally among the eligible residents, and the payment arrives in a single transfer in October.\footnote{The first dividend was paid in June 1982. From 2016 the legislature has set the amount below the formula. In 2008 a one-time energy rebate of \$1,200 was paid together with the dividend.} The dividend per person has ranged from a few hundred to more than two thousand dollars.

The dividend is the largest permanent and universal cash transfer in the United States, and it has become the reference case for the effects of a universal basic income. Its effect on spending is well documented. Households raise their spending on nondurables in the quarter in which the payment arrives and do not respond to news about its size before then \citep{Kueng2018}, a response that an earlier study based on survey income had not detected \citep{Hsieh2003}. Its effect on the labour market is less settled, because two channels push in opposite directions. A cash transfer reduces the labour supply of the household that receives it \citep{Cesarini2017}, while the spending it finances raises the demand for labour in the sectors that sell to local customers. Comparing Alaska with a synthetic control built from other states, \citet{JonesMarinescu2022} find that the introduction of the dividend left the employment rate unchanged and raised part-time work, and they attribute the absence of an employment loss to the demand channel offsetting the supply channel.

Two features of the dividend make the comparison of Alaska with other states informative and a time-series analysis of Alaska alone misleading. The first is that only Alaska pays it, so the other states go through the same national cycle without it. The second is that its size follows the returns of a fund invested in a diversified portfolio of financial assets, which move with the stock market and the national business cycle. The largest dividends came after years of high returns, and those years preceded the recessions of 2001 and 2008. A VAR on Alaska data alone attributes to the dividend the national downturns that followed (Theorem \ref{thm:var}), while the CVAR compares Alaska with states that went through the same cycle without the dividend. The synthetic control of \citet{JonesMarinescu2022} rests on a single event, the introduction of the dividend, with five years of annual data before it, and estimates the average effect of the programme over the following decades. The CVAR uses the variation in the size of more than thirty dividends, estimates the weights on the controls over the whole sample, payment years included, and traces the response of the labour market month by month.

I use monthly data from January 1976 to December 2019, which end before the pandemic. The outcomes are the employment-population ratio, the unemployment rate and the participation rate from the Local Area Unemployment Statistics, and the logarithm of payroll employment in total, in leisure and hospitality and in retail trade from the Current Employment Statistics, which start in January 1990.\footnote{Seasonally adjusted series from the Bureau of Labor Statistics: \texttt{LASST020000000000007}, \texttt{LASST020000000000003} and \texttt{LASST020000000000008} for Alaska, \texttt{SMS02000000000000001}, \texttt{SMS02000007000000001} and \texttt{SMS02000004200000001}, and the analogous codes for the controls.} The policy $W_t$ is the dividend paid in the month, in thousands of 2024 dollars per person, and zero in the other months.\footnote{Amounts from the Permanent Fund Dividend Division, Summary of Applications and Payments, including the 2008 energy rebate, deflated by the consumer price index for all urban consumers (FRED series \texttt{CPIAUCSL}).} The controls are the six states that form the synthetic Alaska of \citet{JonesMarinescu2022} for the employment rate: Utah, Wyoming, Washington, Nevada, Montana and Minnesota. Oil moves both the economy of Alaska and the income of its government, so the VAR includes the monthly change in the logarithm of the price of West Texas Intermediate crude as a second policy $R_t$, ordered first.\footnote{FRED series \texttt{WTISPLC}, spot price of West Texas Intermediate crude oil.}

The surprise in the dividend must also carry no news about movements of employment specific to Alaska (Assumption \ref{assu:A4}). Until 2015 the formula set the dividend from the income of a fund whose diversified portfolio the economy of Alaska does not move. From 2016 the legislature set it with an eye on a budget that depends on oil revenue, a channel that ordering the price of oil first removes from the innovation.

Table \ref{tab:ak-weights} reports the test of the design and the weights. The outcomes of Alaska behave as integrated series, and the test does not reject a stochastic trend specific to Alaska for any outcome, so I estimate the CVAR in changes (Theorem \ref{thm:diff}). I estimate the weights from the regression of the changes of each Alaska outcome on the changes of the six states, with the dividend and the price of oil as regressors, constrained to be non-negative. The weight spreads over several states for every outcome, and no state receives most of it. The bands account for the estimation of the weights, which I repeat in each bootstrap draw on a moving-block resample of the regression. Table \ref{tab:ak-weights} also reports how much of the main common movement of Alaska the weighted donors reproduce, measured on the first principal component of the changes of the other states (Proposition \ref{prop:wchanges}). The estimated weights fall short of it, as the proposition predicts. Equal weights overshoot it for most outcomes, because Alaska follows the common movement less closely than the average of the six states, so the two weightings miss it in opposite directions.

\begin{table}[htbp]
\centering
\caption{Weights on the donor states, reproduction of the common movement and tests of the design\label{tab:ak-weights}}
\footnotesize
\setlength{\tabcolsep}{4pt}
\begin{tabular}{lcccccc}
\toprule
 & Employment rate & Unemployment & Participation & Payroll & Leisure & Retail \\
\midrule
UT & 0.16 & 0.00 & 0.10 & 0.03 & 0.13 & 0.14 \\
WY & 0.12 & 0.14 & 0.12 & 0.11 & 0.12 & 0.07 \\
WA & 0.10 & 0.11 & 0.08 & 0.05 & 0.30 & 0.11 \\
NV & 0.02 & 0.00 & 0.04 & 0.08 & 0.21 & 0.06 \\
MT & 0.01 & 0.24 & 0.03 & 0.00 & 0.14 & 0.09 \\
MN & 0.16 & 0.00 & 0.04 & 0.04 & 0.00 & 0.06 \\
\midrule
Sum & 0.57 & 0.49 & 0.40 & 0.30 & 0.90 & 0.52 \\
\midrule
Common share & 0.42 & 0.53 & 0.20 & 0.34 & 0.17 & 0.27 \\
Reproduced, estimated & 0.98 & 0.86 & 0.57 & 0.61 & 0.58 & 0.63 \\
Reproduced, equal & 1.80 & 1.94 & 1.45 & 1.90 & 0.68 & 1.23 \\
\midrule
$p$, design & 0.42 & 0.39 & 0.55 & 0.35 & 0.68 & 0.59 \\
$p$, Alaska & 0.57 & 0.32 & 0.56 & 0.58 & 0.50 & 0.69 \\
\bottomrule
\end{tabular}
\par\smallskip\parbox{0.95\textwidth}{\footnotesize Columns: outcomes of Alaska. Non-negative weights from the regression of the change in the Alaska outcome on the changes of the six donor states, with a constant and the current value and twelve lags of the dividend and of the change in the oil price, January 1976 (payroll series January 1990) to December 2019. Common share: share of the variance of the changes of the 50 other states explained by their first principal component, the main common movement. Reproduced: ratio of the loading of the weighted donors (estimated or equal weights) to that of Alaska on that component (Proposition \ref{prop:wchanges}); one means that the weights leave none of it in the cycle. $p$, design: test of Section \ref{sec:test} with estimated weights (null: a stochastic trend specific to Alaska). $p$, Alaska: the same test on the Alaska outcome without controls.}
\end{table}

\begin{figure}[htbp]
\centering\safegraphic[width=\textwidth]{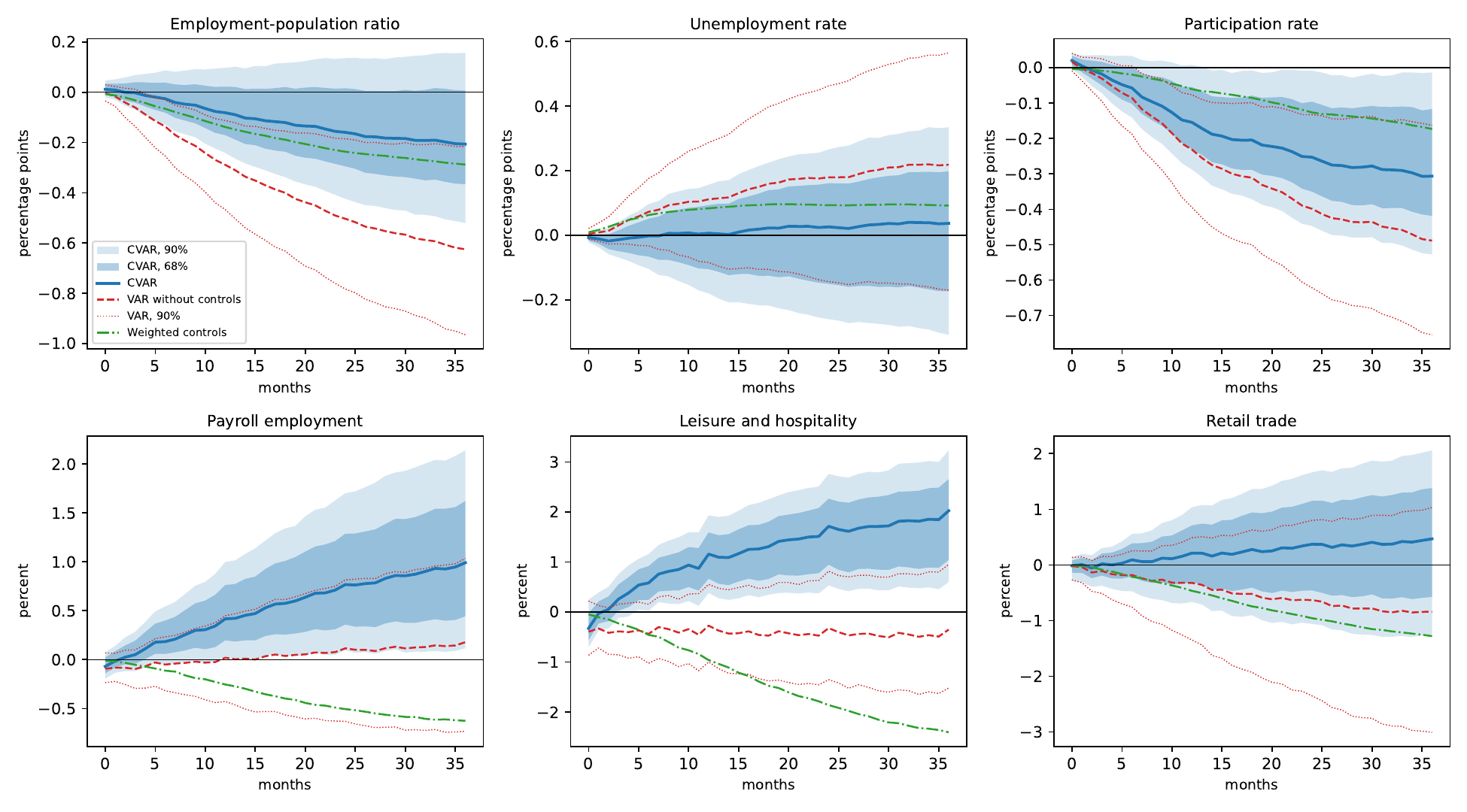}
\caption{Alaska Permanent Fund Dividend: cumulated response to a dividend larger by \$1,000 per person. CVAR with 68 and 90 percent bands, VAR without controls with its 90 percent band, and response of the weighted controls in the VAR with Alaska and the weighted controls. Rates in percentage points, employment in percent. Monthly, January 1976 (payroll series January 1990) to December 2019.\label{fig:ak-main}}
\end{figure}

Figure \ref{fig:ak-main} and Table \ref{tab:ak-main} report the responses to a dividend larger by \$1,000 per person. The CVAR finds no fall in the employment rate and no rise in unemployment at any horizon. The VAR without controls finds a large and significant fall in the employment rate, and in the VAR that contains both Alaska and the weighted controls, the controls, which never receive the dividend, fall with it (Proposition \ref{prop:placebo}). The loss of employment that the VAR attributes to the dividend is the national downturn that followed the largest dividends. Participation falls over the two years after the payment, as an income effect on labour supply predicts.\footnote{Supplementary Appendix \ref{sec:app-eb} re-estimates the main specification on the unadjusted series with month effects, with month effects interacted with a linear trend, and with month effects by decade. The absence of a fall in the employment rate and the rise in leisure and hospitality employment hold in every version, while the responses of participation and of total payroll employment depend on how the seasonal pattern of Alaska is removed.}

Payroll employment rises, and the rise is concentrated in leisure and hospitality, the sector that sells to local customers and where the spending of the dividend lands, while retail trade does not move significantly. For these outcomes the common trends bias the VAR in the opposite direction. After a large dividend the weighted controls lose jobs, and the VAR without controls, which keeps that loss in its response, finds no effect. The common trends therefore produce in the VAR a loss of employment where the dividend has none, and hide the gains in the sector where it has one.

\begin{table}[htbp]
\centering
\caption{Response to a dividend of \$1,000 per person\label{tab:ak-main}}
\footnotesize
\begin{tabular}{lccccc}
\toprule
Months & CVAR & Equal & Impulse causal & VAR without & Weighted \\
 & & weights & effect & controls & controls \\
\midrule
\multicolumn{6}{l}{Employment-population ratio} \\
3 & $-$0.00 & 0.04 & $-$0.00 & $-$0.06 & $-$0.03 \\
 & [$-$0.07, 0.07] & & & [$-$0.14, 0.01] & [$-$0.06, 0.00] \\
12 & $-$0.08 & 0.06 & $-$0.10 & $-$0.29$^{*}$ & $-$0.14$^{*}$ \\
 & [$-$0.24, 0.10] & & & [$-$0.47, $-$0.11] & [$-$0.25, $-$0.02] \\
24 & $-$0.16 & 0.07 & $-$0.14 & $-$0.50$^{*}$ & $-$0.24$^{*}$ \\
 & [$-$0.42, 0.14] & & & [$-$0.78, $-$0.18] & [$-$0.44, $-$0.02] \\
\addlinespace
\multicolumn{6}{l}{Unemployment rate} \\
3 & $-$0.01 & $-$0.04 & $-$0.01 & 0.03 & 0.04$^{*}$ \\
 & [$-$0.07, 0.04] & & & [$-$0.03, 0.09] & [0.00, 0.06] \\
12 & 0.01 & $-$0.10 & 0.01 & 0.11 & 0.08 \\
 & [$-$0.18, 0.16] & & & [$-$0.08, 0.29] & [$-$0.02, 0.17] \\
24 & 0.03 & $-$0.15 & 0.03 & 0.18 & 0.09 \\
 & [$-$0.26, 0.27] & & & [$-$0.14, 0.46] & [$-$0.08, 0.25] \\
\addlinespace
\multicolumn{6}{l}{Participation rate} \\
3 & $-$0.02 & 0.02 & $-$0.02 & $-$0.03 & $-$0.01 \\
 & [$-$0.07, 0.04] & & & [$-$0.09, 0.03] & [$-$0.02, 0.01] \\
12 & $-$0.15 & 0.00 & $-$0.19 & $-$0.23$^{*}$ & $-$0.06$^{*}$ \\
 & [$-$0.29, 0.01] & & & [$-$0.39, $-$0.08] & [$-$0.10, $-$0.02] \\
24 & $-$0.25$^{*}$ & $-$0.02 & $-$0.19 & $-$0.40$^{*}$ & $-$0.13$^{*}$ \\
 & [$-$0.45, $-$0.01] & & & [$-$0.62, $-$0.13] & [$-$0.19, $-$0.05] \\
\addlinespace
\multicolumn{6}{l}{Payroll employment} \\
3 & 0.05 & 0.29 & 0.05 & $-$0.10 & $-$0.05$^{*}$ \\
 & [$-$0.12, 0.28] & & & [$-$0.30, 0.10] & [$-$0.10, $-$0.00] \\
12 & 0.42$^{*}$ & 1.17 & 0.56 & 0.02 & $-$0.25$^{*}$ \\
 & [0.03, 1.00] & & & [$-$0.43, 0.45] & [$-$0.42, $-$0.05] \\
24 & 0.77$^{*}$ & 2.24 & 0.65 & 0.12 & $-$0.51$^{*}$ \\
 & [0.10, 1.70] & & & [$-$0.63, 0.82] & [$-$0.84, $-$0.10] \\
\addlinespace
\multicolumn{6}{l}{Leisure and hospitality} \\
3 & 0.26 & 0.24 & 0.24 & $-$0.39 & $-$0.23$^{*}$ \\
 & [$-$0.12, 0.76] & & & [$-$0.86, 0.16] & [$-$0.44, $-$0.06] \\
12 & 1.16$^{*}$ & 1.11 & 1.69 & $-$0.27 & $-$0.96$^{*}$ \\
 & [0.38, 1.93] & & & [$-$1.00, 0.55] & [$-$1.59, $-$0.33] \\
24 & 1.71$^{*}$ & 1.63 & 1.21 & $-$0.32 & $-$1.86$^{*}$ \\
 & [0.53, 2.76] & & & [$-$1.35, 0.82] & [$-$3.01, $-$0.65] \\
\addlinespace
\multicolumn{6}{l}{Retail trade} \\
3 & 0.02 & 0.13 & 0.02 & $-$0.11 & $-$0.09 \\
 & [$-$0.37, 0.30] & & & [$-$0.51, 0.15] & [$-$0.19, 0.02] \\
12 & 0.21 & 0.64 & 0.25 & $-$0.32 & $-$0.47$^{*}$ \\
 & [$-$0.67, 1.04] & & & [$-$1.35, 0.49] & [$-$0.79, $-$0.15] \\
24 & 0.37 & 1.08 & 0.29 & $-$0.63 & $-$0.96$^{*}$ \\
 & [$-$1.06, 1.66] & & & [$-$2.35, 0.80] & [$-$1.59, $-$0.36] \\
\bottomrule
\end{tabular}
\par\smallskip\parbox{0.95\textwidth}{\footnotesize Cumulated response to an innovation of \$1,000 (2024 dollars) per person in the dividend, in percentage points (rates) or percent (employment), with 90 percent bands below the estimates. CVAR: VAR in the change in the oil price, the dividend and the change in the cycle, with the weights of Table \ref{tab:ak-weights}; bands from a residual bootstrap that re-estimates the weights in each draw on a moving-block resample. Equal weights: the same with weights $1/6$. Impulse causal effect: Theorem \ref{thm:direct}, later dividends held fixed. VAR without controls: the same with the change in the Alaska outcome. Weighted controls: response in the VAR with Alaska and the weighted controls (Proposition \ref{prop:placebo}). $^{*}$: the 90 percent band excludes zero.}
\end{table}

Because the formula averages the income of five years, a dividend larger than predicted is followed by larger dividends in the next years, and the responses in Table \ref{tab:ak-main} include the effect of those later payments. The impulse causal effect of Theorem \ref{thm:direct} holds the later dividends fixed and isolates the effect of a single payment. Employment in leisure and hospitality peaks within a year of the payment and settles at a lower level, and payroll employment settles after a year.

The estimated weights leave part of the common movement of Alaska in the cycle, and Table \ref{tab:ak-adjusted} in the Supplementary Appendix removes the term that Proposition \ref{prop:wchanges} attributes to it, a multiple of the response of the weighted donors with the same sign. The adjustment brings the responses with estimated weights and with equal weights to the same values, although the two weightings miss the common movement in opposite directions. The term works against the main results. After the adjustment the employment rate still does not fall, and the gains in payroll employment and in leisure and hospitality are larger and remain significant, while the fall in participation loses its significance.

The remaining threat to identification is a component of the Alaska economy that the controls do not share and that moves with the dividend. Table \ref{tab:ak-bounds} in the Supplementary Appendix reports the bounds of Proposition \ref{prop:bounds}. The effects on participation, payroll employment and leisure and hospitality remain significant for biases specific to Alaska that are a substantial fraction of the bias from the common trends, which the response of the controls measures and which, for employment, is itself large. The results also hold when the sample ends in 2015, before the legislature set the amounts, and the point estimates are similar, with wider bands, when it ends in 2007, before the dividend, the rebate and the recession of 2008 (Supplementary Appendix \ref{sec:app-eb}).

\section{Extended unemployment benefits\label{sec:ebsum}}

The second application concerns a policy that reacts to the cycle by design. In the United States, the Extended Benefits programme adds weeks of unemployment insurance when the unemployment of a state crosses thresholds set in federal law, so extensions arrive when unemployment is high and rising. I compare each state in which the programme switched on with the seven states in which it never did, leaving out the controls that border it, using monthly unemployment rates from October 2002 to December 2019. The test of Section \ref{sec:test} rejects no more often than its size would produce, and I estimate the CVAR in changes with equal weights on the controls.

A VAR without controls finds that an extension is followed by a fall in unemployment that keeps growing for three years. The states that never extended benefits show almost the same fall in the VAR that contains both (Proposition \ref{prop:placebo}), so the response of the VAR is the recovery from the recession that triggered the extensions. The CVAR finds a response close to zero at every horizon, with bands several times narrower than those of the VAR without controls, in agreement with the evidence that extensions at long potential durations have little effect on unemployment \citep{AcostaEtAl2025}. Supplementary Appendix \ref{sec:eb} reports the institutions, the data, the discussion of the assumptions, the test of the design, the results state by state, the impulse causal effects, the bounds and the comparison with local-projection difference-in-differences.

\section{The Low Carbon Fuel Standard\label{sec:lcfssum}}

The third application turns to a policy whose cost changes every month. California's Low Carbon Fuel Standard requires refiners and importers of gasoline to cover with credits the carbon intensity of their fuel above a declining benchmark, and whether the cost of the credits reaches consumers is disputed \citep{CARB2024,KleinmanCenter2024}. A comparison of the cost with retail prices mixes it with the movements of crude oil and refining margins that affect every region. I compare the retail price of gasoline in California with the prices of three sets of regions that buy crude oil and refined products in the same markets without a comparable standard, using monthly data from January 2014 to July 2026. The test of Section \ref{sec:test} does not reject a trend specific to California with any set of controls, and I estimate the CVAR in changes with equal weights.

With every set of controls, the CVAR finds that the price of gasoline in California rises by about the increase in the cost within a month and stays near that level, as incidence theory predicts for a cost per gallon borne by the suppliers of a good with inelastic demand. A VAR without controls finds a response that keeps rising well above full pass-through, with much wider bands. Supplementary Appendix \ref{sec:lcfs} reports the institutions, the construction of the cost, the data, the discussion of the assumptions and the robustness of the result across samples, sets of controls and treatments of the pandemic years.

\section{Conclusions\label{sec:conclusions}}

This paper has proposed the control VAR, which uses units not exposed to a policy to remove the common trends of the treated unit and estimates the dynamic effect of the policy on the remaining cycle. The causal framework defines the effect of a policy surprise, including the later reaction of the policy, and the assumptions require controls that share the trends of the treated unit and are not affected by the policy. The theorems show what a recursive VAR identifies, that the response of the cycle removes the bias from the reaction of the policy to the common trends, that the response of the controls measures that bias in any VAR, and that the estimated weights enter inference when the controls are cointegrated among themselves and the policy reacts to the past of the treated unit. The simulations show that the CVAR needs long lags and that a bootstrap that re-estimates the weights delivers the nominal coverage. In the applications, a larger Alaska dividend leaves the employment rate unchanged and raises employment in leisure and hospitality, while a VAR without controls attributes to it a loss of employment that the states without the dividend also show. Extended unemployment benefits have a small average effect on unemployment, while a VAR without controls attributes to them large reductions in the states where they coincided with the deepest recessions. The cost of the Low Carbon Fuel Standard passes almost fully to the price of gasoline within a month. A natural extension is a panel version of the CVAR, in which several treated units share the same controls.

\newpage
\appendix
\section{Proofs of the main results\label{sec:proofs-main}}

\begin{lem}[Responses of distributed lags\label{lem:dl}]
Let $u_t$ be orthogonal to every element of $\mathcal F_{t-1}$, with $\E[u_t^2]>0$, let $W_t$ be covariance stationary and measurable with respect to $\mathcal F_t$, let $\psi_k=\E[W_{t+k}u_t]/\E[u_t^2]$, and let $\sum_{j\geq0}|\tau_j|<\infty$. Then, for every $h\geq0$,
\[
\frac{\E\big[\sum_{j\geq0}\tau_jW_{t+h-j}\,u_t\big]}{\E[u_t^2]}=\sum_{j=0}^{h}\tau_j\psi_{h-j},
\qquad
\E\Big[\sum_{j\geq0}\tau_jW_{t-1-j}\,u_t\Big]=0 .
\]
\end{lem}

\paragraph{Proof of Lemma \ref{lem:dl}.} Since $\sum_j|\tau_j|\,\E|W_{t+h-j}u_t|\leq\sum_j|\tau_j|(\E[W_t^2]\E[u_t^2])^{1/2}<\infty$, the sum and the expectation can be interchanged. For $j>h$, and for every term of the second sum, the policy is dated $t-1$ or earlier, so it belongs to $\mathcal F_{t-1}$ and is orthogonal to $u_t$. For $j\leq h$, $\E[W_{t+h-j}u_t]=\psi_{h-j}\E[u_t^2]$. $\square$

\paragraph{Proof of Theorem \ref{thm:var}.} By Assumptions \ref{assu:A1} to \ref{assu:A3}, $Y^1_{t+h}-Y^1_{t-1}=\lambda'(F_{t+h}-F_{t-1})+(e_{t+h}-e_{t-1})+\sum_{j\geq0}\tau_j(W_{t+h-j}-W_{t-1-j})$. Multiply by $\tilde u_t$, take expectations and divide by $\E[\tilde u_t^2]$. By Lemma \ref{lem:dl}, the policy terms give $\sum_{j=0}^h\tau_j\tilde\psi_{h-j}$, which is $\tilde\theta_h$ by \eqref{eq:decomp}. The expectations are finite by Assumption \ref{assu:A5}. $\square$

\paragraph{Proof of Theorem \ref{thm:cycle}.} By Assumption \ref{assu:A2} and $\lambda=\Lambda'b$, the common trends cancel in $Y^1_t(0)-b'Y^0_t=\lambda'F_t+e_t-b'\Lambda F_t-b'v_t=e_t-b'v_t=G^0_t$; by Assumption \ref{assu:A1} the controls do not depend on the policy; and by Assumption \ref{assu:A3}, $G_t=G^0_t+\sum_{j\geq0}\tau_jW_{t-j}$. Multiplying $G_{t+h}$ by $u_t$ and applying Lemma \ref{lem:dl} gives \eqref{eq:cyclebias}, with the policy terms equal to $\theta_h$ by \eqref{eq:decomp}.

For the second statement, let $Z_t$ follow a VAR, possibly of infinite order. The response of $G_{t+h}$ to the recursive innovation of $W_t$, normalised to a unit impact, is the coefficient on $u_t$ in the projection of $G_{t+h}$ on $u_t$, $R_t$ and $\mathcal F_{t-1}$. Since $u_t$ is orthogonal to $R_t$ and to $\mathcal F_{t-1}$, the coefficient is $\E[G_{t+h}u_t]/\E[u_t^2]$ \citep[Proposition 1]{PlagborgMollerWolf2021}. $\square$

\paragraph{Proof of Theorem \ref{thm:ident}.} By Theorem \ref{thm:cycle}, $\gamma_h-\theta_h=\E[G^0_{t+h}u_t]/\E[u_t^2]$, which is zero for every $h\geq0$ by Assumption \ref{assu:A4}. The second equality is the decomposition \eqref{eq:decomp}, which holds under Assumption \ref{assu:A3}. $\square$

\paragraph{Proof of Theorem \ref{thm:direct}.} (a) By Theorem \ref{thm:ident} and \eqref{eq:decomp}, $\gamma_h=\sum_{j=0}^h\tau_j\psi_{h-j}$ for every $h\geq0$, with $\psi_0=1$ because $u_t$ is the part of $W_t$ orthogonal to $R_t$ and $\mathcal F_{t-1}$. The system is triangular with unit diagonal, so $\tau_0=\gamma_0$ and $\tau_h=\gamma_h-\sum_{j=0}^{h-1}\tau_j\psi_{h-j}$ for $h\geq1$. Both $\gamma_h$ and $\psi_h$ are responses of the VAR in $Z_t$ to the innovation of the policy, by the second part of the proof of Theorem \ref{thm:cycle}, applied to $G_t$ and to $W_t$.

(b) Let $\kappa_k=\E[W_{t+k}(0)u_t]/\E[u_t^2]$ be the response of the path the policy would have followed without the surprise, with $\kappa_0=0$. Under Assumption \ref{assu:A3b}, $W_{t+k}=W_{t+k}(0)+\phi_kW_t$, so $\psi_k=\phi_k+\kappa_k$. Substituting in \eqref{eq:decomp} gives \eqref{eq:pathbias}. Under Assumptions \ref{assu:A3} and \ref{assu:A3b}, $Y_{t+h}\langle w\rangle-Y_{t+h}\langle 0\rangle=\sum_{j=0}^h\tau_j\big(W_{t+h-j}(w)-W_{t+h-j}(0)\big)=w\sum_{j=0}^h\tau_j\phi_{h-j}=w\,\theta^\ast_h$. If $\kappa_k=0$ for all $k\geq1$, then $\theta_h=\theta^\ast_h$, and $\gamma_h=\theta^\ast_h$ by Theorem \ref{thm:ident}. $\square$

\paragraph{Proof of Proposition \ref{prop:placebo}.} (a) Since $Y^1_t=G_t+C_t$, the identity holds for the moments, after division by $\E[\breve u_t^2]$. It holds for the Cholesky responses as well, because the VARs in $(R_t',W_t,Y^1_t,C_t)'$ and in $(R_t',W_t,G_t,C_t)'$ are linear transformations of each other that leave the first two variables unchanged, so they have the same policy innovation. Since $G_{t-1}$ belongs to the information set of $\breve u_t$, $\gamma^G_h=\E[G_{t+h}\breve u_t]/\E[\breve u_t^2]$.

(b) By Assumption \ref{assu:A2}, $C_t=b'\Lambda F_t+b'v_t=\lambda'F_t+b'v_t$, and by Assumption \ref{assu:A1} it does not depend on the policy. Substituting in $\gamma^C_h$ gives the expression, which is finite by Assumption \ref{assu:A5}; its second term is zero if $\E[v_{t+j}\breve u_t]=0$ for all $j$.

(c) The proofs of Theorems \ref{thm:cycle} and \ref{thm:ident} apply with $\breve u_t$ in place of $u_t$, because the past of $W_t$ belongs to the information set of $\breve u_t$. $\square$

\paragraph{Proof of Proposition \ref{prop:bounds}.} By Theorem \ref{thm:cycle} applied with the innovation $\breve u_t$, $\gamma^G_h=\breve\theta_h+\breve B_h$, so $\breve\theta_h=\gamma^G_h-\breve B_h$, and Assumption \ref{assu:A4b} gives (a). By Proposition \ref{prop:placebo}(a), $\gamma^{Y^1}_h-\gamma^G_h=\gamma^C_h$, so $|\gamma^{Y^1}_h-\gamma^G_h|=|\gamma^C_h|$, which is at most $\lambda|\gamma^C_h|$ only if $\lambda\geq1$ or $\gamma^C_h=0$. $\square$

\paragraph{Proof of Theorem \ref{thm:est}.} Write $\eta_1=\hat c_1-c_1^\ast$ and $\eta_2=\hat c_2-c_2^\ast$ for the errors in the weights on the trends and on the stationary combinations, and let a tilde denote a series demeaned over the sample, so that $\tilde z_t$ is the demeaned vector $(R_t',W_t,G^\ast_t)'$. Let $M$ collect the population autocovariances of $\tilde z_t$ up to lag $p$, $\tilde M$ their sample counterparts computed with the cycle $G^\ast_t$, and $\hat M$ the same sample autocovariances computed with the estimated cycle $\hat G_t$. Let $\mathcal Q$ be the probability limit of $T^{-1}\sum_t\tilde S_t\tilde S_t'$ after the projection on the other stationary regressors of \eqref{eq:dols}, and $\varepsilon^\infty_t$ the error of \eqref{eq:dols} with all leads and lags.

(a) In the rotated coordinates, \eqref{eq:dols} is a cointegrating regression of $Y^1_t$ on the $r$ integrated regressors $X_{1t}$, whose trends have full rank, and on stationary regressors: $S_t$, the policy terms, the leads and lags of $\Delta X_{1t}$ and the constant. The leads and lags make the error asymptotically orthogonal to the innovations of $X_{1t}$, so the results for dynamic least squares \citep{Saikkonen1991,StockWatson1993}, applied to the canonical form of \citet{SimsStockWatson1990}, give $T\eta_1=O_p(1)$, and $\sqrt{T}$-consistency and asymptotic normality of the coefficients on the stationary regressors, which converge to the population projection coefficients $(a^\ast,c_2^\ast)$; in particular $\sqrt{T}\eta_2=\mathcal Q^{-1}T^{-1/2}\sum_t\tilde S_t\varepsilon^\infty_t+o_p(1)$ without the constraint. Under the constraint $b\geq0$, which is a convex cone in the coordinates $c$ as well, the scaled estimation error converges to the projection of the unconstrained limit on the tangent cone at $c^\ast$ \citep{Andrews1999}. This limit is $O_p(1)$ in both blocks, and it is normal when $c^\ast$ is in the interior.\footnote{With $L$ fixed, least squares remains super-consistent for the coefficients of the integrated regressors \citep{Stock1987}, so $T\eta_1=O_p(1)$ still holds, and parts (a) and (b) hold with $b^\ast$ replaced by the minimiser of the population criterion with $L$ leads and lags.}

(b) The argument has four steps.

Step 1 (smoothness). The least-squares estimator of the VAR and its residual covariance are smooth functions of the sample autocovariances of $\tilde z_t$ up to lag $p$, so $\hat\gamma_h=g_h(\hat M)$ and $\tilde\gamma_h=g_h(\tilde M)$ for a function $g_h$ that is continuously differentiable at $M$.

Step 2 (the effect of the estimated weights on the moments). Since $\hat G_t-G^\ast_t=-(\hat a-a^\ast)-\eta_1'X_{1t}-\eta_2'S_t$, and the constant disappears in the demeaned series,
\[
\tilde{\hat G}_t=\tilde G^\ast_t-\eta_1'\tilde X_{1t}-\eta_2'\tilde S_t .
\]
Two kinds of entries of $\hat M$ change. The first are cross moments of the cycle with another variable of the system, an element $\varsigma_t$ of $\tilde z_{t-k}$ other than the cycle:
\[
\frac1T\sum_t\tilde{\hat G}_t\varsigma_t-\frac1T\sum_t\tilde G^\ast_t\varsigma_t=-\eta_1'\,\frac1T\sum_t\tilde X_{1t}\varsigma_t-\eta_2'\,\frac1T\sum_t\tilde S_t\varsigma_t .
\]
The sample covariance between a demeaned integrated process and a stationary one is $O_p(1)$ \citep{Phillips1988}, so the first term is $O_p(T^{-1})\,O_p(1)=O_p(T^{-1})$. By the law of large numbers, $T^{-1}\sum_t\tilde S_t\varsigma_t=\E[\tilde S_t\varsigma_t]+O_p(T^{-1/2})$, so the second term is $-\eta_2'\E[\tilde S_t\varsigma_t]+O_p(T^{-1})$.

The second kind are the moments of the cycle with itself, at lags $k=0,\dots,p$. Expanding the product,
\begin{align*}
\frac1T\sum_t\tilde{\hat G}_t\tilde{\hat G}_{t-k}-\frac1T\sum_t\tilde G^\ast_t\tilde G^\ast_{t-k}
={}&-\eta_1'\,\frac1T\sum_t\big(\tilde X_{1t}\tilde G^\ast_{t-k}+\tilde X_{1,t-k}\tilde G^\ast_t\big)
-\eta_2'\,\frac1T\sum_t\big(\tilde S_t\tilde G^\ast_{t-k}+\tilde S_{t-k}\tilde G^\ast_t\big)\\
&+\eta_1'\Big(\frac1T\sum_t\tilde X_{1t}\tilde X_{1,t-k}'\Big)\eta_1
+\eta_1'\Big(\frac1T\sum_t\big(\tilde X_{1t}\tilde S_{t-k}'+\tilde X_{1,t-k}\tilde S_t'\big)\Big)\eta_2\\
&+\eta_2'\Big(\frac1T\sum_t\tilde S_t\tilde S_{t-k}'\Big)\eta_2 .
\end{align*}
The orders of the five terms, with $\eta_1=O_p(T^{-1})$ and $\eta_2=O_p(T^{-1/2})$ from part (a), are as follows. The first term is $O_p(T^{-1})\,O_p(1)$, by the same covariance result as above. The second term equals $-\eta_2'\big(\E[\tilde S_t\tilde G^\ast_{t-k}]+\E[\tilde S_{t-k}\tilde G^\ast_t]\big)+O_p(T^{-1})$, by the law of large numbers. In the third term, the sample second moment of a demeaned integrated process grows with the sample, $T^{-1}\sum_t\tilde X_{1t}\tilde X_{1,t-k}'=O_p(T)$, because $T^{-2}\sum_t\tilde X_{1t}\tilde X_{1,t-k}'$ converges to the integral of the outer product of a demeaned Brownian motion \citep{Phillips1987}. The third term is therefore $O_p(T^{-1})\,O_p(T)\,O_p(T^{-1})=O_p(T^{-1})$: the super-consistency of the weights on the trends exactly offsets the growth of the trends. The fourth term is $O_p(T^{-1})\,O_p(1)\,O_p(T^{-1/2})=O_p(T^{-3/2})$, and the fifth is $O_p(T^{-1/2})\,O_p(1)\,O_p(T^{-1/2})=O_p(T^{-1})$. Only the second term is of order $T^{-1/2}$.

Collecting both kinds of entries,
\[
\sqrt{T}(\hat M-\tilde M)=\mathcal C\sqrt{T}\,\eta_2+O_p(T^{-1/2}),
\]
where $\mathcal C$ is a fixed matrix whose rows are minus the population covariances of $\tilde S_t$ with the elements of $\tilde z_t$ at the leads and lags that enter $M$.

Step 3 (delta method). By Step 1 and Step 2,
\[
\sqrt{T}(\hat\gamma_h-\tilde\gamma_h)=\nabla g_h(M)'\mathcal C\sqrt{T}\,\eta_2+o_p(1),
\]
which is \eqref{eq:firststep} with $J_h=\mathcal C'\nabla g_h(M)$. Equivalently, $J_h$ is the derivative at $\delta=0$ of the population response of the cycle $G^\ast_t-\delta'S_t$ to the policy innovation.

Step 4 (joint normality). Up to $o_p(1)$, $\sqrt{T}(\tilde\gamma_h-\gamma_h)$ is a linear function of $\sqrt{T}(\tilde M-M)$, and $\sqrt{T}\,\eta_2=\mathcal Q^{-1}T^{-1/2}\sum_t\tilde S_t\varepsilon^\infty_t+o_p(1)$ by part (a). Both are normalised sums of products of the elements of $(R_t',W_t,G^\ast_t,\Delta X_{1t}',S_t')'$ and of the error of \eqref{eq:dols} at lags up to $p$. These products are functions of at most $p+1$ consecutive observations, so they are strong mixing with coefficients $\alpha(m-p)$ for $m>p$, and they have finite moments of order $2+\epsilon$. The central limit theorem for strong mixing sequences \citep{Herrndorf1984}, with the Cram\'er-Wold device, gives joint asymptotic normality. Proposition \ref{prop:J} gives $J_h$ and conditions under which it is zero.

(c) If $r=K$, there are no stationary combinations of the controls, $S_t$ is absent, and only the terms in $\eta_1$ remain in Step 2. They are $O_p(T^{-1})=o_p(T^{-1/2})$, and the argument uses only $\eta_1=O_p(T^{-1})$, so it holds with and without the constraint. The case $p=p_T\to\infty$ is part (a) of Theorem \ref{thm:joint}. $\square$

\paragraph{Proof of Theorem \ref{thm:joint}.} I keep the notation of the proof of Theorem \ref{thm:est}, with $\tilde z_t$ the demeaned vector of the VAR and $\mathcal Q$ the limit of the second moments of the stationary combinations of the controls, and write $\hat{\mathcal Q}$ for its bootstrap counterpart. For a VAR($p$), $\tilde A(p)$ denotes the least-squares coefficients computed with the cycle $G^\ast_t$, $\mathcal C_p$ the matrix of the population covariances of $\tilde S_{t-l}$ with the elements of $\tilde z_{t-k}$, $l,k\leq p$, and $J_h(p)$ the derivative of the population response of the VAR($p$) with respect to the weights on $S_t$. I use four results on autoregressions of growing order. For a stationary process with a VAR($\infty$) representation satisfying the conditions of the theorem, let $\hat\Gamma_p$ be the sample second-moment matrix of the regressors of a VAR($p$) and $\Gamma_p$ its population counterpart. (LR1) $\|\hat\Gamma_p^{-1}-\Gamma_p^{-1}\|=O_p(pT^{-1/2})$ in operator norm, and $\|\Gamma_p^{-1}\|$ is bounded uniformly in $p$. (LR2) The least-squares coefficients satisfy $\|\hat A(p)-A(p)\|=O_p(p^{1/2}T^{-1/2})$, where $A(p)$ collects the first $p$ coefficients of the VAR($\infty$). (LR3) For any sequence of non-random vectors $\ell_p$ with bounded norm, $\sqrt{T}\ell_p'\mathrm{vec}(\hat A(p)-A(p))=T^{-1/2}\sum_t\ell_p'\mathrm{vec}(x_t(p)\epsilon_t')\Gamma_p^{-1}+o_p(1)$, which is asymptotically normal \citep{LewisReinsel1985}. (IK) The residual bootstrap of the VAR($p$) reproduces (LR1) to (LR3) in probability, and for smooth functions of the slope coefficients and of the innovation covariance it reproduces the limit distribution of the estimator \citep{InoueKilian2002}.

(a) The estimator of the weights does not involve $p$, so Theorem \ref{thm:est}(a) holds unchanged. Write $\hat G_t-G^\ast_t=-(\hat a-a^\ast)-(\hat c_1-c_1^\ast)'X_{1t}-(\hat c_2-c_2^\ast)'S_t$; the constant is absorbed by the constant of the VAR. Replacing $G^\ast_t$ by $\hat G_t$ changes the sample second moments of the regressors and their cross moments with the dependent variables. The term in $X_{1t}$ changes each entry by $(\hat c_1-c_1^\ast)'T^{-1}\sum_t\tilde X_{1,t-l}\tilde z_{t-k}=O_p(T^{-1})\,O_p(1)$, because the sample covariance between a demeaned integrated process and a stationary one is $O_p(1)$ \citep{Phillips1988}, and its quadratic term is $O_p(T^{-1})$. Over the $p$ lags the change in $\hat\Gamma_p$ and in the cross moments is $O_p(pT^{-1})$ in Frobenius norm. The term in $S_t$ changes each entry by $(\hat c_2-c_2^\ast)'\big(\E[\tilde S_{t-l}\tilde z_{t-k}]+O_p(T^{-1/2})\big)$, so the change is $(\hat c_2-c_2^\ast)'\mathcal C_p+O_p(p^{1/2}T^{-1})$, and $\mathcal C_p$ has bounded norm. By (LR1), the coefficients of the VAR in $\hat Z_t$ then satisfy
\[
\hat A(p)-\tilde A(p)=\Gamma_p^{-1}\big[(\hat c_2-c_2^\ast)'\mathcal C_p\big]^{\dagger}+O_p(pT^{-1})+O_p(p^{3/2}T^{-1}),
\]
where $[\cdot]^\dagger$ is the linear map from a perturbation of the moments to the perturbation of the coefficients, evaluated at the population moments. The remainders are $o_p(T^{-1/2})$ because $p^3/T\to0$. The response at a fixed horizon $h$ depends on the first $h$ coefficient matrices and on the innovation covariance through a function that is continuously differentiable with bounded derivatives, so
\[
\sqrt{T}(\hat\gamma_h-\gamma_h)=\sqrt{T}(\tilde\gamma_h-\gamma_h)+J_h(p)'\sqrt{T}(\hat c_2-c_2^\ast)+o_p(1).
\] As $p\to\infty$, the population projections of the VAR($p$) converge to those of the VAR($\infty$), because $\sum_l l\|\Phi_l\|<\infty$, and $J_h(p)\to J_h$, which gives \eqref{eq:firststep}. By (LR3) and the delta method, $\sqrt{T}(\tilde\gamma_h-\gamma_h)$ equals a normalised sum $T^{-1/2}\sum_t\xi_{t,T}$ with $\xi_{t,T}$ a linear function of $\mathrm{vec}(x_t(p)\epsilon_t')$ and of $\epsilon_t\epsilon_t'-\Sigma$, plus $o_p(1)$. By the proof of Theorem \ref{thm:est}(a), $\sqrt{T}(\hat c_2-c_2^\ast)=\mathcal Q^{-1}T^{-1/2}\sum_t\tilde S_t\varepsilon^\infty_t+o_p(1)$. The two sums are sums of a strong mixing triangular array with uniformly bounded moments of order $2+\epsilon$, so the Cram\'er-Wold device and the central limit theorem for strong mixing arrays give their joint asymptotic normality.

(b) Step 1 (the bootstrap world). In the bootstrap world, $Q^\ast_t$ follows the VAR($p$) estimated in step (i), with innovations drawn from the centred residuals. Its coefficients and innovation covariance converge in probability to those of the VAR($p$) approximation of $Q_t$, and hence to those of its VAR($\infty$) representation, by (LR2) and the summability condition. Let $b^\natural$ be the weights to which \eqref{eq:dols} converges in the bootstrap world, and $\gamma^\ast_h$ the response to which the CVAR converges there. Both are the population pseudo-true values of part (a) computed for the process $Q^\ast_t$, so they are continuous functions of its coefficients and innovation covariance, and they converge in probability to $b^\ast$ and $\gamma_h$. Step (v) computes $\gamma^\ast_h$ up to a simulation error that vanishes as the length of the long path grows.

Step 2 (bootstrap expansion). Since $Y^{1\ast}_t-\hat a-\hat b'Y^{0\ast}_t=G^\ast_t$, the bootstrap estimation error of the weights relative to $b^\natural$ is the error of the regression \eqref{eq:dols} of $G^\ast_t$ on the bootstrap regressors, centred at its bootstrap pseudo-true value. The controls $Y^{0\ast}_t$ have the rotated components $X^\ast_{1t}$, integrated, and $S^\ast_t$, stationary, generated jointly with $(R^\ast_t,W^\ast_t,G^\ast_t)$, so the proof of Theorem \ref{thm:est}(a), applied conditionally on the data with (IK) in place of (LR1) to (LR3), gives $\hat c^\ast_1-c^\natural_1=O_{p^\ast}(T^{-1})$ and $\sqrt{T}(\hat c^\ast_2-c^\natural_2)=\hat{\mathcal Q}^{-1}T^{-1/2}\sum_t\tilde S^\ast_t\varepsilon^\ast_t+o_{p^\ast}(1)$, in probability. Part (a), applied in the same way, gives
\[
\sqrt{T}(\hat\gamma^\ast_h-\gamma^\ast_h)=\sqrt{T}(\tilde\gamma^\ast_h-\gamma^\ast_h)+J^\ast_h(p)'\sqrt{T}(\hat c^\ast_2-c^\natural_2)+o_{p^\ast}(1),
\]
where $\tilde\gamma^\ast_h$ is the bootstrap response with the weights $b^\natural$ and $J^\ast_h(p)$ is the derivative of the response with respect to the weights on $S^\ast_t$ in the bootstrap world. By Step 1, $J^\ast_h(p)\to J_h$ in probability.

Step 3 (joint limit). In both worlds the pair formed by the response with the pseudo-true weights and the scaled error of the weights on $S_t$ is, up to negligible terms, a fixed linear functional of the scaled sample autocovariances of $Q_t$ (respectively $Q^\ast_t$) at lags up to $p$, centred at their population values in that world. For a linear process with i.i.d. innovations, the limit covariance of such a functional is a continuous function of the coefficients of the process, of the innovation covariance and of the fourth cumulants of the innovations. In the bootstrap world these are the coefficients of the fitted VAR($p$), the residual covariance and the fourth cumulants of the centred residuals, which converge in probability to their population values \citep{InoueKilian2002}. The bootstrap limit is therefore the normal limit of part (a).

Step 4 (coverage). The shifted draws of step (v) are $\hat\gamma_h+(\hat\gamma^\ast_h-\gamma^\ast_h)$, so their quantiles are $\hat\gamma_h$ plus the quantiles of $\hat\gamma^\ast_h-\gamma^\ast_h$. The limit of $\sqrt{T}(\hat\gamma_h-\gamma_h)$ is normal with mean zero, hence symmetric, and continuous, so the convergence of the distribution functions is uniform and the percentile band of the shifted draws has asymptotic coverage $1-\alpha$. The estimated rank and $\hat D_1$ are held fixed in both worlds; the rank is correct with probability approaching one, and $\hat D_1$ converges at the rate $T^{-1}$, so neither affects the limits. $\square$

\paragraph{Proofs for Section \ref{sec:test}.} The proof of Theorem \ref{thm:po} has three steps. The weights and the policy coefficients of the quasi-differenced regression are $\sqrt{T}$-consistent under the null and under local alternatives (Lemma \ref{lem:orders}). Their estimation changes $S(\bar\alpha)$ and $S(1)$ by amounts of order one, which cancel in the numerator of $P_T$ up to $o_p(1)$, because $\bar\alpha-1=\bar c/T$ (proof of Theorem \ref{thm:po}). It leaves the estimator of the long-run variance consistent (Lemma \ref{lem:lrv}). Lemma \ref{lem:fclt} collects the limits that the three steps use; their bootstrap counterparts, and the proof of Proposition \ref{prop:sieve}, are in Supplementary Appendix \ref{sec:supp-testboot}. Throughout, $\Rightarrow$ denotes weak convergence, $\|\cdot\|$ the Euclidean norm (Frobenius for matrices), and all limits are as $T\to\infty$ with $c$ and $\bar c$ fixed. The statistic is invariant to nonsingular linear transformations of the regressors and of the controls, so I rotate the controls into the $r$ trend directions $x_t=D_1'Y^0_t$ and the $K-r$ stationary combinations $s_t=H'Y^0_t$, and collect the stationary regressors in $w_t=(s_t',P_t')'$ and the stochastic regressors in $V_t=(x_t',w_t')'$. For any variable $z_t$, $z^\alpha_1=z_1$ and $z^\alpha_t=z_t-\alpha z_{t-1}$ for $t\geq2$; $d^\alpha$ is the quasi-differenced constant and $M_\alpha=I-d^\alpha(d^{\alpha\prime}d^\alpha)^{-1}d^{\alpha\prime}$. Under \eqref{eq:testmodel}, $Y^{1\alpha}=\mu d^\alpha+V^\alpha\vartheta+u^\alpha$, with $\vartheta$ collecting $b$ and $\delta(L)$ in the rotated coordinates, so that
\begin{equation}
S(\alpha)=u^{\alpha\prime}M_\alpha u^\alpha-A_\alpha'Q_\alpha^{-1}A_\alpha,\qquad Q_\alpha=V^{\alpha\prime}M_\alpha V^\alpha,\quad A_\alpha=V^{\alpha\prime}M_\alpha u^\alpha,
\label{eq:Sdecomp}
\end{equation}
where the first term, $S_u(\alpha)$, is the sum of squares of the GLS regression of $u_t$ on the constant alone. The test regression is unconstrained, so no boundary issue arises: the restriction $b\geq0$ belongs to the estimator of Section \ref{sec:theorems} and plays no role in the test.

I write the linear processes of Assumption \ref{assu:T} as $v_t=\psi(L)\varepsilon_t$ and $\Delta V_t-E\Delta V_t=\Phi(L)\eta_t$, where $\{(\varepsilon_t,\eta_t')'\}$ is independent and identically distributed with mean zero, finite fourth moments and covariance $\mathrm{diag}(\sigma^2_\varepsilon,\Sigma_\eta)$, the off-diagonal block being zero because $v_t$ and $\Delta V_t$ are uncorrelated at all leads and lags, and $\sum_jj(|\psi_j|+\|\Phi_j\|)<\infty$. Then $\omega=\psi(1)\sigma_\varepsilon>0$ and $\Gamma=E[\Delta V_t\Delta V_t']$ is positive definite.

\begin{lem}\label{lem:fclt}
Under Assumption \ref{assu:T} and $\rho_T=1+c/T$:
\begin{enumerate}
\item[(a)] on $[0,1]$, \[\big(T^{-1/2}u_{[sT]},\,T^{-1/2}x_{[sT]}'\big)\Rightarrow\big(\omega W_c(s),\,B_x(s)'\big),\] where $W_c(s)=\int_0^se^{c(s-r)}dW_0(r)$, $W_0$ is a standard Brownian motion and $B_x$ is a Brownian motion independent of $W_0$;
\item[(b)] $T^{-1}\sum_t\Delta V_t\Delta V_t'\to_p\Gamma$ and $T^{-2}\sum_tx_{t-1}x_{t-1}'\Rightarrow\int_0^1B_xB_x'$;
\item[(c)] $\sum_t\Delta V_tv_t=O_p(T^{1/2})$; $T^{-1}\sum_t\Delta V_tu_{t-1}\Rightarrow\omega\int_0^1W_c\,dB_V$, where $B_V$ is the Brownian motion of the partial sums of $\Delta V_t$; $T^{-1}\sum_tx_{t-1}v_t\Rightarrow\omega\int_0^1B_x\,dW_0$; and $T^{-1}\sum_tw_{t-1}v_t=O_p(T^{-1/2})$.
\end{enumerate}
\end{lem}

\noindent Proof. (a) By the Beveridge-Nelson decomposition $v_t=\psi(1)\varepsilon_t+\tilde v_{t-1}-\tilde v_t$ with $\tilde v_t$ stationary, so $\max_{t\leq T}|T^{-1/2}\sum_{s\leq t}v_s-\psi(1)T^{-1/2}\sum_{s\leq t}\varepsilon_s|=o_p(1)$, and the same holds for $\Delta x_t$ with $\Phi_x(1)$ \citep[Theorem 3.4]{PhillipsSolo1992}. The partial sums of $(\varepsilon_t,\eta_t')'$ satisfy a multivariate invariance principle with a block-diagonal covariance, so the limits are jointly Gaussian with zero cross-covariance, hence independent. Writing $u_t=\sum_{s=1}^t\rho_T^{t-s}v_s+\rho_T^tu_0$, the near-integrated limit $T^{-1/2}u_{[sT]}\Rightarrow\omega W_c(s)$ follows from \citet[Lemma 1]{Phillips1987}, and $\rho_T^tT^{-1/2}u_0\to_p0$ uniformly in $t$ by Assumption \ref{assu:T}(iii).

(b) The first limit is the weak law of large numbers for sample second moments of linear processes with absolutely summable coefficients \citep[Theorem 3.7, Remark 3.9(i) and Section 4(i)]{PhillipsSolo1992}; the second follows from (a) and the continuous mapping theorem.

(c) For the first statement, each element of $\sum_t(\Delta V_t-E\Delta V_t)v_t$ is a sample cross-covariance at lag zero of two elements of the vector linear process $(v_t,\Delta V_t')'$, whose innovations are i.i.d. with finite fourth moments and whose coefficients satisfy $\sum_jj(|\psi_j|+\|\Phi_j\|)<\infty$. Its population value is zero, because $v_t$ and $\Delta V_t$ are uncorrelated at all leads and lags, and the central limit theorem for sample covariances of linear processes \citep[Theorem 3.8, Remark 3.9(i) and Section 4(i)]{PhillipsSolo1992} makes the normalised sum $O_p(1)$, so the sum is $O_p(T^{1/2})$; the mean of $\Delta V_t$ multiplies $\sum_tv_t=O_p(T^{1/2})$. The second and third statements are the convergence of sample covariances between a near-integrated process and a stationary linear process to stochastic integrals \citep{Phillips1987,Phillips1988}; the one-sided long-run covariance corrections $\sum_{j\geq1}E[v_{t-j}\Delta V_t']$ and $\sum_{j\geq0}E[\Delta x_{t-j}v_t]$ vanish because the cross-covariances are zero at all lags. The last statement follows as the first, since $w_{t-1}$ is stationary. $\square$

\begin{lem}\label{lem:orders}
Under the conditions of Lemma \ref{lem:fclt}, $T^{-1}Q_1\to_p\Gamma$, $Q_{\bar\alpha}-Q_1=O_p(1)$, $A_1=O_p(T^{1/2})$ and $A_{\bar\alpha}-A_1=O_p(1)$. Consequently $\hat\vartheta(\bar\alpha)-\vartheta=Q_{\bar\alpha}^{-1}A_{\bar\alpha}=O_p(T^{-1/2})$.
\end{lem}

\noindent Proof. For $t\geq2$, $V^{\bar\alpha}_t=\Delta V_t-(\bar c/T)V_{t-1}$ and $u^{\bar\alpha}_t=\Delta u_t-(\bar c/T)u_{t-1}$, with $\Delta u_t=v_t+(c/T)u_{t-1}$. Before the projection on the constant,
\[
\textstyle\sum_{t\geq2}V^{\bar\alpha}_tV^{\bar\alpha\prime}_t-\sum_{t\geq2}\Delta V_t\Delta V_t'=-\frac{\bar c}{T}\sum(\Delta V_tV_{t-1}'+V_{t-1}\Delta V_t')+\frac{\bar c^2}{T^2}\sum V_{t-1}V_{t-1}'.
\]
By Lemma \ref{lem:fclt}, the blocks of the right-hand side involving $x_{t-1}$ are $O_p(1)$: $T^{-1}\sum\Delta V_tx_{t-1}'$ converges to a stochastic integral plus a long-run covariance and $T^{-2}\sum x_{t-1}x_{t-1}'$ to $\int B_xB_x'$; the blocks involving only $w_{t-1}$ are $O_p(T^{-1/2})$ and $O_p(T^{-1})$. Similarly
\[
\textstyle\sum_{t\geq2}V^{\bar\alpha}_tu^{\bar\alpha}_t-\sum_{t\geq2}\Delta V_t\Delta u_t=-\frac{\bar c}{T}\sum(\Delta V_tu_{t-1}+V_{t-1}\Delta u_t)+\frac{\bar c^2}{T^2}\sum V_{t-1}u_{t-1}=O_p(1),
\]
using $T^{-1}\sum V_{t-1}\Delta u_t=T^{-1}\sum V_{t-1}v_t+cT^{-2}\sum V_{t-1}u_{t-1}=O_p(1)$ and $T^{-2}\sum x_{t-1}u_{t-1}\Rightarrow\omega\int B_xW_c$. Finally $\sum\Delta V_t\Delta u_t=\sum\Delta V_tv_t+(c/T)\sum\Delta V_tu_{t-1}=O_p(T^{1/2})$ and $T^{-1}\sum_{t\geq2}\Delta V_t\Delta V_t'\to_p\Gamma$. The first observation contributes $V_1V_1'$ and $V_1u_1$, which are $O_p(1)$ for both values of $\alpha$. The projection on the quasi-differenced constant subtracts $(d^{\alpha\prime}d^\alpha)^{-1}(d^{\alpha\prime}V^\alpha)'(d^{\alpha\prime}u^\alpha)$ and the analogous term from $Q_\alpha$. For $\alpha=1$ it removes only the first observation. For $\alpha=\bar\alpha$, $d^{\bar\alpha\prime}d^{\bar\alpha}=1+(T-1)\bar c^2/T^2\to1$, $d^{\bar\alpha\prime}V^{\bar\alpha}=V_1-(\bar c/T)\sum_{t\geq2}V^{\bar\alpha}_t=V_1-(\bar c/T)(V_T-V_1)+O_p(1)=O_p(1)$, and in the same way $d^{\bar\alpha\prime}u^{\bar\alpha}=O_p(1)$. Both corrections are therefore $O_p(1)$, which proves the four orders; the last statement follows from $Q_{\bar\alpha}^{-1}=O_p(T^{-1})$ and $A_{\bar\alpha}=O_p(T^{1/2})$. $\square$

\begin{lem}\label{lem:lrv}
Let $\hat\omega^2(k)=\hat s^2_k/(1-\sum_{j=1}^k\hat\beta_{j,k})^2$, where $\hat s^2_k$ and $\hat\beta_{j,k}$ are the residual variance and lag coefficients of the regression of $\Delta\hat u_t$ on $\hat u_{t-1}$ and $\Delta\hat u_{t-1},\dots,\Delta\hat u_{t-k}$, and $\hat u_t$ is the residual in levels of the regression with $\alpha=\bar\alpha$; let $\tilde\omega^2(k)$ be the same estimator computed from the GLS-demeaned cycle $\tilde u_t$. Under the conditions of Lemma \ref{lem:fclt} and $k_T\to\infty$ with $k_T^3/T\to0$, $\max_{k\leq k_T}|\hat\omega^2(k)-\tilde\omega^2(k)|=o_p(1)$. If, in addition, the lag order satisfies $\mathbb P(\underline k_T\leq\hat k\leq k_T)\to1$, where $\underline k_T\to\infty$ with $T^{1/2}\sum_{j>\underline k_T}|\beta_j|\to0$ for the coefficients $\beta_j$ of the autoregressive representation of $v_t$, or $\underline k_T$ is at least the order of that representation when it is finite, then $\hat\omega^2(\hat k)\to_p\omega^2$ under the null and under local alternatives.
\end{lem}

\noindent Proof. Let $\mathcal M_k$ be the $(k+1)\times(k+1)$ second-moment matrix, divided by $T$, of the regressors $(T^{-1/2}\hat u_{t-1},\Delta\hat u_{t-1},\dots,\Delta\hat u_{t-k})$ of the augmented Dickey-Fuller regression, $m_k$ the vector of their moments with $\Delta\hat u_t$, and $\tilde{\mathcal M}_k$ and $\tilde m_k$ the same objects computed from $\tilde u_t$.

Step 1 (the estimated cycle). The difference $\chi_t=\hat u_t-\tilde u_t=-V_t'(\hat\vartheta-\vartheta)-(\hat\mu-\tilde\mu)$ is small. By Lemma \ref{lem:orders}, $\hat\vartheta-\vartheta=O_p(T^{-1/2})$, and the GLS estimates of the constant differ by $(d^{\bar\alpha\prime}d^{\bar\alpha})^{-1}d^{\bar\alpha\prime}V^{\bar\alpha}(\hat\vartheta-\vartheta)=O_p(T^{-1/2})$. In changes, $\Delta\chi_t=-\Delta V_t'(\hat\vartheta-\vartheta)$ for $t\geq2$, so $T^{-1}\sum_t(\Delta\chi_t)^2=O_p(T^{-1})$. In levels, $\max_t|\chi_t|=O_p(1)$ because $\max_t\|x_t\|=O_p(T^{1/2})$, so $T^{-2}\sum_t\chi_t^2=O_p(T^{-1})$.

Step 2 (moments of the augmented Dickey-Fuller regression). By the Cauchy-Schwarz inequality and Step 1, every element of $\mathcal M_k-\tilde{\mathcal M}_k$ and of $m_k-\tilde m_k$ that involves lagged changes is bounded by
\[
2\Big(T^{-1}\sum_t\Delta\tilde u_t^2\Big)^{1/2}\Big(T^{-1}\sum_t\Delta\chi_t^2\Big)^{1/2}+T^{-1}\sum_t\Delta\chi_t^2=O_p(T^{-1/2}),
\]
and the elements that involve the lagged level satisfy the same bound, with $T^{-2}\sum_t\tilde u_t^2=O_p(1)$ and $T^{-2}\sum_t\chi_t^2=O_p(T^{-1})$. The bounds do not depend on the lag indices, so $\max_{k\leq k_T}\|\mathcal M_k-\tilde{\mathcal M}_k\|=O_p(k_TT^{-1/2})$ and $\max_{k\leq k_T}\|m_k-\tilde m_k\|=O_p(k_T^{1/2}T^{-1/2})$.

Step 3 (coefficients and residual variance). The smallest eigenvalue of $\tilde{\mathcal M}_k$ is bounded away from zero uniformly in $k\leq k_T$ with probability tending to one \citep{Berk1974}, and the coefficient vectors $\tilde\beta_{\cdot,k}$ are bounded uniformly in $k$, by the summability of the coefficients of $v_t$. Hence $\max_{k\leq k_T}\|\hat\beta_{\cdot,k}-\tilde\beta_{\cdot,k}\|=O_p(k_TT^{-1/2})$, and the sums of the lag coefficients satisfy
\[
\max_{k\leq k_T}\Big|\sum_{j=1}^k\hat\beta_{j,k}-\sum_{j=1}^k\tilde\beta_{j,k}\Big|\leq k_T^{1/2}\,O_p(k_TT^{-1/2})=O_p\big((k_T^3/T)^{1/2}\big)=o_p(1).
\]
The residual variance is the second moment of $\Delta\hat u_t$ minus $m_k'\mathcal M_k^{-1}m_k$, so by the same bounds it differs from its counterpart for $\tilde u_t$ by $O_p(k_TT^{-1/2})=o_p(1)$, uniformly in $k\leq k_T$. Since $1-\sum_j\tilde\beta_{j,k}$ is bounded away from zero for $\underline k_T\leq k\leq k_T$, the continuous mapping theorem gives the first statement.

Step 4 (the selected order). For $\underline k_T\leq k\leq k_T$, the autoregressive estimator computed from the GLS-demeaned cycle is consistent for $\omega^2$ uniformly in $k$, under the null and under local alternatives, by the results for autoregressions of growing order with a unit or near-unit root in the level term \citep{Berk1974,SaidDickey1984,PerronNg1998}. On the event $\underline k_T\leq\hat k\leq k_T$, $|\hat\omega^2(\hat k)-\omega^2|\leq\max_{k\leq k_T}|\hat\omega^2(k)-\tilde\omega^2(k)|+\max_{\underline k_T\leq k\leq k_T}|\tilde\omega^2(k)-\omega^2|$, and both terms are $o_p(1)$.\footnote{\citet[Section 6.2]{NgPerron2001} note that, under standard asymptotics, the order selected by the modified information criterion grows at a logarithmic rate \citep{HannanDeistler1988}, and their results for the tests are stated for a consistent estimator of the long-run variance. The condition on $\hat k$ is the one under which that consistency holds; with a finite autoregression for $v_t$ it requires only that the criterion does not select fewer lags than the true order.} $\square$

\paragraph{Proof of Theorem \ref{thm:po}.} Write $E_Q=Q_{\bar\alpha}-Q_1$ and $E_A=A_{\bar\alpha}-A_1$ for the changes that quasi-differencing at $\bar\alpha$ instead of $1$ makes to the moments of \eqref{eq:Sdecomp}. By \eqref{eq:Sdecomp},
\[
P_T=\frac{S_u(\bar\alpha)-\bar\alpha S_u(1)}{\hat\omega^2}-\frac{A_{\bar\alpha}'Q_{\bar\alpha}^{-1}A_{\bar\alpha}-\bar\alpha A_1'Q_1^{-1}A_1}{\hat\omega^2}.
\]
The first term is $P^u_T$ up to the estimator of the long-run variance, which is consistent for $\omega^2$ by Lemma \ref{lem:lrv}. For the second, $E_Q,E_A=O_p(1)$, $A_1=O_p(T^{1/2})$ and $T^{-1}Q_1\to_p\Gamma$ by Lemma \ref{lem:orders}. Then $Q_{\bar\alpha}^{-1}-Q_1^{-1}=-Q_1^{-1}E_QQ_{\bar\alpha}^{-1}=O_p(T^{-2})$ and
\[
A_{\bar\alpha}'Q_{\bar\alpha}^{-1}A_{\bar\alpha}-\bar\alpha A_1'Q_1^{-1}A_1=(1-\bar\alpha)A_1'Q_1^{-1}A_1+2E_A'Q_1^{-1}A_1+E_A'Q_1^{-1}E_A+A_{\bar\alpha}'(Q_{\bar\alpha}^{-1}-Q_1^{-1})A_{\bar\alpha},
\]
whose four terms are $O_p(T^{-1})$, $O_p(T^{-1/2})$, $O_p(T^{-1})$ and $O_p(T^{-1})$. Hence $P_T=P^u_T+o_p(1)$ under the null and local alternatives. For (a), a constant satisfies Condition B of \citet{ERS1996}, so $P^u_T\Rightarrow\bar c^2\int_0^1W_c(s)^2ds-\bar c\,W_c(1)^2$ (their Theorem 2), and this limit involves neither $K$, nor $r$, nor the policy terms. For (b), $P^u_T$ is the statistic of the point-optimal test computed from the counterfactual cycle with known weights and policy coefficients. By Theorems 1 and 2 of \citet{ERS1996}, that test is asymptotically equivalent to the Neyman-Pearson test against $c=\bar c$, and its power at $\bar c$ equals the envelope $\Pi(\bar c)$, an upper bound for the power of any test of a unit root in the cycle under Gaussian errors. Since $P_T$ does not use the weights and has the same limit, it belongs to the class of tests that do not know them and attains at $\bar c$ the bound of that class, which cannot exceed $\Pi(\bar c)$. $\square$

\paragraph{Proof of Theorem \ref{thm:diff}.} The decomposition in the proof of Theorem \ref{thm:cycle} uses only $\lambda=\Lambda'b$, so it holds under Assumption \ref{assu:A2d}, and $G_{t+h}-G_{t-1}=(G^0_{t+h}-G^0_{t-1})+\sum_{j\geq0}\tau_j(W_{t+h-j}-W_{t-1-j})$. Multiplying by $u_t$ and applying both parts of Lemma \ref{lem:dl} gives the first statement of (a). For the second, $G_{t+h}-G_{t-1}=\sum_{j=0}^h\Delta G_{t+j}$, and the second part of the proof of Theorem \ref{thm:cycle}, applied to the stationary variable $\Delta G_t$ of $Z^\Delta_t$, gives $\E[\Delta G_{t+j}u_t]/\E[u_t^2]=\gamma^\Delta_j$. For (b), Assumption \ref{assu:A4d}(ii) sets the second term to zero, so $\Gamma_h=\sum_{j=0}^h\tau_j\psi_{h-j}$ with $\psi_0=1$, the system solved in the proof of Theorem \ref{thm:direct}. For (c), Assumption \ref{assu:A2d} with $\lambda\neq\Lambda'b$ gives $Y^1_t(0)-b'Y^0_t=(\lambda-\Lambda'b)'F_t+e_t-b'v_t$, so the cumulated change of the cycle contains the additional term $(\lambda-\Lambda'b)'(F_{t+h}-F_{t-1})$, and the argument of part (a) applies to the rest. $\square$

\paragraph{Proof of Proposition \ref{prop:wchanges}.} Let $\xi_t$ and $\zeta_t$ denote $\Delta Y^0_t$ and $\Delta Y^1_t$ after the projection on the policy terms, and let $\Xi=\Lambda'D^{-1}\Lambda$, the precision of the controls' changes as signals of the common trends. By Assumption \ref{assu:A2d}, $\xi_t=\Lambda\Delta F_t+\Delta v_t$ and $\zeta_t=\lambda'\Delta F_t+\Delta e_t$, so that, under the stated covariances, $\operatorname{Var}(\xi_t)=\Lambda\Sigma_F\Lambda'+D$ and $\operatorname{Cov}(\xi_t,\zeta_t)=\Lambda\Sigma_F\lambda$. The changes are stationary and ergodic by Assumption \ref{assu:A4d}(i), so the sample moments converge and the least-squares weights converge to $w^\ast=\operatorname{Var}(\xi_t)^{-1}\operatorname{Cov}(\xi_t,\zeta_t)$. The identity $(\Lambda\Sigma_F\Lambda'+D)^{-1}\Lambda=D^{-1}\Lambda(I_r+\Sigma_F\Xi)^{-1}$ gives $\Lambda'w^\ast=\Xi(I_r+\Sigma_F\Xi)^{-1}\Sigma_F\lambda=\Xi\Sigma_F(I_r+\Xi\Sigma_F)^{-1}\lambda=\big(I_r-(I_r+\Xi\Sigma_F)^{-1}\big)\lambda$, hence $\lambda-\Lambda'w^\ast=(I_r+\Xi\Sigma_F)^{-1}\lambda$. The cycle built with $w^\ast$ loads on $F_t$ with $\lambda-\Lambda'w^\ast$, and part (c) of Theorem \ref{thm:diff} applies with $b=w^\ast$. With $r=1$, $\Xi\Sigma_F=s$, so $(I_r+\Xi\Sigma_F)^{-1}=1/(1+s)$, and the response of the weighted controls to the innovation is $w^{\ast\prime}\Lambda$ times the response of the trend, which equals $s/(1+s)$ times $\lambda$ times that response, so the added term is $1/s$ times the response of the weighted controls. $\square$

\section{Data sources\label{sec:data}}

\begin{table}[H]
\centering\small
\caption{Data sources.\label{tab:data}}
\begin{tabular}{p{5cm}p{9cm}}
\toprule
Series & Source\\
\midrule
State unemployment rates & Bureau of Labor Statistics, Local Area Unemployment Statistics, seasonally adjusted (FRED series \texttt{TXUR} for Texas and the analogous codes for other states)\\
Alaska and donor-state labour market outcomes & Bureau of Labor Statistics, Local Area Unemployment Statistics (employment-population ratio, unemployment and participation rates) and Current Employment Statistics (payroll employment, leisure and hospitality, retail trade), seasonally adjusted and unadjusted\\
Permanent Fund Dividend & Alaska Department of Revenue, Permanent Fund Dividend Division, Summary of Applications and Payments\\
Oil price and consumer prices & FRED series \texttt{WTISPLC} (West Texas Intermediate) and \texttt{CPIAUCSL} (consumer price index)\\
Weeks of EB by state & Department of Labor trigger notices; replication files of \citet{BooneEtAl2021}\\
Weeks of EUC by state & Replication files of \citet{BooneEtAl2021}, openICPSR project 118525\\
Retail regular gasoline prices & Energy Information Administration, weekly retail gasoline prices for California, the East Coast, Midwest, Gulf Coast and Rocky Mountain regions, Texas and Florida\\
LCFS credit prices & California Air Resources Board, LCFS credit transfer reports, monthly average price\\
LCFS cost per gallon reported by refiners & California Energy Commission, estimated gasoline price breakdown and margins (reports under Senate Bill 1322), May 2024 onward\\
LCFS benchmarks and carbon intensities & California Code of Regulations, title 17, section 95484\\
\bottomrule
\end{tabular}
\end{table}

\bibliographystyle{apalike}
\bibliography{bibliography}

\end{document}


\maketitle

This supplement has three parts. Sections \ref{sec:identification} to \ref{sec:spillnl} give further theory: what a VAR without controls identifies, which completes Section \ref{sec:var} (Section \ref{sec:identification}); conditioning on the common shock as an alternative to differencing (Section \ref{sec:proxy}); three results on the inference of Section \ref{sec:estimation} (Section \ref{sec:supp-inf}); the assumption, the bootstrap and an extension of the test of Section \ref{sec:test} (Section \ref{sec:supp-testboot}); and the extensions to spillovers and to effects that are not linear in the policy (Section \ref{sec:spillnl}). Section \ref{sec:montecarlo} reports the Monte Carlo simulations, and Section \ref{sec:screen} applies the test of the design to the states of the United States. Sections \ref{sec:app-eb} to \ref{sec:lcfs} complete the applications: further results for Alaska, and the full applications to extended benefits and to the Low Carbon Fuel Standard that the paper summarises. Each section introduces the notation it needs. References to sections, equations, assumptions and results without the prefix S are to the paper.

\section{What a VAR without controls identifies\label{sec:identification}}

This section characterises the impact coefficient of a recursive VAR for a dummy, a continuous and a non-negative policy, with an outcome $V_t$ that may be the treated outcome or the cycle $G_t$ of Section \ref{sec:causal}. The notation is local to this section: $\theta_t(w)$ below is a conditional effect given the past, not the estimand of Section \ref{sec:causal}.

Consider a policy $W_t$ and an outcome of interest $V_t$ observed for $t=1,\dots,T$. In a VAR without controls the outcome is a variable such as consumer prices, and in the CVAR of Section \ref{sec:theorems} it is the gap between a treated unit and its weighted controls. Every period the policy is assigned, the outcome is observed, and a potential outcome process describes the outcome under each assignment path \citep{RambachanSheppard2021}. A dynamic causal effect is the difference between two such paths at a point in time, and only one path is observed.

Let $\mathcal{F}_{t-1}$ be the observed past. The potential outcome $V_t(w)$ is the outcome at $t$ if the policy at $t$ were set to $w$ and the past were as observed. At horizon $h$, $V_{t+h}(w)$ is the outcome at $t+h$ if the policy at $t$ were set to $w$ and the policy after $t$ followed its own, possibly reacting, process. This is the object an impulse response estimates. The direct effect of the policy, which holds the future policy path fixed, is a different object, and Appendix \ref{sec:montecarlo} shows that the two differ when the policy reacts to outcomes.

For a dummy policy, the average treatment effect and the average treatment effect on the treated given the past are
\begin{equation}
\mathrm{ATE}_t=\E[V_t(1)-V_t(0)\mid\mathcal{F}_{t-1}],\qquad \mathrm{ATT}_t=\E[V_t(1)-V_t(0)\mid W_t=1,\mathcal{F}_{t-1}].
\label{eq:ate-att}
\end{equation}
They differ by a selection term, the difference between treated and untreated periods in the outcome that would have been observed without the policy. For a continuous policy, let
\begin{equation}
\beta_t(w)=\E[V_t(w)-V_t(0)\mid\mathcal{F}_{t-1}],\qquad \theta_t(w)=\E[V_t(w)-V_t(0)\mid W_t=w,\mathcal{F}_{t-1}].
\label{eq:beta-theta}
\end{equation}
The average causal response (ACR) is the derivative $\beta_t'(w)$ \citep{AngristImbensACR1995}. Its conditional counterpart, the average causal response on the treated (ACRT), is the derivative of the effect with respect to the dose among the periods in which that dose is received,
\begin{equation}
\mathrm{ACRT}_t(w)=\partial_l\,\E[V_t(l)-V_t(0)\mid W_t=w,\mathcal{F}_{t-1}]\big|_{l=w}.
\label{eq:acrt}
\end{equation}
Defining the effects given the past is what allows non-stationary outcomes to be handled without filtering, and it means that what a VAR identifies is a weighted average of these effects over the possible pasts.

Let $Z_t$ be a stationary vector, in differences if the variables are non-stationary, that contains the policy $W_t$ first and the outcome $V_t$. The VAR($p$) is $Z_t=c+\sum_{l=1}^pA_lZ_{t-l}+\epsilon_t$, with residual covariance $\Sigma$. A Cholesky decomposition with the policy first, normalised so that the policy has unit impact, gives the response of the outcome to a unit policy innovation. Its impact is $\gamma=\Sigma_{VW}/\Sigma_{WW}$, and the response at horizon $h$ propagates it through the coefficients $A_l$.

Let $\nu_t=\E[W_t\mid\mathcal{F}_{t-1}]$ and $u_t=W_t-\nu_t$ be the predictable part and the innovation of the policy, and let
\begin{equation}
\begin{aligned}
\mu_t(w)&=\E[V_t\mid W_t=w,\mathcal{F}_{t-1}],\qquad m_t=\E[V_t(0)\mid\mathcal{F}_{t-1}],\\
\omega_t(w)&=\E\big[(W_t-\nu_t)\mathbf{1}\{W_t\geq w\}\mid\mathcal{F}_{t-1}\big].
\end{aligned}
\label{eq:objects}
\end{equation}
The weights $\omega_t$ are non-negative and integrate to $\var(W_t\mid\mathcal{F}_{t-1})$. Let $\mu_t^\ast$ equal $\mu_t$ on the support of $W_t$ given the past and be linear across the gaps of that support. For a dummy policy, $\mu_t^\ast$ is the line between $\mu_t(0)$ and $\mu_t(1)$ and $\omega_t=p_t(1-p_t)$, with $p_t=\mathbb{P}(W_t=1\mid\mathcal{F}_{t-1})$.

\begin{assumption}[Consistency and no anticipation\label{assu:C1}]
$V_t=V_t(W_t)$, and the potential outcome at $t$ depends on the policy only up to $t$.
\end{assumption}

\begin{assumption}[Policy equation\label{assu:C3}]
$\E[W_t\mid\mathcal{F}_{t-1}]=c+\sum_{l=1}^pa_l'Z_{t-l}$, $Z_t$ is covariance stationary with finite second moments, and $\E[\var(W_t\mid\mathcal{F}_{t-1})]>0$.
\end{assumption}

\begin{assumption}[Baseline independence given the past\label{assu:C2}]
$\E[V_t(0)\mid W_t,\mathcal{F}_{t-1}]=\E[V_t(0)\mid\mathcal{F}_{t-1}]=m_t$ almost surely.
\end{assumption}

\begin{assumption}[Strong parallel trends\label{assu:C5}]
$\E[V_t(w)-V_t(0)\mid W_t,\mathcal{F}_{t-1}]=\beta_t(w)$ for all $w$.
\end{assumption}

Assumption \ref{assu:C1} is the consistency condition of the Rubin causal model adapted to time series, and it rules out anticipation. Assumption \ref{assu:C3} says that the lags of the system span the predictable part of the policy. Assumption \ref{assu:C2} says that, given the past, the policy innovation is unrelated to the outcome the economy would have had without the policy. It is the conditional counterpart of the independence condition of \citet{RambachanSheppard2021}, and it fails when the policy reacts to the contemporaneous state of the economy, as a monetary policy that follows a Taylor rule does. Assumption \ref{assu:C5} is the strong parallel trends condition of \citet{callawaysantannabacon2021}. It rules out that the effect of a dose is systematically related to the dose that was received. It does not restrict the shape of the dose-response, which may be non-linear.

The assumptions are stated on the observed outcome given the observed past, not on the residuals of the VAR. The following lemma shows that nothing is lost by doing so.

\begin{lem}[From residuals to outcomes\label{lem:outcome}]
Under Assumption \ref{assu:C3},
\begin{equation}
\gamma=\frac{\Sigma_{VW}}{\Sigma_{WW}}=\frac{\E[V_tu_t]}{\E[u_t^2]}=\frac{\E[\cov(V_t,W_t\mid\mathcal{F}_{t-1})]}{\E[\var(W_t\mid\mathcal{F}_{t-1})]}.
\label{eq:gamma-bridge}
\end{equation}
The first column of the Cholesky factor is therefore the coefficient of a regression of the outcome on the policy innovation, and it does not depend on the specification of the equation of the outcome. At horizon $h$, the second and third expressions hold with $V_{t+h}$ in place of $V_t$, and they equal the response of the VAR when the VAR is the linear projection representation of $Z_t$, as a VAR of infinite order is \citep{PlagborgMollerWolf2021}.
\end{lem}

If Assumption \ref{assu:C3} fails, the coefficient contains the additional term
\[
\Delta_{AR}=\frac{\cov(m_t,u_t^W)}{\var(u_t^W)},
\]
where $u_t^W$ is the residual of the policy equation. It vanishes if the policy equation is correct and also if $m_t$ is linear in the same lags (proof of Lemma \ref{lem:outcome}).

\begin{thm}[Decomposition of the VAR coefficient\label{thm:weights}]
Under Assumption \ref{assu:C3}, if $\mu_t^\ast$ is absolutely continuous and $\E\int|\mu_t^{\ast\prime}(w)|\omega_t(w)\,dw<\infty$, then
\begin{equation}
\gamma=\frac{\E\left[\int\mu_t^{\ast\prime}(w)\,\omega_t(w)\,dw\right]}{\E\left[\int\omega_t(w)\,dw\right]},
\label{eq:weights}
\end{equation}
a weighted average, with non-negative weights that integrate to one, of the slope of the conditional mean of the outcome given the policy and the past. No causal assumption is used.
\end{thm}

\begin{thm}[Dummy policy\label{thm:dummy}]
Let $W_t\in\{0,1\}$.
\begin{enumerate}
\item[(a)] Under Assumptions \ref{assu:C1} and \ref{assu:C3},
\begin{equation}
\begin{aligned}
\gamma&=\frac{\E[p_t(1-p_t)\,(\mathrm{ATT}_t+\Delta_t^S)]}{\E[p_t(1-p_t)]},\\
\Delta_t^S&=\E[V_t(0)\mid W_t=1,\mathcal{F}_{t-1}]-\E[V_t(0)\mid W_t=0,\mathcal{F}_{t-1}],
\end{aligned}
\label{eq:dummy-a}
\end{equation}
where $\Delta_t^S$ is the selection bias.
\item[(b)] If Assumption \ref{assu:C2} also holds, $\Delta_t^S=0$ and $\gamma=\E[p_t(1-p_t)\,\mathrm{ATT}_t]/\E[p_t(1-p_t)]$, a weighted average of the effect on the treated given the past. If Assumption \ref{assu:C5} holds as well, $\mathrm{ATT}_t=\mathrm{ATE}_t$ and $\gamma$ is the same weighted average of the average treatment effect.
\end{enumerate}
\end{thm}

The weights favour the periods in which the policy is least predictable from the past. If the effect does not depend on the past, $\gamma$ is the ATT, and the ATE under strong parallel trends.

\begin{thm}[Continuous policy\label{thm:continuous}]
Let the policy be continuously distributed given the past, with $\mu_t$ absolutely continuous and the integrability condition of Theorem \ref{thm:weights}.
\begin{enumerate}
\item[(a)] Under Assumptions \ref{assu:C1} and \ref{assu:C3},
\begin{equation}
\gamma=\frac{\E\left[\int\big(\mathrm{ACRT}_t(w)+S_t(w)+B_t'(w)\big)\,\omega_t(w)\,dw\right]}{\E\left[\int\omega_t(w)\,dw\right]},
\label{eq:cont-a}
\end{equation}
where $S_t(w)=\partial_{w'}\E[V_t(w)-V_t(0)\mid W_t=w',\mathcal{F}_{t-1}]|_{w'=w}$ is the selection on the effect and $B_t(w)=\E[V_t(0)\mid W_t=w,\mathcal{F}_{t-1}]$ is the baseline outcome of the periods that receive dose $w$, whose slope is the selection on the baseline, both assumed differentiable.
\item[(b)] If Assumptions \ref{assu:C2} and \ref{assu:C5} also hold, $B_t'\equiv0$, $S_t\equiv0$, $\mathrm{ACRT}_t=\beta_t'$, and $\gamma$ is the weighted average of the ACR, $\gamma=\E[\int\beta_t'\omega_t]/\E[\int\omega_t]$. If $W_t\mid\mathcal{F}_{t-1}\sim\mathcal{N}(\nu_t,\sigma_t^2)$ the weights are $\omega_t(w)=\sigma_t^2\phi(w;\nu_t,\sigma_t^2)$, but normality is not needed.
\end{enumerate}
\end{thm}

The third case is a non-negative policy, such as an outage measured in megawatts. The support of $W_t$ given the past is $\{0\}\cup[d_L,d_U]$, with $d_L>0$ the smallest positive dose, and $\mathbb{P}(W_t=0\mid\mathcal{F}_{t-1})>0$ (Assumption \ref{assu:nonneg}).

\begin{assumption}[Non-negative policy\label{assu:nonneg}]
$W_t\geq0$, $\mathbb{P}(W_t=0\mid\mathcal{F}_{t-1})>0$, and the support of $W_t$ given the past is $\{0\}\cup[d_L,d_U]$.
\end{assumption}

\begin{thm}[Non-negative policy\label{thm:nonneg}]
Let Assumptions \ref{assu:C1}, \ref{assu:C3} and \ref{assu:nonneg} hold, with $\mu_t$ absolutely continuous on $[d_L,d_U]$ and the integrability condition of Theorem \ref{thm:weights}.
\begin{enumerate}
\item[(a)] The coefficient is
\begin{equation}
\begin{aligned}
\gamma&=\frac{\E\left[q_{0,t}\dfrac{\mu_t(d_L)-\mu_t(0)}{d_L}+\displaystyle\int_{d_L}^{d_U}\mu_t'(w)\,\omega_t(w)\,dw\right]}{\E[\var(W_t\mid\mathcal{F}_{t-1})]},\\
q_{0,t}&=d_L\,\nu_t\,\mathbb{P}(W_t=0\mid\mathcal{F}_{t-1}).
\end{aligned}
\label{eq:nonneg-a}
\end{equation}
\item[(b)] If Assumptions \ref{assu:C2} and \ref{assu:C5} also hold, the first component is the average treatment effect of the smallest dose against no policy per unit of dose, $\mathrm{ATE}_t(d_L,0)/d_L$, and the second is the ACR.
\end{enumerate}
\end{thm}

The coefficient is a mixture of an extensive margin and the marginal effects of the dose. The weights are known, but the two components answer different questions, and a researcher interested in one of them needs a design that separates them, for instance a dummy for the extensive margin.

\paragraph{Proof of Lemma \ref{lem:outcome}.} Step 1 (the residual of the outcome equation). Let $\Pi_V$ be the linear projection of $V_t$ on $1,Z_{t-1},\dots,Z_{t-p}$ and $u_t^V=V_t-\Pi_V$, so that $\Sigma_{VW}=\E[u_t^Vu_t^W]$. Under Assumption \ref{assu:C3}, the residual of the policy equation is the innovation, $u_t^W=W_t-\nu_t=u_t$. The projection $\Pi_V$ is $\mathcal{F}_{t-1}$-measurable and $\E[u_t\mid\mathcal{F}_{t-1}]=0$, so $\E[\Pi_Vu_t]=0$ and
\[
\Sigma_{VW}=\E[V_tu_t],\qquad \Sigma_{WW}=\E[u_t^2].
\]

Step 2 (the conditional form). Since $\nu_t$ is $\mathcal{F}_{t-1}$-measurable, iterated expectations give
\[
\begin{aligned}
\E[V_tu_t]&=\E\big[\E[V_t(W_t-\nu_t)\mid\mathcal{F}_{t-1}]\big]=\E[\cov(V_t,W_t\mid\mathcal{F}_{t-1})],\\
\E[u_t^2]&=\E[\var(W_t\mid\mathcal{F}_{t-1})],
\end{aligned}
\]
which proves \eqref{eq:gamma-bridge}.

Step 3 (horizon $h$). The argument of Step 2 applies to $\E[V_{t+h}u_t]$ unchanged. The response of the VAR at horizon $h$ to a unit policy innovation is the coefficient on $u_t$ in the linear projection of $V_{t+h}$ on $u_t$ and the past when the VAR is the projection representation of $Z_t$, and that coefficient is $\E[V_{t+h}u_t]/\E[u_t^2]$ because $u_t$ is orthogonal to the past \citep{PlagborgMollerWolf2021}.

Step 4 (a misspecified policy equation). If Assumption \ref{assu:C3} fails, $u_t^W$ is the linear projection error of the policy, which has mean zero and is orthogonal to the lags but is not a martingale difference, and Step 1 gives $\gamma=\E[V_tu_t^W]/\E[(u_t^W)^2]$. Under Assumption \ref{assu:C2}, $\E[V_t\mid W_t,\mathcal{F}_{t-1}]=m_t+\theta_t(W_t)$, so
\[
\E[V_tu_t^W]=\E[\theta_t(W_t)\,u_t^W]+\cov(m_t,u_t^W),
\]
which gives the additional term $\Delta_{AR}=\cov(m_t,u_t^W)/\var(u_t^W)$. It is zero if $m_t$ is linear in the lags, because $u_t^W$ is orthogonal to them. $\square$

\paragraph{Proof of Theorem \ref{thm:weights}.} Step 1 (a covariance identity). Let $X$ be a random variable with $\E|X|<\infty$ and $\varphi$ an absolutely continuous function with $\int|\varphi'(x)|\,\E\big[|X-\E X|\mathbf{1}\{X\geq x\}\big]\,dx<\infty$. Writing $\varphi(X)-\varphi(c)=\int\varphi'(x)\big(\mathbf{1}\{X\geq x\}-\mathbf{1}\{c\geq x\}\big)\,dx$ for a constant $c$, taking the covariance with $X$ and exchanging the order of integration (Fubini),
\[
\begin{aligned}
\cov(\varphi(X),X)&=\int\varphi'(x)\,\cov(\mathbf{1}\{X\geq x\},X)\,dx\\
&=\int\varphi'(x)\,\E\big[(X-\E X)\mathbf{1}\{X\geq x\}\big]\,dx .
\end{aligned}
\]
No restriction on the distribution of $X$ is needed, so the identity covers discrete and mixed policies \citep{Yitazaki1996}.

Step 2 (the weights). For a variable $W$ with conditional mean $\nu$ and variance $\sigma^2$ given $\mathcal{F}$, let $\omega(w)=\E[(W-\nu)\mathbf{1}\{W\geq w\}\mid\mathcal{F}]$. For $w\geq\nu$, every value of $(W-\nu)\mathbf{1}\{W\geq w\}$ is non-negative. For $w<\nu$, since $\E[W-\nu\mid\mathcal{F}]=0$, $\omega(w)=-\E[(W-\nu)\mathbf{1}\{W<w\}\mid\mathcal{F}]$, and every value of $(W-\nu)\mathbf{1}\{W<w\}$ is non-positive. Hence $\omega\geq0$. Integrating the two pieces and exchanging the order of integration,
\[
\int_\nu^\infty\omega(w)\,dw=\E\big[((W-\nu)^+)^2\mid\mathcal{F}\big],\qquad \int_{-\infty}^\nu\omega(w)\,dw=\E\big[((W-\nu)^-)^2\mid\mathcal{F}\big],
\]
which sum to $\sigma^2$.

Step 3 (the coefficient). By Lemma \ref{lem:outcome}, $\gamma=\E[\cov(V_t,W_t\mid\mathcal{F}_{t-1})]/\E[\var(W_t\mid\mathcal{F}_{t-1})]$. Given $\mathcal{F}_{t-1}$, $\cov(V_t,W_t\mid\mathcal{F}_{t-1})=\cov(\mu_t^\ast(W_t),W_t\mid\mathcal{F}_{t-1})$, because $\E[V_t\mid W_t,\mathcal{F}_{t-1}]=\mu_t(W_t)$ and $W_t$ lies in the support, where $\mu_t^\ast=\mu_t$. Step 1, applied conditionally on $\mathcal{F}_{t-1}$ with $\varphi=\mu_t^\ast$, gives the numerator $\E[\int\mu_t^{\ast\prime}(w)\,\omega_t(w)\,dw]$, and Step 2 gives the denominator. The integrability condition of the theorem justifies the exchanges of integration and expectation. $\square$

\paragraph{Proof of Theorem \ref{thm:dummy}.} Step 1 (the weights). For $W_t\in\{0,1\}$, $\nu_t=p_t$, and for $w\in(0,1]$, $\omega_t(w)=\E[(W_t-p_t)\mathbf{1}\{W_t=1\}\mid\mathcal{F}_{t-1}]=p_t(1-p_t)$, while $\omega_t(w)=0$ outside $[0,1]$. The function $\mu_t^\ast$ is the line between $\mu_t(0)$ and $\mu_t(1)$, so Theorem \ref{thm:weights} gives
\[
\gamma=\frac{\E[p_t(1-p_t)\,(\mu_t(1)-\mu_t(0))]}{\E[p_t(1-p_t)]} .
\]

Step 2 (the contrast). By Assumption \ref{assu:C1},
\[
\begin{aligned}
\mu_t(1)&=\E[V_t(1)\mid W_t=1,\mathcal{F}_{t-1}]=\mathrm{ATT}_t+\E[V_t(0)\mid W_t=1,\mathcal{F}_{t-1}],\\
\mu_t(0)&=\E[V_t(0)\mid W_t=0,\mathcal{F}_{t-1}],
\end{aligned}
\]
so $\mu_t(1)-\mu_t(0)=\mathrm{ATT}_t+\Delta_t^S$, which is (a).

Step 3 (part (b)). Under Assumption \ref{assu:C2}, both conditional means of $V_t(0)$ equal $m_t$, so $\Delta_t^S=0$. Under Assumption \ref{assu:C5}, $\mathrm{ATT}_t=\beta_t(1)=\mathrm{ATE}_t$. $\square$

\paragraph{Proof of Theorem \ref{thm:continuous}.} Step 1 (the conditional mean). By Assumption \ref{assu:C1}, $V_t=V_t(W_t)$, so
\[
\mu_t(w)=\E[V_t(w)\mid W_t=w,\mathcal{F}_{t-1}]=B_t(w)+\theta_t(w),
\]
with $B_t(w)=\E[V_t(0)\mid W_t=w,\mathcal{F}_{t-1}]$ and $\theta_t$ as in \eqref{eq:beta-theta}.

Step 2 (the slope of the effect). Let $g_t(l,w')=\E[V_t(l)-V_t(0)\mid W_t=w',\mathcal{F}_{t-1}]$, so that $\theta_t(w)=g_t(w,w)$. By the chain rule along $l=w'=w$,
\[
\theta_t'(w)=\partial_lg_t(l,w)\big|_{l=w}+\partial_{w'}g_t(w,w')\big|_{w'=w}=\mathrm{ACRT}_t(w)+S_t(w).
\]

Step 3 (part (a)). By Steps 1 and 2, $\mu_t'(w)=\mathrm{ACRT}_t(w)+S_t(w)+B_t'(w)$. The policy is continuously distributed given the past, so $\mu_t^\ast=\mu_t$, and Theorem \ref{thm:weights} gives \eqref{eq:cont-a}.

Step 4 (part (b)). Under Assumption \ref{assu:C2}, $B_t(w)=m_t$ for every $w$, so $B_t'\equiv0$. Under Assumption \ref{assu:C5}, $g_t(l,w')=\beta_t(l)$ does not depend on $w'$, so $S_t\equiv0$ and $\mathrm{ACRT}_t=\beta_t'$. The decomposition of the slope into the effect on the treated and selection terms is the continuous-dose decomposition of \citet{callawaysantannabacon2021}.

Step 5 (the Gaussian case). If $W_t\mid\mathcal{F}_{t-1}\sim\mathcal{N}(\nu_t,\sigma_t^2)$, then $\omega_t(w)=\int_w^\infty(x-\nu_t)\,\phi(x;\nu_t,\sigma_t^2)\,dx=\sigma_t^2\,\phi(w;\nu_t,\sigma_t^2)$, because $(x-\nu_t)\phi(x;\nu_t,\sigma_t^2)=-\sigma_t^2\,\partial_x\phi(x;\nu_t,\sigma_t^2)$. $\square$

\paragraph{Proof of Theorem \ref{thm:nonneg}.} Step 1 (the extended conditional mean). The support of $W_t$ given the past is $\{0\}\cup[d_L,d_U]$, so $\mu_t^\ast$ equals $\mu_t$ on $[d_L,d_U]$ and is the line between $(0,\mu_t(0))$ and $(d_L,\mu_t(d_L))$ on the gap $(0,d_L)$, with slope $(\mu_t(d_L)-\mu_t(0))/d_L$.

Step 2 (the weights). For $w<0$, $\mathbf{1}\{W_t\geq w\}=1$ and $\omega_t(w)=\E[W_t-\nu_t\mid\mathcal{F}_{t-1}]=0$, so no weight falls below zero. On the gap, $\mathbf{1}\{W_t\geq w\}=\mathbf{1}\{W_t>0\}$, so $\omega_t$ is constant there, and since $\E[W_t-\nu_t\mid\mathcal{F}_{t-1}]=0$,
\[
\begin{aligned}
\omega_t(w)&=\E[(W_t-\nu_t)\mathbf{1}\{W_t>0\}\mid\mathcal{F}_{t-1}]=-\E[(W_t-\nu_t)\mathbf{1}\{W_t=0\}\mid\mathcal{F}_{t-1}]\\
&=\nu_t\,\mathbb{P}(W_t=0\mid\mathcal{F}_{t-1}),\qquad 0<w<d_L .
\end{aligned}
\]

Step 3 (part (a)). By Theorem \ref{thm:weights}, the numerator is the integral of $\mu_t^{\ast\prime}\omega_t$ over the gap plus that over $[d_L,d_U]$. The first is the constant slope of Step 1 times the constant weight of Step 2 times the length $d_L$ of the gap,
\[
d_L\,\nu_t\,\mathbb{P}(W_t=0\mid\mathcal{F}_{t-1})\cdot\frac{\mu_t(d_L)-\mu_t(0)}{d_L}=q_{0,t}\,\frac{\mu_t(d_L)-\mu_t(0)}{d_L},
\]
and the second is $\int_{d_L}^{d_U}\mu_t'(w)\,\omega_t(w)\,dw$, which gives \eqref{eq:nonneg-a}.

Step 4 (part (b)). Under Assumptions \ref{assu:C1}, \ref{assu:C2} and \ref{assu:C5}, $\mu_t(d_L)-\mu_t(0)=\big(m_t+\beta_t(d_L)\big)-m_t=\mathrm{ATE}_t(d_L,0)$, and on $[d_L,d_U]$ Step 4 of the proof of Theorem \ref{thm:continuous} gives $\mu_t'=\beta_t'$. $\square$

\section{Conditioning on the common shock\label{sec:proxy}}

The CVAR cancels the common shock by differencing, which requires the two units to load on it in the same ratio at impact. An alternative estimates the contemporaneous loading from the data. The VAR is estimated on the policy, the treated outcome and the controls, with the controls ordered first, so that the policy innovation is purged of the part that is correlated with the contemporaneous innovations of the controls. The identifying condition is then independence of the policy innovation from the treated unit's own innovation given the common shock, which is weaker than Assumption \ref{assu:C2} for the outcome. Let $e^W_t$, $e^1_t$, $e^{P_1}_t$ and $e^{P_2}_t$ be the innovations of the policy, of the treated outcome and of two proxies for the common shock, each a control or an average of controls.

\begin{assumption}[Common-shock representation\label{assu:P}]
There are mutually uncorrelated shocks $\eta_t$, $\omega_t$, $v_t$, $a_{1t}$, $a_{2t}$, with $\sigma_\eta^2>0$ and $\sigma_\omega^2>0$, and constants $\lambda$, $\beta$, $\theta$ and $\ell_2\neq0$, such that
\begin{equation}
e^W_t=\lambda\eta_t+\omega_t,\qquad e^1_t=\beta\eta_t+v_t+\theta\,e^W_t,\qquad e^{P_1}_t=\eta_t+a_{1t},\qquad e^{P_2}_t=\ell_2\eta_t+a_{2t}.
\label{eq:common-shock}
\end{equation}
The policy does not react to the treated unit's own shock $v_t$ or to the proxy noises, and the controls do not respond to the policy.
\end{assumption}

Here $\eta_t$ is the common shock, with the scale chosen so that the first proxy loads on it with coefficient one, $v_t$ is the shock specific to the treated unit, and $a_{jt}$ is the idiosyncratic innovation of proxy $j$. The parameter $\theta$ is the response of the treated outcome to a unit policy innovation.

\begin{thm}[Conditioning on the common shock\label{thm:proxy}]
Under Assumption \ref{assu:P}:
\begin{enumerate}
\item[(a)] The coefficient on $e^W_t$ in the regression of $e^1_t$ on $e^W_t$ and $e^{P_1}_t$ converges to
\begin{equation}
\theta+\frac{\beta\lambda\kappa}{\sigma_\omega^2+\lambda^2\kappa},\qquad \kappa=\frac{\sigma_\eta^2\sigma_{a_1}^2}{\sigma_\eta^2+\sigma_{a_1}^2}.
\label{eq:proxy-bias}
\end{equation}
The bias is zero only if $\beta\lambda=0$ or $\sigma_{a_1}^2=0$. If the proxy is the average of $K$ controls with equal loadings and uncorrelated idiosyncratic innovations of variance $\sigma_a^2$, then $\sigma_{a_1}^2=\sigma_a^2/K$ and the bias is $O(1/K)$.
\item[(b)] Two-stage least squares of $e^1_t$ on $e^W_t$ and $e^{P_1}_t$, with $e^W_t$ and $e^{P_2}_t$ as instruments, is consistent for $\theta$.
\item[(c)] The response of the treated outcome at horizon $h$ to a unit policy innovation is the second element of $\Psi_h\,b$, where $\Psi_h$ are the moving-average coefficients of the VAR and $b=(1,\theta,0,0)'$ is the impact vector. It is consistent when $\theta$ and the VAR coefficients are.
\end{enumerate}
\end{thm}

Part (a) is the errors-in-variables bias of conditioning on one noisy control. Part (b) uses one control as an instrument for the other, which requires their idiosyncratic innovations to be uncorrelated, and fails for controls that share regional shocks. The estimator runs a VAR on the policy, the treated outcome and the two proxies, takes the residuals, computes the two-stage least squares coefficient and propagates $b$ through the VAR. Standard errors come from the residual bootstrap applied to the whole procedure.

\paragraph{Proof of Theorem \ref{thm:proxy}.} Write $S=\sigma_\eta^2+\sigma_{a_1}^2$ and $\kappa=\sigma_\eta^2\sigma_{a_1}^2/S$.

Step 1 (part (a): the regression with one proxy). By the Frisch-Waugh-Lovell theorem, the coefficient on $e^W_t$ in the regression of $e^1_t$ on $e^W_t$ and $e^{P_1}_t$ is $\cov(e^1_t,\tilde e^W_t)/\var(\tilde e^W_t)$, where $\tilde e^W_t$ is the residual of the projection of $e^W_t$ on $e^{P_1}_t$. The projection coefficient is $c=\cov(e^W_t,e^{P_1}_t)/\var(e^{P_1}_t)=\lambda\sigma_\eta^2/S$, so
\[
\tilde e^W_t=e^W_t-c\,e^{P_1}_t=(\lambda-c)\,\eta_t+\omega_t-c\,a_{1t},\qquad \lambda-c=\frac{\lambda\sigma_{a_1}^2}{S}.
\]
Its variance and its covariance with the common shock are
\[
\var(\tilde e^W_t)=(\lambda-c)^2\sigma_\eta^2+\sigma_\omega^2+c^2\sigma_{a_1}^2=\sigma_\omega^2+\lambda^2\kappa,\qquad
\cov(\eta_t,\tilde e^W_t)=(\lambda-c)\sigma_\eta^2=\lambda\kappa .
\]
Since $v_t$ is uncorrelated with $\tilde e^W_t$, and $\cov(e^W_t,\tilde e^W_t)=\var(\tilde e^W_t)$ because $\tilde e^W_t$ is a projection residual of $e^W_t$,
\[
\cov(e^1_t,\tilde e^W_t)=\beta\lambda\kappa+\theta\var(\tilde e^W_t).
\]
Dividing by $\var(\tilde e^W_t)$ gives \eqref{eq:proxy-bias}. The bias is zero if and only if $\beta\lambda\kappa=0$, and $\kappa=0$ if and only if $\sigma_{a_1}^2=0$. If the proxy is the average of $K$ controls with equal loadings and uncorrelated idiosyncratic innovations of variance $\sigma_a^2$, its noise has variance $\sigma_a^2/K$, and $\kappa\leq\sigma_{a_1}^2=O(1/K)$.

Step 2 (part (b): two proxies). Substituting $\eta_t=e^{P_1}_t-a_{1t}$,
\[
e^1_t=\theta\,e^W_t+\beta\,e^{P_1}_t+(v_t-\beta a_{1t}).
\]
The error $v_t-\beta a_{1t}$ is uncorrelated with $e^W_t$ and with $e^{P_2}_t$, because the shocks are mutually uncorrelated. The matrix of cross moments of the instruments $(e^W_t,e^{P_2}_t)$ with the regressors $(e^W_t,e^{P_1}_t)$ is
\[
\begin{pmatrix}\lambda^2\sigma_\eta^2+\sigma_\omega^2&\lambda\sigma_\eta^2\\ \ell_2\lambda\sigma_\eta^2&\ell_2\sigma_\eta^2\end{pmatrix},
\]
with determinant $\ell_2\sigma_\eta^2\sigma_\omega^2\neq0$. The instruments are therefore relevant and exogenous, and two-stage least squares is consistent for $(\theta,\beta)$.

Step 3 (part (c): the dynamic response). Under Assumption \ref{assu:P}, a unit innovation in the policy with the common shock held fixed moves the policy, the treated outcome and the two proxies on impact by $b=(1,\theta,0,0)'$, because the controls do not respond to the policy. The response at horizon $h$ is the second element of $\Psi_hb$, a continuous function of $\theta$ and of the coefficients of the VAR, so it is consistent when they are. $\square$

\section{Inference: known weights, the first-step term and the wild bootstrap\label{sec:supp-inf}}

This section gives three results on the inference of Section \ref{sec:estimation}: the validity of the residual bootstrap when the weights on the controls are known, the coefficient with which the error of the weights on the stationary combinations of the controls enters the response, and the failure of the wild bootstrap at impact.

\begin{thm}[Residual bootstrap\label{thm:resid}]
Let $Z_t$ follow a stationary VAR($p$) with i.i.d. innovations with finite fourth moments and positive definite covariance, and let $\gamma_h$ be the response at a fixed horizon $h$. Then the residual bootstrap distribution of $\sqrt{T}(\gamma_h^\ast-\hat\gamma_h)$ converges to the limit of the distribution of $\sqrt{T}(\hat\gamma_h-\gamma_h)$ in probability, and the percentile band has asymptotic coverage $1-\alpha$.
\end{thm}

\begin{prop}[The first-step term\label{prop:J}]
Let the conditions of Theorem \ref{thm:est}(b) hold, and let $\pi_{G,l}$ be the coefficient of $G^\ast_{t-l}$ in the projection of $W_t$ on $R_t$ and $p$ lags of $Z_t$. For any $\delta\in\mathbb{R}^{K-r}$,
\begin{equation}
J_h'\delta=\frac{-\E[S_{t+h}u_t]'\delta+\E[G^\ast_{t+h}\dot u_t(\delta)]-2\gamma_h\E[u_t\dot u_t(\delta)]}{\E[u_t^2]},
\label{eq:J}
\end{equation}
where $\dot u_t(\delta)$ is the derivative of the policy innovation with respect to the weights on $S_t$, given in Appendix \ref{sec:proofs-main}. In particular $J_h=0$ for all $h$ if (i) $\pi_{G,l}=0$ for $l=1,\dots,p$ and (ii) $\E[S_{t+j}u_t]=0$ for all $j$.
\end{prop}

\paragraph{Proof of Theorem \ref{thm:resid}.} Let $n$ be the number of observations used in the VAR, $A$ the matrix of its coefficients, $\Sigma$ the covariance matrix of its innovations, and $\zeta=(\mathrm{vec}(A)',\mathrm{vech}(\Sigma)')'$ the vector of its parameters, with asymptotic covariance $V$. Let $q^\ast_\alpha$ be the $\alpha$ quantile of the bootstrap distribution of $\gamma^\ast_h$, so that the percentile band is $[q^\ast_{\alpha/2},q^\ast_{1-\alpha/2}]$.

Step 1 (the estimator). Under the conditions of the theorem, the least-squares estimator satisfies
\[
\sqrt n(\hat\zeta-\zeta)\Rightarrow N(0,V),
\]
where $V$ depends on $A$, $\Sigma$ and the fourth moments of the innovations \citep[Chapter 12]{KilianLutkepohl2017}. The response at horizon $h$ is $\gamma_h=g_h(\zeta)$ for a function $g_h$ that is continuously differentiable at $\zeta$, because it is a polynomial in the first $h$ coefficient matrices multiplied by a column of the Cholesky factor of $\Sigma$, normalised by its diagonal element. By the delta method,
\[
\sqrt n(\hat\gamma_h-\gamma_h)\Rightarrow N\big(0,\nabla g_h(\zeta)'V\nabla g_h(\zeta)\big).
\]

Step 2 (the bootstrap estimator). Conditionally on the data, the bootstrap sample is generated by the VAR with coefficients $\hat A$ and innovations drawn with replacement from the centred residuals, whose covariance and fourth moments converge in probability to those of the innovations. The bootstrap central limit theorem for autoregressions \citep{Bose1988,KilianLutkepohl2017} gives
\[
\sqrt n(\hat\zeta^\ast-\hat\zeta)\Rightarrow N(0,V)\quad\text{in probability},
\]
and the delta method, applied conditionally on the data with $\nabla g_h(\hat\zeta)\to_p\nabla g_h(\zeta)$, gives the same normal limit for $\sqrt n(\gamma_h^\ast-\hat\gamma_h)$ in probability.

Step 3 (coverage). Since the limit in Steps 1 and 2 is continuous, the convergence of the distribution functions is uniform, and
\[
\begin{aligned}
\mathbb P\big(q^\ast_{\alpha/2}\leq\gamma_h\leq q^\ast_{1-\alpha/2}\big)&=\mathbb P\big(\sqrt n(\hat\gamma_h-q^\ast_{1-\alpha/2})\leq\sqrt n(\hat\gamma_h-\gamma_h)\leq\sqrt n(\hat\gamma_h-q^\ast_{\alpha/2})\big)\\
&\to1-\alpha,
\end{aligned}
\]
because the bootstrap quantiles of $\sqrt n(\gamma^\ast_h-\hat\gamma_h)$ converge to those of the normal limit, and that limit is symmetric about zero. $\square$

\paragraph{Proof of Proposition \ref{prop:J}.} Fix a direction $\delta$ and, for a scalar $\varepsilon$, let $G_t(\varepsilon)=G^\ast_t-\varepsilon\delta'S_t$ be the cycle with the weights on $S_t$ moved by $\varepsilon\delta$. Let $x_t(\varepsilon)$ be the regressors of the policy equation, $R_t$ and $p$ lags of $(R_t',W_t,G_t(\varepsilon))'$, let $\pi(\varepsilon)$ be the coefficients of the population projection of $W_t$ on them and $u_t(\varepsilon)=W_t-x_t(\varepsilon)'\pi(\varepsilon)$ its residual, so that $x_t(0)=x_t$, $\pi(0)=\pi$ and $u_t(0)=u_t$. A dot denotes the derivative at $\varepsilon=0$.

Step 1 (the response as a function of $\varepsilon$). By the second part of the proof of Theorem \ref{thm:cycle},
\[
\gamma_h(\varepsilon)=\frac{\E[G_{t+h}(\varepsilon)u_t(\varepsilon)]}{\E[u_t(\varepsilon)^2]},
\]
and by Step 3 of the proof of Theorem \ref{thm:est}, $J_h'\delta$ is the derivative of $\gamma_h(\varepsilon)$ at $\varepsilon=0$.

Step 2 (the derivatives of the ingredients). The cycle gives $\dot G_{t+h}=-\delta'S_{t+h}$. The regressors give $\dot x_t$, the vector with $-\delta'S_{t-l}$ in the position of $G^\ast_{t-l}$, $l=1,\dots,p$, and zero elsewhere. The innovation gives
\[
\dot u_t=-\dot x_t'\pi-x_t'\dot\pi=\sum_{l=1}^p\pi_{G,l}\,\delta'S_{t-l}-x_t'\dot\pi ,
\]
since the only non-zero elements of $\dot x_t$ multiply the coefficients $\pi_{G,l}$. The projection coefficients satisfy the normal equations $\E[x_t(\varepsilon)u_t(\varepsilon)]=0$ for every $\varepsilon$, and differentiating them gives $\E[\dot x_tu_t]+\E[x_t\dot u_t]=0$. Substituting $\dot u_t$ and solving,
\[
\dot\pi=\E[x_tx_t']^{-1}\Big(\E[\dot x_tu_t]+\sum_{l=1}^p\E[x_tS_{t-l}']\delta\,\pi_{G,l}\Big).
\]
This is the expression $\dot u_t(\delta)$ referred to in the statement.

Step 3 (quotient rule). Differentiating $\gamma_h(\varepsilon)$,
\[
\dot\gamma_h=\frac{\E[\dot G_{t+h}u_t]+\E[G^\ast_{t+h}\dot u_t]}{\E[u_t^2]}-\frac{\E[G^\ast_{t+h}u_t]}{\E[u_t^2]}\,\frac{2\E[u_t\dot u_t]}{\E[u_t^2]},
\]
and substituting $\dot G_{t+h}=-\delta'S_{t+h}$ and $\E[G^\ast_{t+h}u_t]/\E[u_t^2]=\gamma_h$ gives \eqref{eq:J}.

Step 4 (conditions for $J_h=0$). Under (i), $\pi_{G,l}=0$, so $\dot u_t=-x_t'\dot\pi$ and $\dot\pi=\E[x_tx_t']^{-1}\E[\dot x_tu_t]$. The elements of $\E[\dot x_tu_t]$ are $-\delta'\E[S_{t-l}u_t]$, which are zero under (ii). Hence $\dot\pi=0$ and $\dot u_t=0$, so the second and third terms of \eqref{eq:J} vanish, and the first is zero under (ii). $\square$

A natural alternative to the residual bootstrap of Section \ref{sec:estimation} multiplies the residuals by Rademacher weights. The wild bootstrap reproduces the sampling distribution of the slope coefficients of an autoregression \citep{goncalveskilian2004}, and it is known to fail for functions of the innovation covariance \citep{bruggemann2016inference}. The following result makes the failure explicit for the impact response of the CVAR, which depends only on that covariance.

\begin{thm}[Wild bootstrap at impact\label{thm:wild}]
Replace step (ii) by $\epsilon^\ast_t=\xi_t\hat\epsilon_t$ with $\xi_t$ i.i.d. Rademacher, and let $\tilde\Sigma^\ast$ be the residual covariance matrix of the VAR re-estimated on the bootstrap series. Then
\begin{equation}
\tilde\Sigma^\ast=\hat\Sigma-Q^\ast,\qquad Q^\ast=(\tilde B^\ast-\hat B)'\,\frac{X^{\ast\prime}X^\ast}{N}\,(\tilde B^\ast-\hat B)\succeq0,
\label{eq:wild}
\end{equation}
where $\hat B$ and $\tilde B^\ast$ are the original and the bootstrap coefficient estimates, $X^\ast$ the bootstrap regressors and $N$ the number of observations. If the bootstrap process is stationary, $Q^\ast=O_p(N^{-1})$. The bootstrap impact response, a smooth function of $\tilde\Sigma^\ast$, therefore differs from the estimate by $O_p(N^{-1})$ while the sampling error of the estimate is $O_p(N^{-1/2})$, so the coverage of the percentile band at impact tends to zero as $N$ grows.
\end{thm}

Theorem \ref{thm:wild} is exact for the identity in \eqref{eq:wild} and asymptotic for the rate. The impact response is not constant across wild bootstrap draws, because the VAR is re-estimated in each draw, but its variation is of smaller order than the sampling variation.

\paragraph{Proof of Theorem \ref{thm:wild}.} The bootstrap series is $Z^\ast_t=\hat c+\sum_l\hat A_lZ^\ast_{t-l}+\xi_t\hat\epsilon_t$. Let $E^\ast$ be the $N\times n$ matrix of the bootstrap errors $\xi_t\hat\epsilon_t'$, $X^\ast$ the matrix of the regressors, and $\tilde B^\ast=(X^{\ast\prime}X^\ast)^{-1}X^{\ast\prime}Z^\ast$ the least-squares estimate.

Step 1 (the bootstrap covariance). The bootstrap residuals are $E^\ast-X^\ast(\tilde B^\ast-\hat B)$, and the normal equations $X^{\ast\prime}(E^\ast-X^\ast(\tilde B^\ast-\hat B))=0$ give
\[
\begin{aligned}
\tilde\Sigma^\ast&=\frac1N\big(E^\ast-X^\ast(\tilde B^\ast-\hat B)\big)'\big(E^\ast-X^\ast(\tilde B^\ast-\hat B)\big)\\
&=\frac1NE^{\ast\prime}E^\ast-(\tilde B^\ast-\hat B)'\frac{X^{\ast\prime}X^\ast}{N}(\tilde B^\ast-\hat B).
\end{aligned}
\]
Since the same weight $\xi_t$ multiplies every element of $\hat\epsilon_t$ and $\xi_t^2=1$,
\[
\frac1NE^{\ast\prime}E^\ast=\frac1N\sum_t\xi_t^2\,\hat\epsilon_t\hat\epsilon_t'=\hat\Sigma
\]
in every draw, which gives \eqref{eq:wild}. The matrix $Q^\ast$ is a quadratic form in a positive semidefinite matrix, hence positive semidefinite.

Step 2 (the order of $Q^\ast$). If the bootstrap process is stationary, $X^{\ast\prime}X^\ast/N=O_p(1)$ and $\tilde B^\ast-\hat B=(X^{\ast\prime}X^\ast)^{-1}X^{\ast\prime}E^\ast=O_p(N^{-1/2})$ in probability, by the central limit theorem for the wild bootstrap of autoregressions \citep{goncalveskilian2004,bruggemann2016inference}. Hence $Q^\ast=O_p(N^{-1})$.

Step 3 (the impact response). The impact response is a smooth function of the innovation covariance matrix alone, $\gamma_0=\Sigma_{GW\cdot R}/\Sigma_{WW\cdot R}$, the ratio of the partial covariances given the innovations of $R_t$, which is continuously differentiable where $\Sigma_{WW\cdot R}>0$. By Step 1 and Step 2, every bootstrap draw satisfies $\gamma^\ast_0-\hat\gamma_0=O_p(N^{-1})$, so the percentile band has length $O_p(N^{-1})$ and lies within $O_p(N^{-1})$ of $\hat\gamma_0$. If $\sqrt N(\hat\gamma_0-\gamma_0)$ converges to a non-degenerate normal distribution, the probability that $\gamma_0$ lies in a set of length $O_p(N^{-1})$ around $\hat\gamma_0$ tends to zero. $\square$

\section{The test of the design: assumption, bootstrap and correlated innovations\label{sec:supp-testboot}}\label{sec:supp-test}

This section states the assumption under which Theorem \ref{thm:po} holds and the sieve bootstrap that computes the p-value of the test, with its proof, and then extends the test to innovations of the cycle that are correlated with those of the controls.

\begin{assumption}[Orthogonal innovations\label{assu:T}]
(i) The policies are stationary under the null and under the alternative, and $(\Delta Y^{0\prime}_t,P_t')'$ is a stationary linear process with absolutely summable coefficients, independent and identically distributed innovations with finite fourth moments, and $r\leq K$ stochastic trends. (ii) The innovations of $v_t$ are uncorrelated with the changes of the controls and of the policies at all leads and lags. (iii) The initial condition is small, $T^{-1/2}u_0\to_p0$.
\end{assumption}

To see what the assumption asks of an application, consider a regional government that pays its residents a dividend once a year, with an amount set by conditions that may themselves respond to the economy, as in the application to Alaska. Part (i) asks the dividend to be a stationary policy, as a bounded payment made once a year is, and to remain one under both hypotheses, whether the controls reproduce every trend of the region or the region also carries a trend of its own, since the test is computed under both. Part (ii) concerns the shocks specific to the region, the ones that would give it a trend of its own. It asks them to be unrelated to the shocks that move the controls and to the innovations of the dividend, and the second half is a restriction on the policy: the government may set the dividend in response to developments the region shares with its controls, but it may not respond to the past of the component specific to the region, such as its own labour market outcomes.\footnote{When the policy reacts to past changes of the treated unit and the cycle is serially correlated, the changes of the policy are correlated with the current innovations of the cycle and the null distribution shifts by a term proportional to that correlation. In the simulations of Supplementary Appendix \ref{sec:montecarlo}, with a reaction to the twelve-month change of the treated unit, the shift makes the test conservative.} Part (iii) is the condition under which point-optimal unit root tests are optimal \citep{ERS1996}: it excludes an initial condition drawn from the stationary distribution of a process far from the unit root, which affects the power of these tests \citep{MullerElliott2003}.\footnote{When the innovations of $v_t$ are correlated with the changes of the controls, the limit of $P_T$ depends on the long-run correlation, as for the residual-based tests \citep{Pesavento2004}, and power falls. Supplementary Appendix \ref{sec:supp-test} gives a version of the test that uses the correlation and remains valid, at a cost in power when the correlation is absent.}

The sieve bootstrap works as follows. Let $\hat e_t$ be the residuals of the regression with $\alpha=1$, fit an autoregression of order $p_T$ to them, resample its residuals, rebuild $Y^{1\ast}_t$ by cumulating the fitted changes and the bootstrap errors with the controls and the policy terms held at their observed values, and compute $P^\ast_T$.

\begin{prop}[Sieve bootstrap\label{prop:sieve}]
Let Assumption \ref{assu:T} hold under the null, with $v_t$ an invertible linear process, and let $p_T\to\infty$ with $p_T=o\big((T/\log T)^{1/2}\big)$. Then the bootstrap distribution of $P^\ast_T$ converges in probability to the null limit in Theorem \ref{thm:po}(a), and the bootstrap test has asymptotic size equal to its nominal level.
\end{prop}

In applications I set $p_T=\lceil 16\,(T/200)^{1/3}\rceil$, which satisfies the rate of Proposition \ref{prop:sieve} and gives sixteen lags with two hundred monthly observations. The simulations of Supplementary Appendix \ref{sec:montecarlo} show that data-driven choices of the order select autoregressions too short to capture the dynamics of a differenced persistent cycle, that shorter fixed orders leave the test oversized and longer ones cost power, and that this rule delivers close to the nominal size while the test keeps power well above the residual-based tests.

The proof uses the notation of the proofs for Section \ref{sec:test} in the Appendix of the paper. For the bootstrap, $\hat\kappa$ denotes the coefficients of the regression in differences, of $\Delta Y^1_t$ on the constant and $\Delta V_t$, $\hat e_t$ its residuals, and $P^{u\ast}_T$ the statistic computed from the bootstrap cycle $u^\ast_t$ with the constant as the only regressor.

\begin{lem}\label{lem:boot}
Let $E^\ast$, $\mathrm{Var}^\ast$ and $P^\ast$ refer to the bootstrap distribution conditional on the data, and let $e^\ast_t=\sum_{j=1}^{p_T}\hat\phi_je^\ast_{t-j}+\varepsilon^\ast_t$ be the sieve process, with $\varepsilon^\ast_t$ drawn independently from the centred residuals of the autoregression of order $p_T$ fitted to the residuals $\hat e_t$ of the regression in differences, and $u^\ast_t=\sum_{s\leq t}e^\ast_s$. Under the null, Assumption \ref{assu:T} and $p_T\to\infty$ with $p_T=o((T/\log T)^{1/2})$:
\begin{enumerate}
\item[(a)] $T^{-1/2}u^\ast_{[sT]}\Rightarrow\omega W_0(s)$ in $P$-probability;
\item[(b)] $\sum_t\Delta V_te^\ast_t=O_{p^\ast}(T^{1/2})$, $T^{-1}\sum_t\Delta V_tu^\ast_{t-1}=O_{p^\ast}(1)$, $T^{-1}\sum_tx_{t-1}e^\ast_t=O_{p^\ast}(1)$ and $T^{-1}\sum_tw_{t-1}e^\ast_t=O_{p^\ast}(T^{-1/2})$, in $P$-probability.
\end{enumerate}
\end{lem}

\noindent Proof. (a) Under the null, $\hat e_t=v_t-\Delta V_t'(\hat\kappa-\vartheta)$, up to the constant, with $\hat\kappa-\vartheta=O_p(T^{-1/2})$, because the regression in differences regresses $v_t$ on stationary regressors uncorrelated with it; hence $T^{-1}\sum_t(\hat e_t-v_t)^2=O_p(T^{-1})$. The autoregressive coefficients fitted to $\hat e_t$ and to $v_t$ then differ by $O_p(p_TT^{-1/2})$ in norm by the eigenvalue argument of Lemma \ref{lem:lrv}, which is $o_p(1)$ under the rate on $p_T$, and the invariance principle for the sieve bootstrap of \citet[Theorem 3.3]{Park2002} applies to the sieve fitted to $\hat e_t$, as in \citet{ChangPark2003}, with long-run variance converging to $\omega^2$.
(b) Conditional on the data the regressors are fixed, $E^\ast[e^\ast_t]=0$, and with probability tending to one the spectral density of $e^\ast_t$ is bounded by a constant $C$ \citep{Park2002}. For any fixed weights $a_t$, $\mathrm{Var}^\ast(\sum_ta_te^\ast_t)\leq2\pi C\sum_ta_t^2$. Taking $a_t=\Delta V_t$ gives $O_p(T)$, hence the first statement. Writing $\sum_t\Delta V_tu^\ast_{t-1}=\sum_se^\ast_s(V_T-V_s)$, with $\sum_s\|V_T-V_s\|^2=O_p(T^2)$ by Lemma \ref{lem:fclt}(a), gives the second. Taking $a_t=x_{t-1}$, with $\sum\|x_{t-1}\|^2=O_p(T^2)$, gives the third, and $a_t=w_{t-1}$, with $\sum\|w_{t-1}\|^2=O_p(T)$, the fourth. $\square$

\paragraph{Proof of Proposition \ref{prop:sieve}.} The bootstrap sample is $Y^{1\ast}_t=Y^1_1+\sum_{2\leq s\leq t}(\hat\kappa'\Delta V_s+e^\ast_s)$, with the regressors held at their observed values. Since the bootstrap statistic is invariant to the coefficients of the regressors, $P^\ast_T$ is the statistic computed from $Y^{1\ast}$ with cycle $u^\ast_t$, whose initial condition is zero. The decomposition \eqref{eq:Sdecomp} holds with $u^\ast$ in place of $u$ and the observed regressors, and Lemma \ref{lem:boot} provides, in $P^\ast$-probability and with probability tending to one, the orders of Lemma \ref{lem:fclt} on which Lemma \ref{lem:orders} rests, with $c=0$. The argument of Lemma \ref{lem:orders} and of the proof of Theorem \ref{thm:po} therefore applies conditionally, so $P^\ast_T=P^{u\ast}_T+o_{p^\ast}(1)$ in probability. The proof of Lemma \ref{lem:lrv} applies conditionally as well, with $e^\ast_t$ a finite autoregression of order $p_T$ whose coefficients converge to those of the autoregressive representation of $v_t$, so the bootstrap long-run variance estimator converges to $\omega^2$ in $P^\ast$-probability. By Lemma \ref{lem:boot}(a) and the continuous mapping theorem, $P^{u\ast}_T$ converges in distribution, in probability, to $\bar c^2\int_0^1W_0^2-\bar c\,W_0(1)^2$, the null limit of Theorem \ref{thm:po}(a), which has a continuous distribution. The bootstrap p-value is therefore asymptotically uniform under the null, and the test has asymptotic size equal to its nominal level \citep{Park2003}. $\square$

When the innovations of the cycle are correlated with those of the controls, the null distribution and the power of the test of Section \ref{sec:test} depend on the correlation, as for every test of the null of no cointegration \citep{Pesavento2004}. In the error-correction form of \eqref{eq:testmodel},
\[
\Delta Y^1_t=(\rho-1)(Y^1_{t-1}-\mu-b'Y^0_{t-1})+\kappa'\Delta Y^0_t+\eta_t,\qquad \kappa=b+\omega,
\]
where $\omega$ is the coefficient of the projection of the innovations of the cycle on those of the controls. The test of Section \ref{sec:test} imposes $\kappa=b$, which is the restriction implied by $\omega=0$. Two conditional versions use the correlation. CPO$_C$ adds $\Delta Y^0_t$, not quasi-differenced, to the regressions at $\bar\alpha$ and at 1, with a free coefficient. CPO$_K$ fixes the coefficient at its estimate $\hat\kappa$ from the regression of $\Delta Y^1_t$ on $\Delta Y^0_t$ and the changes of the policy terms, where it is identified under the null, and regresses $Y^1_t-\alpha Y^1_{t-1}-\hat\kappa'\Delta Y^0_t$ on the quasi-differenced constant, on $(1-\alpha)Y^0_{t-1}$ and on the quasi-differenced policy terms.

\begin{figure}[H]
\centering\safegraphic[width=\textwidth]{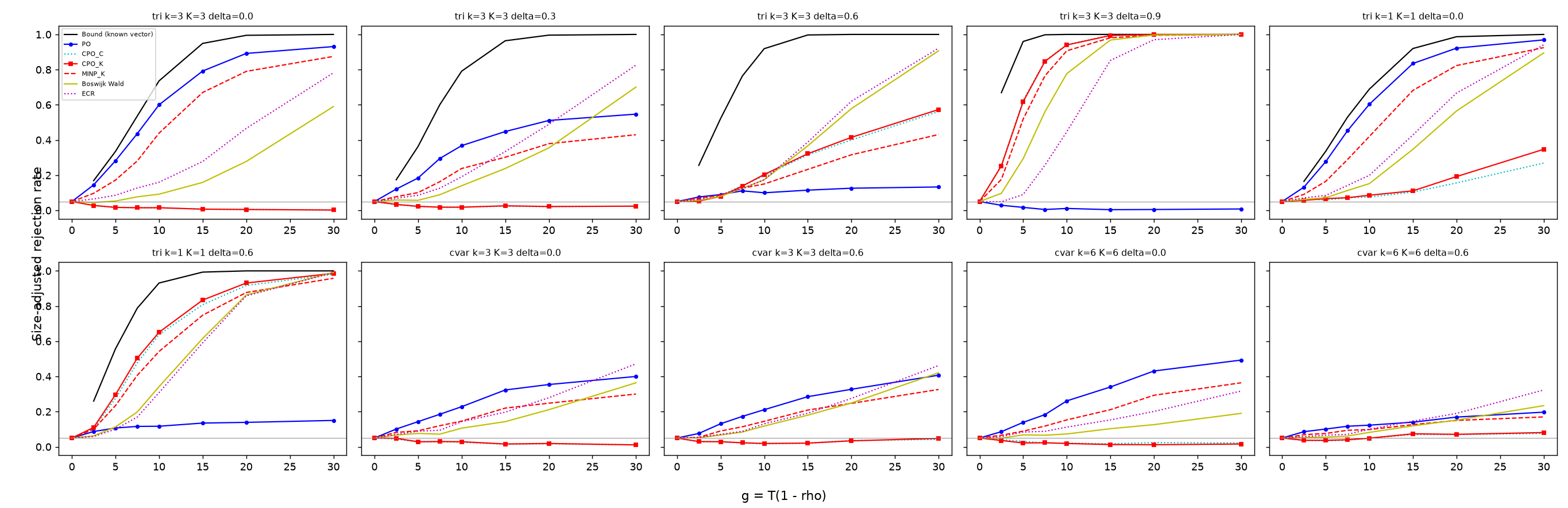}
\caption{Conditional and adaptive tests, $T=200$: size-adjusted rejection rates against $g=T(1-\rho)$.\label{fig:supp-cpo}}
\end{figure}

Table \ref{tab:supp-cpo} and Figure \ref{fig:supp-cpo} compare them, with $\bar c=-7$ for every point-optimal statistic, as in Section \ref{sec:test}. Both conditional tests are the most powerful feasible tests when the correlation is strong, but without correlation they reject less often under the alternative than under the null when there are several controls. The reason is common to both. Once the changes of the controls carry their own coefficient, the coefficient on the lagged levels of the controls is no longer tied to the coefficient on their changes, and under the null the lagged levels can fit part of the treated unit's own random walk. The test of Section \ref{sec:test} avoids the problem because the controls enter only quasi-differenced, so that the coefficient on their lagged levels is pinned down by the co-movement of the changes. Using the correlation therefore requires information on the weights, which is the role of the weighting over cointegrating vectors in \citet{ElliottPesavento2009}.

The adaptive test rejects when the smaller of the p-values of the test of Section \ref{sec:test} and of CPO$_K$ is below the 5 percent quantile of its null distribution, both computed under uncorrelated innovations, or from the sieve bootstrap. It holds its size, is never far from the better of the two tests, and gives up part of the power of the test of Section \ref{sec:test} when the correlation is absent. In the CVAR designs the test of Section \ref{sec:test} remains the most powerful even with correlation, so the adaptive test is the alternative for applications in which the treated unit is suspected to share specific shocks with a few of the controls.

\safeinput{table15.tex}

\section{Spillovers and non-linear effects\label{sec:spillnl}}

This section extends the CVAR to policies that spill over to the controls and to effects that are not linear in the policy surprise. Supplementary Appendix \ref{sec:montecarlo} reports simulations for both.

Part (ii) of Assumption \ref{assu:A1} excludes spillovers from the policy to the controls. In many empirical applications, the use of control units is the subject of debate, because a policy applied to one unit may spill over to its neighbours. This concern has given rise to a literature built around exposure mappings \citep{Manski2013,AronowSamii2017,Leung2020,forastiere2021identification,VazquezBare2023}. An exposure mapping is a function that links each control unit to the treated unit, usually calibrated on the distance between the two. Works in this literature generally assume that the exposure mapping is known, which turns spillovers into a problem akin to an omitted variable bias, and several papers show that including the exposure term in the main regression removes the bias that results from its omission \citep{DelgadoFlorax2015,Clarke2017,CaoDowd2019,Butts2021}.

The main difference between difference-in-differences or synthetic control and the setting considered here is that in macroeconomics it is often hard to find several counterfactuals with varying degrees of exposure to the spillover. One could argue that the economy of Georgia can act as a counterfactual for what happens in Tennessee, but it is harder to argue that New York can act as a counterfactual for Oklahoma, because the two economies are exposed in different ways to sectors such as agriculture and finance. Geographical distance affects both the size of the spillovers and the similarity of two economies. Accounting for spillovers is therefore first a task of the modeller, who must show that the treated unit and the controls are similarly exposed to the common trends that the CVAR removes. When this is done and two or more controls are available, the spillover can be specified through a spatial weight based on the squared distance between the treated unit and each control, as in spatial econometrics \citep{Anselin1988,LeSagePace2009}, and the effect of the policy can be recovered from the variation in exposure across controls. I follow this literature in making the following assumption.

\begin{assumption}[Exposure mapping\label{assu:S}]
For each control $k$ and every path, $Y^0_{kt}(w_{-\infty:t})-Y^0_{kt}(0)=\pi x_k\,[Y^1_t(w_{-\infty:t})-Y^1_t(0)]$, where $x_k\geq0$ is a known exposure of control $k$ to the treated unit, such as $x_k=(d_1/d_k)^2$ with $d_k$ the distance between the treated unit and control $k$, and $\pi$ is unknown. For each control, the cycle $G_{kt}=Y^1_t-b_kY^0_{kt}$ satisfies Assumptions \ref{assu:A2}, \ref{assu:A4} and \ref{assu:A5} with the weight $b_k$, and Assumption \ref{assu:A1} holds with its part (ii) replaced, for the controls, by the condition above.
\end{assumption}

In the application, the exposure of a control state is its squared distance from Alaska relative to the closest control, so that Washington, the state geographically closest to Alaska \citep{Kueng2018}, has the largest exposure and Minnesota the smallest. The parameter $\pi$ is the share of the effect on the treated state that reaches a control at the reference distance.

Let $Z_t=(R_t',W_t,G_{1t},\dots,G_{Kt})'$, let $u_t$ be the innovation of the policy in the recursive VAR in $Z_t$, defined as in \eqref{eq:innovation}, and let $g_{k,h}=\E[G_{k,t+h}u_t]/\E[u_t^2]$ be the response of the cycle with control $k$.

\begin{thm}[Spillovers\label{thm:spill}]
Under Assumptions \ref{assu:A3} and \ref{assu:S}, for every horizon $h\geq0$:
\begin{enumerate}
\item[(a)] $g_{k,h}=(1-b_k\pi x_k)\,\theta_h$ for every control $k$, and the response of the cycle with weights $b$ is $(1-\pi\,b'x)\,\theta_h$.
\item[(b)] If a control has $x_j=0$, then $\theta_h=g_{j,h}$.
\item[(c)] If the numbers $c_k=b_kx_k$ are not all equal, the linear projection of $g_{k,h}$ on a constant and $c_k$ across controls has intercept $\theta_h$ and slope $-\pi\theta_h$.
\end{enumerate}
\end{thm}

Part (a) shows that a spillover attenuates the response of the CVAR by the share of the effect that reaches the controls, weighted by their exposure. Parts (b) and (c) show how to remove the attenuation. The estimator for (c) estimates the weight of each control separately by \eqref{eq:dols}, estimates one VAR in the policies and the $K$ cycles, so that all the responses share the same policy innovation, and regresses the responses $\hat g_{k,h}$ on a constant and $\hat b_kx_k$ across controls, horizon by horizon. The intercept estimates the effect of the policy surprise, and minus the slope divided by the intercept estimates $\pi$. Bands follow from the bootstrap of Section \ref{sec:estimation} applied to the whole procedure. The simulations in Supplementary Appendix \ref{sec:montecarlo} show that the response of the cycle with the closest control is attenuated when spillovers are present, and that the exposure estimator removes the attenuation.

The second extension relaxes the linearity of Assumption \ref{assu:A3}, under which the effect of a dividend of two thousand dollars is twice that of one thousand, and the effect of a fall in the dividend is the opposite of that of a rise. When the effect is not linear, the response of the CVAR is a weighted average of marginal effects (Supplementary Appendix \ref{sec:identification}), and it does not give the response to a surprise of a given size. Let $\beta_h(\delta)$ be the response of the cycle at horizon $h$ to a policy innovation of size $\delta$.

\begin{assumption}[Innovation-driven non-linearity\label{assu:N}]
(i) The innovations $u_t$ are i.i.d. with finite variance and independent of $\mathcal{F}_{t-1}$ and of the counterfactual cycle $\{G^0_s\}_{s\geq t}$. (ii) The effect on the cycle is additive over dates, $G_{t+h}=G^0_{t+h}+\sum_{j\geq0}\beta_j(u_{t+h-j})$ with $\beta_j(0)=0$. (iii) For the horizons of interest $\beta_h(u)=\psi(u)'b^{(h)}$ for a dictionary $\psi=(\psi_1,\dots,\psi_L)'$ with $\psi_1(u)=u$, finite second moments and a non-singular covariance matrix.
\end{assumption}

In the application, part (iii) with $\psi(u)=(u,u^2)'$ allows the effect of an unexpectedly large dividend to grow more than proportionally with its size, and $\psi(u)=(u^+,u^-)'$ allows dividends above and below expectations to have effects of different size.

\begin{thm}[Non-linear impulse response\label{thm:nonlin}]
Let Assumptions \ref{assu:A1}, \ref{assu:A2} and \ref{assu:N} hold.
\begin{enumerate}
\item[(a)] The response of the linear CVAR at horizon $h$ is
\[
\gamma_h=\frac{\cov(\beta_h(u_t),u_t)}{\var(u_t)},
\]
and, when $\beta_h$ is absolutely continuous,
\[
\gamma_h=\frac{\int\beta_h'(w)\omega(w)\,dw}{\int\omega(w)\,dw},
\]
a weighted average of the derivative of $\beta_h$ with non-negative weights. It equals $\beta_h(\delta)/\delta$ for every size $\delta$ only if $\beta_h$ is linear.
\item[(b)] The regression of $G_{t+h}$ on a constant and $\psi(u_t)$ has coefficient vector $b^{(h)}$ on $\psi(u_t)$, so the non-linear impulse response $\beta_h(\delta)=\psi(\delta)'b^{(h)}$ is identified for every $\delta$ in the support. The same holds when $p$ lags of $Z_t$ are added to the regression, and when $u_t$ is replaced by the residual $\hat u_t$ of the policy equation.
\item[(c)] Linearity is the hypothesis $b_l^{(h)}=0$ for all $l\geq2$.
\end{enumerate}
\end{thm}

The estimator is a local projection of the cycle on a dictionary of the estimated policy innovation. Because the non-linearity enters only through the contemporaneous innovation, no transformed lags appear, which is the case in which linear local projections are consistent for the average derivative \citep{GoncalvesHerreraKilianPesavento2021}. The simulations in Supplementary Appendix \ref{sec:montecarlo} show that the linear CVAR misstates the response to surprises of opposite sign when the effect is quadratic, and that the dictionary estimator recovers it.

\paragraph{Proof of Theorem \ref{thm:spill}.} Let $\Delta_t=Y^1_t-Y^1_t(0)$ be the effect of the policy on the treated unit, and let $G^0_{kt}=Y^1_t(0)-b_kY^0_{kt}(0)$ be the counterfactual cycle formed with control $k$.

Step 1 (the cycles). By Assumption \ref{assu:S}, the outcome of control $k$ is its counterfactual plus the share of the effect that reaches it,
\[
Y^0_{kt}=Y^0_{kt}(0)+\pi x_k\Delta_t .
\]
Substituting in the cycle formed with control $k$,
\[
G_{kt}=Y^1_t(0)+\Delta_t-b_k\big(Y^0_{kt}(0)+\pi x_k\Delta_t\big)=G^0_{kt}+(1-b_k\pi x_k)\,\Delta_t .
\]
With the weights $b$ on all the controls, the same substitution gives
\[
G_t=Y^1_t-b'Y^0_t=G^0_t+(1-\pi\,b'x)\,\Delta_t ,
\]
where $G^0_t=Y^1_t(0)-b'Y^0_t(0)$.

Step 2 (the responses). Multiply the first display of Step 1, dated $t+h$, by $u_t$, take expectations and divide by $\E[u_t^2]$:
\[
g_{k,h}=\frac{\E[G^0_{k,t+h}u_t]}{\E[u_t^2]}+(1-b_k\pi x_k)\,\frac{\E[\Delta_{t+h}u_t]}{\E[u_t^2]} .
\]
The first term is zero by Assumption \ref{assu:A4} for the cycle formed with control $k$, and the second ratio is $\theta_h$ by \eqref{eq:estimand}. Hence $g_{k,h}=(1-b_k\pi x_k)\theta_h$. The same argument applied to the last display of Step 1 gives $(1-\pi b'x)\theta_h$ for the cycle with weights $b$, which proves (a).

Step 3 (identification). If $x_j=0$, then $g_{j,h}=\theta_h$, which is (b). For (c), write $c_k=b_kx_k$. By (a),
\[
g_{k,h}=\theta_h-\pi\theta_h\,c_k ,\qquad k=1,\dots,K ,
\]
so the points $(c_k,g_{k,h})$ lie on a line with intercept $\theta_h$ and slope $-\pi\theta_h$. When the $c_k$ are not all equal, the linear projection of $g_{k,h}$ on a constant and $c_k$ across controls fits the line exactly and returns its intercept and slope. All the responses are defined with the same innovation $u_t$, because they come from one VAR in $Z_t$. $\square$

\paragraph{Proof of Theorem \ref{thm:nonlin}.} Step 1 (the conditional mean of the cycle). By Assumption \ref{assu:N}(ii), at horizon $h$
\[
G_{t+h}=G^0_{t+h}+\beta_h(u_t)+\sum_{j\neq h}\beta_j(u_{t+h-j}) .
\]
By Assumption \ref{assu:N}(i), $u_t$ is independent of $G^0_{t+h}$ and of the innovations at every other date, so
\[
\E[G_{t+h}\mid u_t]=c_h+\beta_h(u_t),
\]
where the constant $c_h$ collects the means of the other terms.

Step 2 (the linear response). By the second part of the proof of Theorem \ref{thm:cycle} and Step 1,
\[
\gamma_h=\frac{\E[G_{t+h}u_t]}{\E[u_t^2]}=\frac{\cov(\beta_h(u_t),u_t)}{\var(u_t)} .
\]
When $\beta_h$ is absolutely continuous, write $\beta_h(u)=\int_0^u\beta_h'(w)\,dw$. Exchanging the order of integration (Fubini),
\[
\cov(\beta_h(u_t),u_t)=\int\beta_h'(w)\,\omega(w)\,dw,\qquad
\omega(w)=\begin{cases}\E[u_t\mathbf{1}\{u_t\geq w\}], & w\geq0,\\ -\E[u_t\mathbf{1}\{u_t<w\}], & w<0,\end{cases}
\]
and $\omega(w)\geq0$ because $\E[u_t]=0$. The same identity with $\beta_h(u)=u$ gives $\var(u_t)=\int\omega(w)\,dw$ \citep{Yitazaki1996}, which proves the weighted-average form. The weights do not depend on $t$ because the innovations are identically distributed. If $\gamma_h=\beta_h(\delta)/\delta$ for every $\delta$, then $\beta_h(\delta)=\gamma_h\delta$, so $\beta_h$ is linear. This proves (a).

Step 3 (the dictionary regression). By Assumption \ref{assu:N}(iii) and Step 1, $G_{t+h}=c_h+\psi(u_t)'b^{(h)}+\eta_{t+h}$ with $\E[\eta_{t+h}\mid u_t]=0$. The regressors $\psi(u_t)$ have a non-singular covariance matrix, so the population regression of $G_{t+h}$ on a constant and $\psi(u_t)$ has coefficient vector $b^{(h)}$ on $\psi(u_t)$, and $\beta_h(\delta)=\psi(\delta)'b^{(h)}$ is identified for every $\delta$ in the support.

Step 4 (lags as controls). The lags of $Z_t$ are measurable with respect to $\mathcal F_{t-1}$, so they are independent of $u_t$ by Assumption \ref{assu:N}(i). In the regression with a constant, $\psi(u_t)$ and the lags are therefore uncorrelated, and by the Frisch-Waugh-Lovell theorem adding the lags leaves the coefficients on $\psi(u_t)$ unchanged.

Step 5 (the estimated innovation). When the policy equation spans the predictable part of the policy, $\hat u_t-u_t=x_t'(\pi-\hat\pi)$ with $\hat\pi-\pi=O_p(T^{-1/2})$. If each $\psi_l$ is Lipschitz, or continuously differentiable with a derivative whose growth is bounded by a polynomial and $u_t$ and $x_t$ have the corresponding moments, then by the mean value theorem and the Cauchy-Schwarz inequality
\[
\frac1T\sum_t\big\|\psi(\hat u_t)-\psi(u_t)\big\|^2=o_p(1),
\]
so the sample moments of the regression with $\psi(\hat u_t)$ differ from those with $\psi(u_t)$ by $o_p(1)$, and the least-squares coefficients converge to $b^{(h)}$. Both dictionaries of the application, $(u,u^2)'$ and $(u^+,u^-)'$, satisfy the condition.

Part (c) follows from (b), since $\beta_h$ is linear if and only if the coefficients on $\psi_2,\dots,\psi_L$ are zero. $\square$

\section{Monte Carlo simulations\label{sec:montecarlo}}

\begin{figure}[H]
\centering\safegraphic[width=\textwidth]{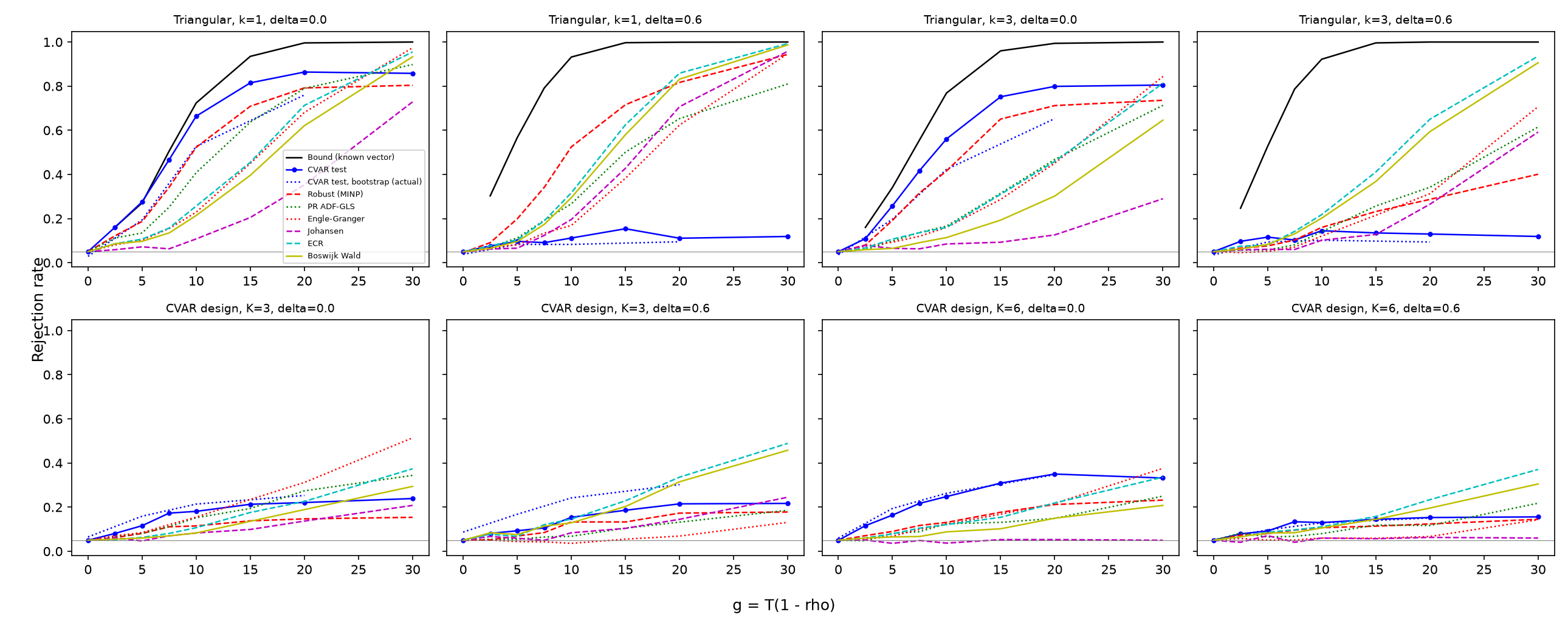}
\caption{Test of the design, $T=200$: rejection rates against $g=T(1-\rho)$. Size-adjusted power for all tests except the bootstrap, whose curve shows actual rejection rates.\label{fig:mc-test-T200}}
\end{figure}

\begin{figure}[H]
\centering\safegraphic[width=\textwidth]{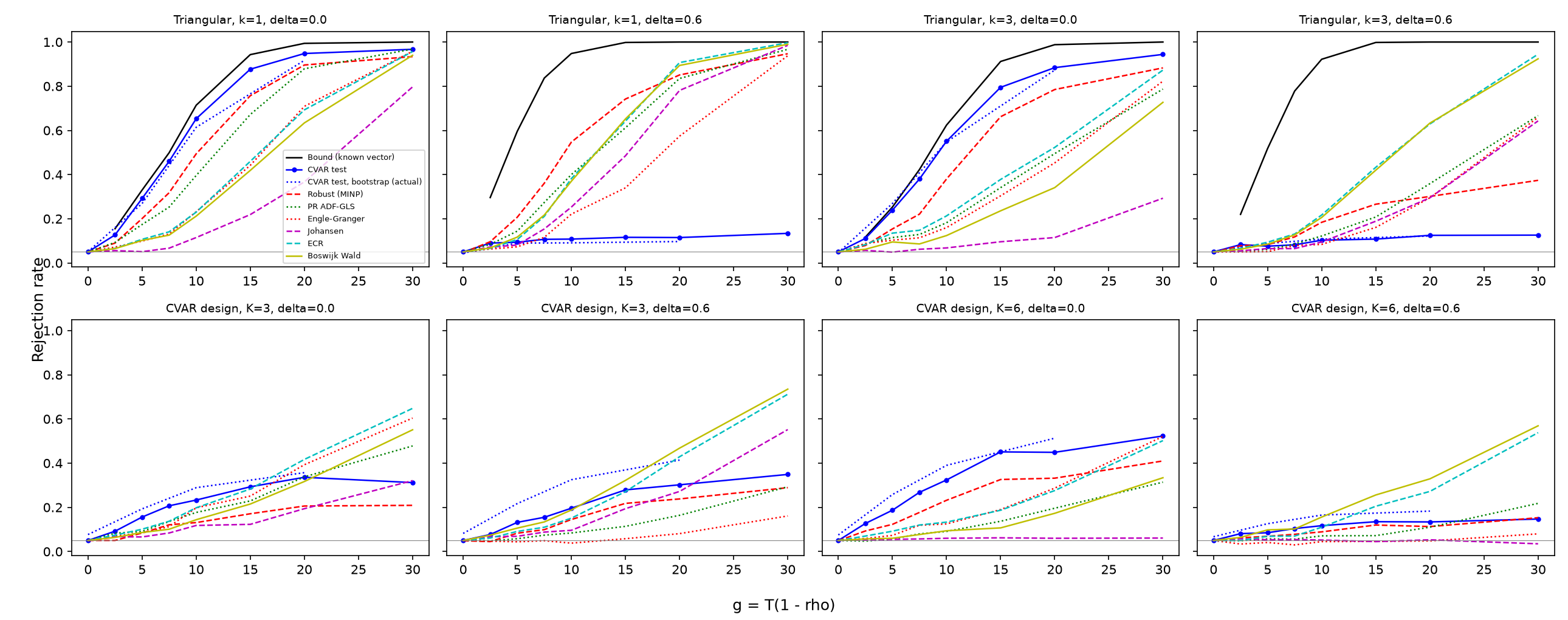}
\caption{Test of the design, $T=400$. As in Figure \ref{fig:mc-test-T200}.\label{fig:mc-test-T400}}
\end{figure}

The simulations answer six questions, in the order of the paper: how far the response of a VAR without controls is from that of the CVAR (Table \ref{tab:mc-var}), how the lag length affects the response (Table \ref{tab:mc-lags}), whether the bootstraps that estimate the weights cover the true response (Tables \ref{tab:mc-boot}, \ref{tab:mc-grid} and \ref{tab:mc-joint}), how the extensions to spillovers and non-linear effects and the diagnostics of stationarity perform (Tables \ref{tab:mc-spill}, \ref{tab:mc-nonlin} and \ref{tab:mc-stat}), how the test of the design behaves (Tables \ref{tab:mc-test-size} to \ref{tab:mc-test-feedback} and Figures \ref{fig:mc-test-T200} and \ref{fig:mc-test-T400}), and how the CVAR in changes compares with the alternatives (Table \ref{tab:mc-diff}). The first three use one data generating process, with the notation of the paper, calibrated to the application to unemployment benefits; the extensions use variants of it, and the test of the design and the CVAR in changes use the designs described with their results. The common trends follow
\begin{equation}
\Delta F_t=0.3\,\Delta F_{t-1}+\eta_t,
\end{equation}
with $\eta_t\sim N(0,I_r)$. The controls and the treated unit without the policy are
\begin{equation}
Y^0_t=\Lambda F_t+v_t,\qquad Y^1_t(0)=\lambda'F_t+e_t,\qquad \lambda=\Lambda'b,
\end{equation}
with $v_{kt}=0.5\,v_{k,t-1}+\xi_{kt}$, $e_t=0.5\,e_{t-1}+\zeta_t$, standard normal innovations and equal weights $b=\iota_K/K$. When $r=K$, $\Lambda=I_K+0.3\,\iota_K\iota_K'$. When $r<K$, the first $r$ rows of $\Lambda$ are $I_r+0.3\,\iota_r\iota_r'$ and the remaining rows are $(1/r+0.3)\iota_r'$, so that the controls have $K-r$ cointegrating relations among themselves. The two policies react to the common shocks,
\begin{align}
R_t&=0.7\,R_{t-1}+0.5\,\iota_r'\eta_t/\sqrt{r}+\varepsilon^R_t,\\
W_t&=0.5\,W_{t-1}+0.5\,\iota_r'\eta_t/\sqrt{r}+0.3\,R_t+\phi\,G_{t-1}+\varepsilon^W_t,
\end{align}
where the second policy also reacts to the lagged cycle with coefficient $\phi$, as extended benefits react to the recent unemployment of a state. The effect of the policy is
\begin{equation}
Y^1_t=Y^1_t(0)+E_t,\qquad E_t=0.8\,E_{t-1}+0.2\,W_t.
\end{equation}
The calibration closest to the application has $T=200$, $K=6$, $r=5$ and $\phi=0.3$: six controls with one cointegrating relation among them, as the trace test finds for three of the five treated states. The true responses $\theta_h$ and $\gamma_h$ are computed from a sample of 100,000 observations. Each experiment has 500 replications with 499 bootstrap draws, and the bands are 90 percent percentile bands.

The first simulation compares the response of a VAR without controls in $(R_t,W_t,Y^1_t)$ with the CVAR, both with twelve lags, in the calibration closest to the application. Table \ref{tab:mc-var} reports the bias, the root mean squared error and the coverage of the 90 percent bands for the effect of a policy surprise $\theta_h$. The VAR without controls nearly doubles the effect on impact, and its bias persists for more than a year, because the policy reacts to the common shocks and the response inherits their path. Its bands rarely cover the true effect in the first months, and they reach the nominal coverage only at long horizons, where they are much wider than those of the CVAR. The CVAR is close to unbiased at every horizon, with coverage close to the nominal level and a much smaller root mean squared error.

\safeinput{tab_mc_var.tex}

The second simulation compares the CVAR with the lag length selected by the Akaike criterion and fixed at 8 and 12 lags. Table \ref{tab:mc-lags} reports the results. The Akaike criterion selects very few lags, and the response is biased toward zero beyond the first months, with coverage far below the nominal level after a year. With 12 lags the bias disappears up to a year, and the coverage is close to the nominal level at every horizon up to a year. The Ljung-Box test on the residuals of the equation of the cycle never signals the misspecification, at any of the three lag lengths.

\safeinput{tab_mc_lags.tex}

The third simulation compares bootstraps with 12 lags that treat the weights in different ways: one with the weights known, one that holds the estimated weights fixed, one that re-estimates them in every draw with the controls held at their observed values, and the joint bootstrap of Theorem \ref{thm:joint}, which regenerates the controls and re-estimates the weights. The grid compares the first three across seventeen designs, which combine $T=200$ and $400$, three and six controls, controls with and without cointegrating relations among them, policies with and without feedback from the lagged cycle, and a design in which one of the true weights is zero, at the boundary of the constraint. Table \ref{tab:mc-boot} reports the average coverage across designs, and Table \ref{tab:mc-grid} reports every design. Re-estimating the weights with the controls held fixed matches the coverage obtained with the weights known at every horizon up to a year, in every design, including the one at the boundary, which Theorem \ref{thm:joint} excludes, and holding the weights fixed lowers the coverage, most in the designs with six controls. Table \ref{tab:mc-joint} adds the joint bootstrap in the design with six controls and one cointegrating relation among them, with feedback from the lagged cycle, at both sample sizes and with a stronger feedback. Its coverage is close to that obtained with the weights known at every horizon up to a year in all three designs, although its VAR includes the controls, and its bands are of similar width. The recentred version of Theorem \ref{thm:joint} and the version without recentring have coverage within simulation error of each other and bands of the same width, so recentring costs nothing in these designs. In these designs the first-step term is small relative to the sampling error of the VAR, so the bootstraps that hold the controls fixed also come close to the nominal coverage, and the gap of the bootstrap with fixed weights disappears at $T=400$. Two years after the surprise, all bootstraps, including the one with the weights known, under-cover in the designs with feedback, where the response decays slowly and twelve lags leave some truncation, as Theorem \ref{thm:joint} anticipates for a lag length that does not grow with the sample.

\safeinput{tab_mc_boot.tex}

\safeinput{tab_mc_grid.tex}

\safeinput{tab_mc_joint.tex}

The remaining simulations turn to the extensions and to the test of the design. The spillover experiment uses five controls that share one common trend with the treated unit, at increasing distance from it, and lets each control receive the share $\pi x_k$ of the effect on the treated unit, with $x_k=(d_1/d_k)^2$. Table \ref{tab:mc-spill} shows that without spillovers all estimators are close to unbiased, and the CVAR with all controls is the most precise. With spillovers, the CVAR with the nearest control is strongly attenuated and the CVAR with all controls moderately so, while the exposure estimator of Theorem \ref{thm:spill} remains unbiased, with a precision close to that of the most distant control.

The non-linear experiment lets the effect of the policy depend on $u+\kappa u^2$. Table \ref{tab:mc-nonlin} shows that when the effect is linear the linear CVAR is unbiased and more precise than the dictionary estimator. When it is quadratic, the linear CVAR understates the response to large surprises and misstates the asymmetry between positive and negative surprises, while the dictionary estimator of Theorem \ref{thm:nonlin} is close to unbiased for surprises of both signs and of different size.

The stationarity experiment replaces the idiosyncratic component of the treated unit with an autoregression of coefficient $\rho$. Table \ref{tab:mc-stat} shows that the Engle-Granger critical values are not valid for the cycle of the CVAR: because the weights are estimated with the policy terms and the constraint, the ADF test rejects a true unit root far more often than its nominal level. With simulated critical values the test has the correct size by construction, and its power against a persistent but stationary cycle is low with two hundred observations and high with four hundred. The KPSS test on the cycle has a size close to the nominal level at the standard critical value, but little power. Netting out the policy terms raises the power of both tests at simulated critical values.

\safeinput{tab_mc_spill.tex}

\safeinput{tab_mc_nonlin.tex}

\safeinput{tab_mc_stat.tex}

The simulations for Section \ref{sec:test} use two designs. In the triangular design, $Y^1_t=0.5+b'x_t+u_t$, the $k$ regressors $x_t$ are random walks with standard normal innovations, $b$ has elements $1/k$, $u_t=\rho u_{t-1}+v_t$ with $u_0=0$, and the innovations of $v_t$ have total correlation $\delta$ with those of $x_t$, spread equally across the regressors. In the CVAR design, the treated unit and the controls follow the data generating process of this section, with persistent idiosyncratic components and two policies that react to the common shocks, and the treated unit receives a further component $u_t$ whose innovations have correlation $\delta$ with the average change of the controls. In both designs $\rho=1-g/T$, so that $g=0$ is the null hypothesis of a trend specific to the treated unit and $g>0$ a local alternative under which the CVAR is valid.

Table \ref{tab:mc-test-size} reports the rejection rates under the null. In the triangular designs the test with the critical values of \citet{ERS1996} has close to the nominal size for one and three regressors, as Theorem \ref{thm:po}(a) implies. In the CVAR design, the persistent idiosyncratic components make the asymptotic critical value too permissive in samples of this length, and the sieve bootstrap with the lag rule of Section \ref{sec:test} brings the rejection rate close to the nominal level. In the triangular designs, where the cycle is serially uncorrelated, the bootstrap is conservative, because the rule fits a longer autoregression than the dynamics require. The residual-based test of \citet{EngleGranger1987} and the system test of \citet{Johansen1991} reject a true null far more often than their nominal level in the CVAR design, and the latter almost always with six controls. Table \ref{tab:mc-test-power} and Figures \ref{fig:mc-test-T200} and \ref{fig:mc-test-T400} report power. Without correlation between the innovations, the CVAR test is close to the bound computed with the true weights, the gap narrows as the sample grows, and the test is far more powerful than every competitor. With correlation the test loses its power, which the adaptive test of Appendix \ref{sec:supp-test} recovers.

Table \ref{tab:mc-test-lags} compares orders of the autoregression in the sieve bootstrap. Short orders leave the test oversized and long orders cost power; orders around twenty with four hundred observations, which the rule of Section \ref{sec:test} delivers, control size in every design while keeping most of the power. Table \ref{tab:mc-test-feedback} reports designs in which the policy reacts to the treated unit. A reaction to the twelve-month change of the treated unit, as in the triggers of extended benefits, leaves the size of the test close to the case without feedback and makes it conservative when the cycle is serially correlated. The augmented statistic, which adds lags of $\Delta Y^1_t$ to both regressions, is oversized and less powerful, so the paper does not use it.

\safeinput{table11.tex}

\safeinput{table12.tex}

\safeinput{table13.tex}

\safeinput{table14.tex}

Finally, the simulations for Theorem \ref{thm:diff} use a data generating process with one common trend, $F_t=F_{t-1}+\eta_t$, $K$ controls $Y^0_{kt}=F_t+v_{kt}$ with $v_{kt}=0.5v_{k,t-1}+\sigma_v z_{kt}$, and a treated unit $Y^1_t=F_t+e_t+\sum_{j=0}^{2}\tau_jW_{t-j}$ with $\tau=(0.5,0.3,0.1)$. The policy reacts to the common shock, $W_t=0.7W_{t-1}+c\,\eta_t+\varepsilon_t$, and all innovations are independent standard normal. The component specific to the treated unit is a random walk, $e_t=e_{t-1}+\sigma_e\zeta_t$, in the design with a trend of the treated unit, and $e_t=0.8e_{t-1}+\sigma_e\zeta_t$ in the stationary design. The true cumulated response is $\Gamma_h=\sum_{j=0}^{\min(h,2)}\tau_j0.7^{h-j}$, and a VAR without controls adds $c/(1+c^2)$ to it at every horizon. Four designs vary one feature at a time: a trend of the treated unit with four controls, $\sigma_v=0.5$, $\sigma_e=0.3$ and $c=1$; a stationary cycle with $\sigma_e=0.5$; a single control with $\sigma_v=2$, which adds more variance to the cycle than the common trend it removes; and no reaction of the policy to the common shock, $c=0$. Table \ref{tab:mc-diff} compares the CVAR in changes with equal weights, the VAR without controls in changes and the CVAR in levels with estimated weights. The CVAR in changes is close to unbiased in every design, and its bands have close to nominal coverage, slightly below it at intermediate horizons when the treated unit has a trend of its own. The VAR without controls carries the common-trend term whenever the policy reacts to the common shock, and its bands then almost never contain the true response. When the policy does not react, it is unbiased and its coverage is close to nominal. The CVAR in levels with estimated weights gives accurate point estimates in every design, but when the treated unit has a trend of its own its bands at impact contain the true response in a minority of replications, and its coverage is below the nominal level at the other horizons, because the weights come from a regression without a cointegrating relation. When the cycle is stationary it is valid and, at long horizons, more precise than the CVAR in changes. This is the trade-off that the test of Section \ref{sec:test} resolves. The width of the bands depends on how much variance the controls add to the cycle. With four controls whose idiosyncratic components are small, the bands of the CVAR in changes are considerably narrower than those of the VAR without controls, also in the design in which the policy does not react to the common shock and the VAR without controls is unbiased. With a single control whose idiosyncratic component is large, the cycle is more volatile than the outcome of the treated unit, and the bands of the CVAR are wider at impact and of similar width at longer horizons.

\safeinput{tab_mc_diff.tex}

\section{How often do units share all their stochastic trends?\label{sec:screen}}

The CVAR in levels, like any comparison in levels with untreated units, requires that the weighted controls reproduce every stochastic trend of the treated unit. I check how often the data support this requirement by applying the test of Section \ref{sec:test} to the states of the United States, with no policy terms, each state compared with the states it borders. For each state and variable the table reports three tests: the state alone, which should not reject when its series has a stochastic trend; the state against the average of its neighbours; and the state against its neighbours with weights estimated inside the test. The variables are the employment-population ratio and the participation rate from the Local Area Unemployment Statistics, the logarithm of payroll employment from the Current Employment Statistics, the logarithm of the state coincident index of the Federal Reserve Bank of Philadelphia, the logarithm of total personal income from the Bureau of Economic Analysis, and the logarithm of the all-transactions house price index of the Federal Housing Finance Agency, each from the start of its series to the end of 2019. There are no consumer price indices for states, so for prices each of the four Census regions is compared with the other three, and New York, Chicago and Los Angeles, the metropolitan areas with monthly indices over the whole period, with each other.

\safeinput{tab_screen.tex}

Every series behaves as integrated on its own: the test of the unit alone almost never rejects. The gap between a unit and its neighbours, by contrast, rejects about as often as the nominal size of the test for output, income, house prices and consumer prices, and only somewhat more often for the two labour force rates, whose bounded range limits how far a state can drift from its neighbours. Even consumer prices, observed monthly for more than eighty years in the three metropolitan areas, show no rejection, in line with the slow convergence of regional price levels documented for the United States. The few states whose labour force rates keep a stationary gap with their neighbours include Indiana and Illinois, which share the labour market of Chicago.

\safeinput{tab_screen_power.tex}

How much a non-rejection in Table \ref{tab:screen} says depends on the power of the test against gaps that return to their mean slowly. Table \ref{tab:screen-power} reports it in simulated panels built like the state data. The test keeps its nominal size, and its power depends on the persistence of the gap relative to the length of the sample, as the theory of Section \ref{sec:test} implies: it rejects most of the time when the gap has a half-life of about a year and four decades of monthly data are available, and much less often when the half-life is several years. A sieve bootstrap built from the residuals of the autoregression in levels recovers the power of the asymptotic critical values but rejects a true null more than twice as often as its nominal level, which weakens the protection against estimating in levels a treated unit with a trend of its own. The rare rejections of Table \ref{tab:screen} therefore leave two readings open, a stochastic trend specific to most units or gaps that return to their mean over many years, and the data cannot tell them apart. Under either reading the assumption that the controls reproduce every trend of the treated unit cannot be supported from the data, which is why the applications use the CVAR in changes.

\section{Alaska: further results\label{sec:app-eb}}

Tables \ref{tab:ak-adjusted} and \ref{tab:ak-bounds} complete the application to Alaska in Section \ref{sec:alaska}: the responses adjusted for the common movement that the estimated weights leave in the cycle, and the bounds of Proposition \ref{prop:bounds}.

\safeinput{tab_alaska_adjusted.tex}

\safeinput{tab_alaska_bounds.tex}

Three further checks concern the sample, the donor pool and the treatment of seasonality. Table \ref{tab:ak-samples} re-estimates the main specification on samples ending in December 2015, before the legislature set the amounts of the dividend, and in December 2007, before the dividend, the energy rebate and the recession of 2008, with the weights re-estimated in each sample. The shorter samples contain fewer dividends, and the payroll series start in 1990.

\safeinput{tab_alaska_samples.tex}

Table \ref{tab:ak-pools} estimates the weights from the 12 other states of the Census West region and from all the other states and the District of Columbia. The larger pools reproduce a larger share of the main common movement of Alaska, and the full pool more than all of it, with weights on states far from Alaska in geography and economic structure. The responses with the larger pools are close to those of the main specification and of its adjusted version in Table \ref{tab:ak-adjusted}.

\safeinput{tab_alaska_pools.tex}

The main specification uses seasonally adjusted series. The dividend is paid every October, so its timing coincides with the seasonal pattern of Alaska, and only the variation in its size identifies its effect. The seasonal adjustment removes the average October pattern together with any part of a persistent dividend effect that its moving seasonal factors absorb. Month effects in the unadjusted series remove only the seasonal pattern that they specify. Table \ref{tab:ak-seasonal} reports the main specification with month effects added to the adjusted series and with the unadjusted series and month effects, constant, interacted with a linear trend, and by decade, in the regression of the weights and in the VAR. Month effects leave the results on the adjusted series unchanged. With the unadjusted series, the employment rate rises and unemployment falls in the first three months after the payment, leisure and hospitality employment rises, and the responses of participation and of total payroll employment lose their significance.

\safeinput{tab_alaska_seasonal.tex}

\section{Extended unemployment benefits\label{sec:eb}}

This section reports in full the application that Section \ref{sec:ebsum} of the paper summarises. In the United States, unemployment insurance replaces part of the income of workers who lose their jobs, for up to 26 weeks in most states. Since 1970 the Extended Benefits programme (EB) has added weeks of benefits during recessions, and in most recessions Congress has also created a temporary federal programme, most recently Emergency Unemployment Compensation (EUC), which ran from July 2008 to December 2013. Extensions switch on through trigger rules that take as inputs the recent unemployment of each state.

Whether extensions, designed to support workers, also prolong periods of high unemployment is a long-running debate, and neither the theory nor the evidence has settled it. A large microeconometric literature finds that longer potential durations lengthen individual unemployment spells \citep{SchmiederVonWachter2016}. What holds for one worker need not hold for the labour market as a whole, because the channels that operate in the aggregate push in different directions. Extensions transfer income to households that spend most of it, and can act as fiscal stimulus in a recession \citep{McKayReis2016,Kekre2023}. Longer benefits can also improve the outside option of workers in wage bargaining and discourage firms from posting vacancies \citep{HagedornEtAl2019}. Which of these channels dominates is an empirical question.

The empirical question has a clear endogeneity problem. Benefits are extended precisely when the economy enters a recession and unemployment is high and rising, so a comparison of unemployment with and without extensions mixes their effect with the recession that triggered them. The literature has addressed this reverse causality in three ways. Border designs compare neighbouring counties on opposite sides of a state line \citep{BooneEtAl2021,DieterleEtAl2020}, and are exposed to workers who commute across the border and carry the effect of the policy into the comparison county \citep{DieterleEtAl2020}. \citet{ChodorowReichEtAl2019} use revisions of the data that set off the triggers, which produce variation that is short-lived. \citet{AcostaEtAl2025}, building on \citet{Rothstein2011}, compare states that faced the same trigger conditions but had adopted different optional triggers. Most estimates for the Great Recession are close to zero, with exceptions that find large effects \citep{HagedornEtAl2019,JohnstonMas2018}. \citet{AcostaEtAl2025} find that a standard extension raises unemployment when the duration of benefits is short to begin with, and has virtually no effect when durations are already long, as they were in the Great Recession.

The CVAR addresses the reverse causality directly. A national recession raises unemployment in every state, and the triggers switch on the extensions in the states whose unemployment crosses the thresholds set in federal law, while the others go through the same recession without them. A VAR without controls cannot solve this problem. It does not control for the common trends, so by Theorem \ref{thm:var} its response contains the reaction of the extensions to the national cycle, including the recovery that follows the downturn in which they switched on, and it attributes that recovery to the extensions even if they have no effect at all. The endogeneity of the policy is exactly what a VAR on the data of the treated state alone leaves unaddressed. Comparing each state that received extensions with states that did not removes that reaction, and choosing comparison states that do not border it addresses the spillovers that affect border designs. I estimate the CVAR and a VAR without controls on the same data, state by state and on average.

The assumptions of Section \ref{sec:assumptions} take a specific form for extended benefits. Extensions switch on and off through weekly trigger notices, so at a monthly frequency workers and firms have little scope to respond to an extension before it starts (Assumption \ref{assu:A1}(i)). The triggers are averages of insured unemployment over the previous thirteen weeks, so an extension responds to the recent past of the state, which the VAR conditions on, and the data that set them off are measured in real time with errors that are unrelated to economic conditions \citep{ChodorowReichEtAl2019}; both features support Assumption \ref{assu:A4}. The triggers also compare current unemployment with its average over the previous two years, so the policy is a function of changes in unemployment and remains stationary even when a state has a trend of its own, as part (i) of Assumption \ref{assu:T} requires.

To compare each state with its controls, I use monthly data from October 2002 to December 2019, which end before the pandemic programmes. The outcome is the state unemployment rate from the Local Area Unemployment Statistics.\footnote{Seasonally adjusted series of the Local Area Unemployment Statistics, retrieved from FRED, series \texttt{TXUR} for Texas and the analogous code for the other states: \url{https://fred.stlouisfed.org/series/TXUR}.} The policy $W_t$ is the number of weeks of EB available in the state, averaged over the weeks of the month and divided by ten, built from the trigger notices of the Department of Labor.\footnote{Trigger notices of the Department of Labor, \url{https://oui.doleta.gov/unemploy/claims_arch.asp}.} It coincides with the series in the replication files of \citet{BooneEtAl2021} where the two overlap. The other policy $R_t$ is the number of weeks of EUC from the same replication files, and zero outside the period in which EUC existed. Appendix \ref{sec:data} lists the sources.

The controls are the seven states in which EB never switched on: South Dakota, North Dakota, Iowa, Nebraska, Wyoming, Utah and Oklahoma. Every other state with at least twelve months of EB in the sample is a treated state, so that no state is selected by its results. For each treated state I drop the controls that border it, which removes, for example, Nebraska and Oklahoma for Kansas, and Oklahoma for Texas.

\begin{figure}[htbp]
\centering\safegraphic[width=0.85\textwidth]{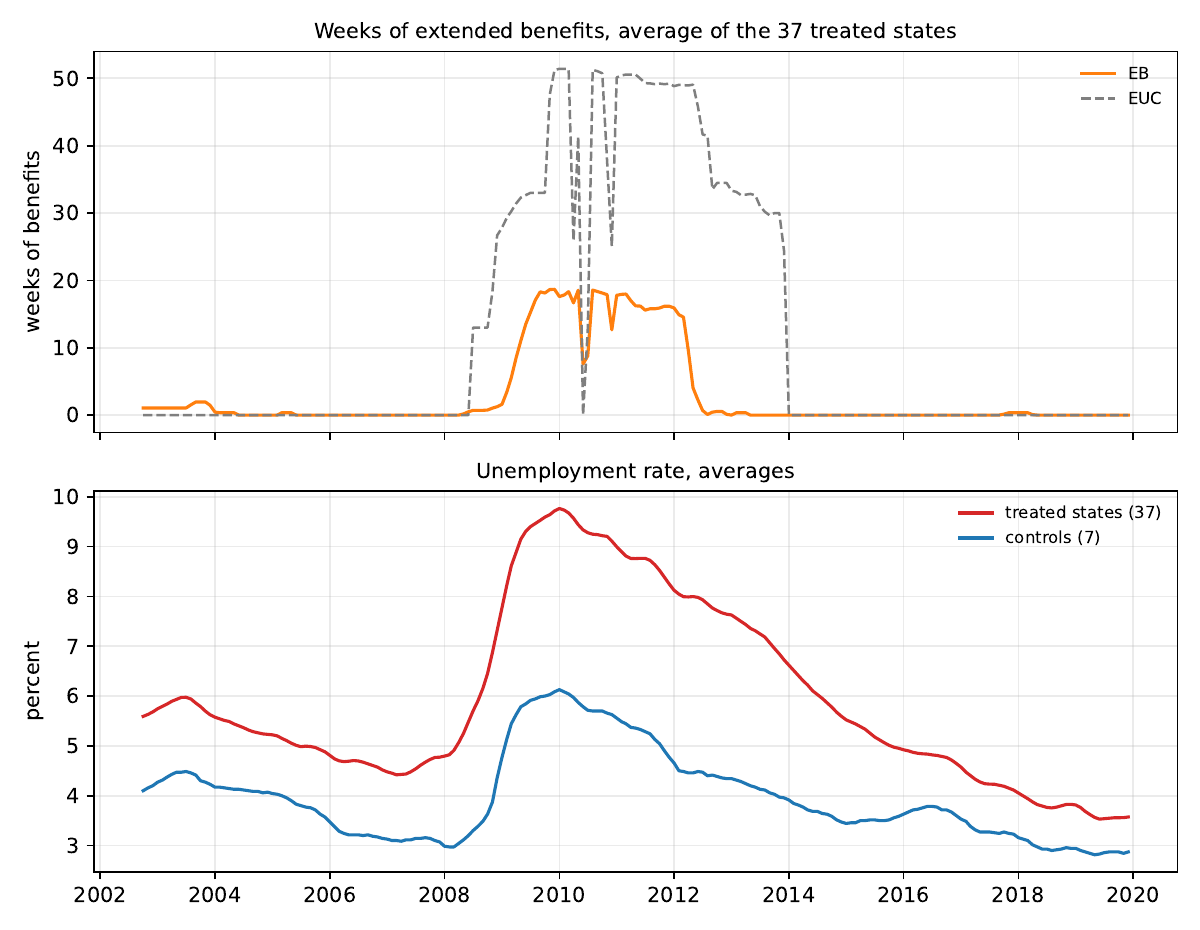}
\caption{Extended benefits: policy and outcomes. Top: weeks of EB and of EUC available, averaged over the treated states. Bottom: unemployment rate, average of the treated states and average of the controls, the states in which EB never switched on. Monthly, October 2002 to December 2019.\label{fig:eb-data}}
\end{figure}

Figure \ref{fig:eb-data} shows the policies and the outcome. Nearly all the weeks of EB fall between 2009 and 2012, while EUC was also in force. The lower panel compares the average unemployment rate of the treated states with that of the controls: the two share the movements of the national cycle, which the CVAR removes, and the treated states rise further in the Great Recession, which is the component that the test of the design attributes to the treated states.

Before estimating the CVAR, the test of Section \ref{sec:test} checks whether each treated state shares its trends with its controls, and its outcome decides which version of the CVAR to estimate. Table \ref{tab:eb-tests} reports it, together with tests for a unit root in each unemployment rate. The unemployment rates behave as integrated series over the sample: the test of \citet{ERS1996} rejects a unit root for few of them, so the framework of common stochastic trends applies. The test of the design rejects no more often than its size would produce by chance, with weights estimated by \eqref{eq:dols} or set equal, with or without level shifts, and for every number of lags of the policies. The controls are rural states of the Great Plains and the Mountain West, where the Great Recession was mild, and the depth of the recession in each treated state leaves in its unemployment a slow-moving component that the controls do not reproduce. Following Section \ref{sec:test}, I estimate the CVAR in changes, with equal weights on the controls of each treated state (Theorem \ref{thm:diff}). The arguments given for Assumption \ref{assu:A4} carry over to Assumption \ref{assu:A4d}, because the triggers respond to the recent past of insured unemployment, which the VAR conditions on through the past changes of the cycle.

\safeinput{tab_eb_tests.tex}

The CVAR in changes then takes a simple form. For each treated state $s$ with controls $k=1,\dots,K_s$, the cycle is
\begin{equation}
G_{st}=Y_{st}-\frac{1}{K_s}\sum_{k=1}^{K_s}Y^0_{kt},
\end{equation}
and the VAR is
\begin{equation}
\begin{pmatrix}R_{st}\\W_{st}\\\Delta G_{st}\end{pmatrix}=c_s+\sum_{l=1}^{p_s}A_{sl}\begin{pmatrix}R_{s,t-l}\\W_{s,t-l}\\\Delta G_{s,t-l}\end{pmatrix}+\epsilon_{st},
\end{equation}
with EUC ordered first, so that the EB innovation is net of the contemporaneous movement in EUC. The response is the cumulated response $\Gamma_h$ of \eqref{eq:cumcycle} to an EB innovation of ten weeks. The VAR without controls replaces $\Delta G_{st}$ with $\Delta Y_{st}$. The lag order of each model is the smallest order up to twelve at which the test of \citet{LjungBox1978} with twelve lags does not reject the absence of autocorrelation at 5 percent in the residuals of the equations for EB and for the outcome.\footnote{The equation for EUC is excluded from the rule, because EUC follows a schedule set by federal legislation.} The weeks of EB are zero in most months and 13 or 20 when the programme is on, with intermediate monthly averages in the months in which it switches. By Theorem \ref{thm:nonneg}, the response therefore mixes the effect of switching EB on with the effect of moving from 13 to 20 weeks, with weights that depend on how often each occurs. The weights on the controls are known, so the bands come from the residual bootstrap of Theorem \ref{thm:resid}, with the responses cumulated in each draw. For the average across states, every state uses the same lag order, the larger of the median orders selected for the two models, and the bootstrap draws the same months for every state, so that the bands of the average account for the controls that the states share.

\begin{figure}[htbp]
\centering\safegraphic[width=\textwidth]{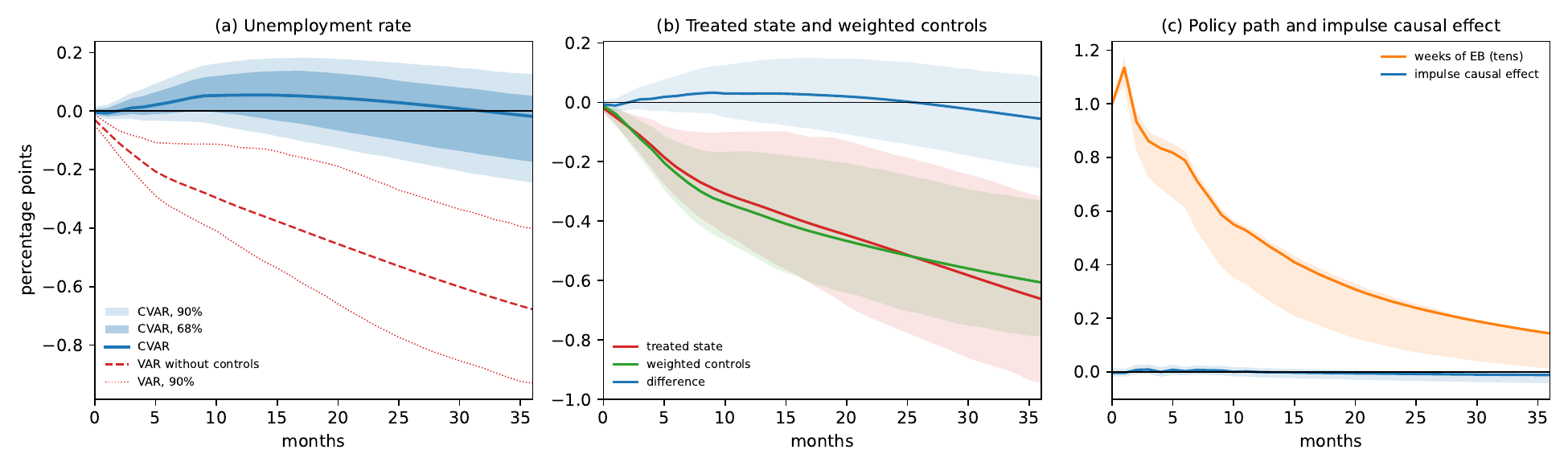}
\caption{Response of the unemployment rate to ten additional weeks of EB, average across states. (a) Cumulated response of the CVAR, with 68 and 90 percent bands, and of the VAR without controls, with its 90 percent band. (b) Response of the treated state and of the weighted controls in the VAR with both, and of their difference (Proposition \ref{prop:placebo}), with 90 percent bands. (c) Response of the weeks of EB, in tens, to their own innovation, and impulse causal effect of one week of EB from Theorem \ref{thm:direct}, with 90 percent bands. Percentage points.\label{fig:eb-avg}}
\end{figure}

Figure \ref{fig:eb-avg} and Table \ref{tab:eb-avg} report the results, averaged across states. The VAR without controls finds that an extension is followed by a fall in unemployment that keeps growing for three years. The CVAR finds a response close to zero at every horizon, and from six months on its band excludes the estimate of the VAR without controls. Its band is also narrower at every horizon.

\safeinput{tab_eb_average.tex}

\begin{figure}[htbp]
\centering\safegraphic[width=\textwidth]{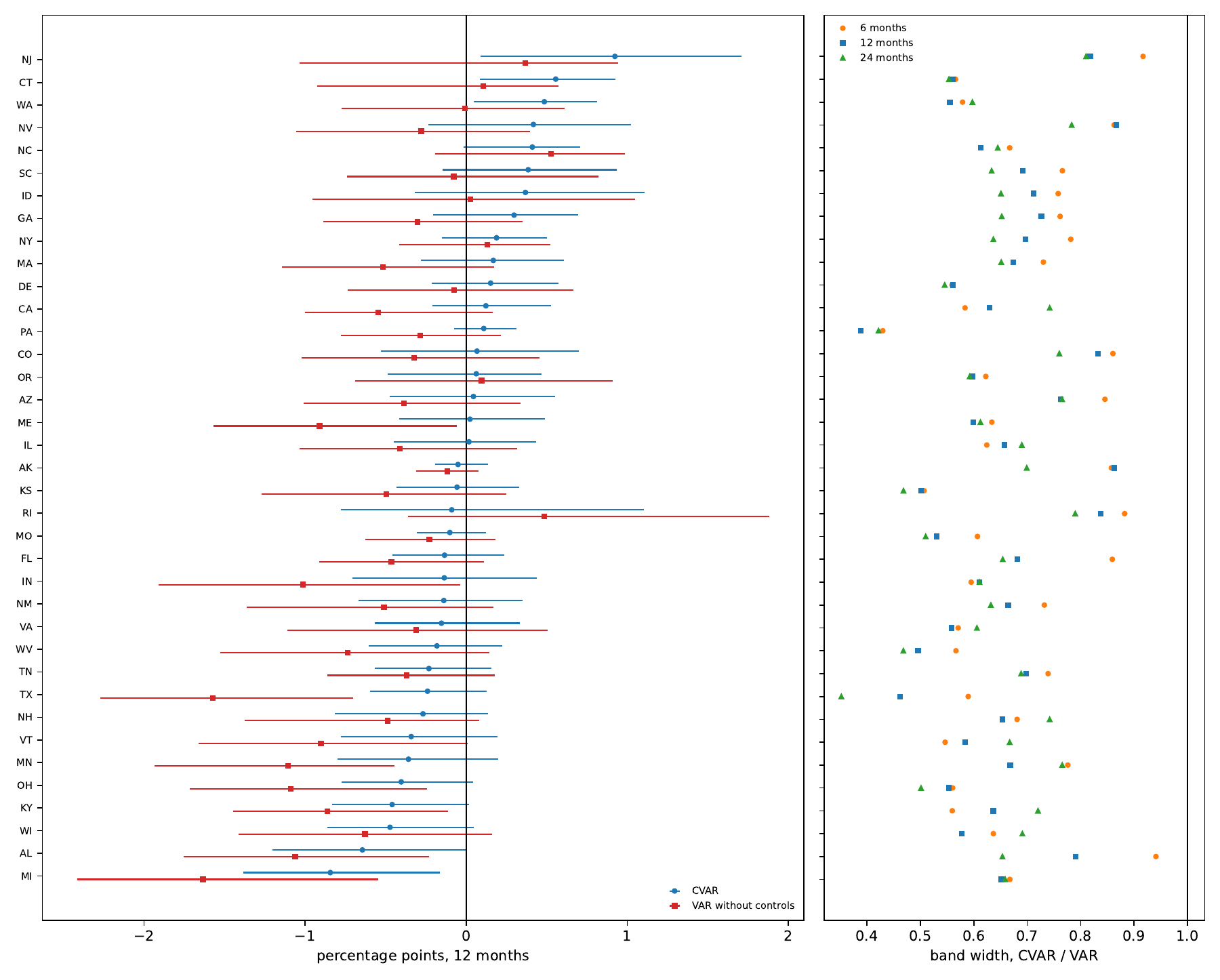}
\caption{Response of the unemployment rate to ten additional weeks of EB, by state. Left: response after twelve months, CVAR and VAR without controls, with 90 percent bands. Right: width of the band of the CVAR relative to that of the VAR without controls, at six, twelve and twenty-four months. Lag orders chosen state by state.\label{fig:eb-states}}
\end{figure}

Section \ref{sec:app-eb} reports local-projection difference-in-differences estimates on the same panel, with the never-triggered states and with clean controls as comparison units (Table \ref{tab:eb-lpdid}). With clean controls, which include the treated states in the years in which they had no extended benefits, the difference-in-differences estimate is close to zero at every horizon, in agreement with the CVAR, and its bands are several times wider. Figure \ref{fig:eb-states} shows that the pattern of the CVAR holds state by state. The VAR without controls finds a fall in unemployment in most states. The responses of the CVAR are spread around zero, falls and increases in roughly equal number, with few bands that exclude zero. From six months on, the band of the CVAR is narrower than that of the VAR without controls in every state, and at impact the two are of similar width. Section \ref{sec:app-eb} reports every state.

Panel (b) of Figure \ref{fig:eb-avg} shows where the response of the VAR without controls comes from. The weighted controls, states that never extended benefits, respond to the extensions of the treated states about as much as the treated states themselves, with bands that exclude zero from six months on, and the difference between the two responses stays close to zero at every horizon. By Proposition \ref{prop:placebo}, the response of the controls measures the bias from the common trends: the extensions switch on in the depth of a recession that also raises unemployment in the controls, and the recovery that follows lowers unemployment in every state. Table \ref{tab:eb-bounds} reports the bounds of Proposition \ref{prop:bounds}. The bounds reach the response of the treated state in the VAR with the controls added only when the bias specific to the treated states is allowed to be of the same order as the bias that the controls measure, so the conclusion that extensions had little effect does not rest on the exact form of Assumption \ref{assu:A4d}.

\safeinput{tab_eb_bounds.tex}

Panel (c) of Figure \ref{fig:eb-avg} reports the decomposition of Theorem \ref{thm:direct}. A surprise extension is followed by additional weeks of EB for more than two years, through the persistence of the programme and the reaction of the triggers. Once these weeks are divided out of the response of the cycle, the impulse causal effect of one week of benefits, holding the later weeks fixed, is close to zero at every horizon.

Nearly all the variation in EB in the sample occurs between 2009 and 2012, when EUC was in force in every state and potential durations were already long. The estimates of the CVAR therefore speak to extensions at long durations, and they agree with the finding of \citet{AcostaEtAl2025} that such extensions have virtually no effect on unemployment. The VAR without controls reaches the opposite conclusion, a large fall in unemployment, which is also opposite in sign to the microeconometric evidence, and the response of the controls shows that it reflects the recovery from the Great Recession.

A different comparison comes from difference-in-differences. Table \ref{tab:eb-lpdid} estimates the effect of extended benefits with the local-projection difference-in-differences of \citet{dube2023local}. An event is a month in which EB switches on in a state after a year without it, and the comparison units are either the never-triggered states or every state-month without EB from a year before the event to the horizon. The regression includes month fixed effects, lags of the change in the unemployment rate and the changes in the weeks of EUC. To compare the coefficients with the responses of the CVAR, the table divides them by the average size of a switch-on in tens of weeks. The two estimators answer different questions: the difference-in-differences estimates the effect of switching the programme on, including the weeks that follow, against states in which it stays off, while the CVAR estimates the response to a surprise in the weeks of benefits, conditional on the past of the cycle and of both programmes. With the never-triggered states as comparison units, the difference-in-differences finds that unemployment rises after EB switches on, significantly at impact, after six months and after three years. The sign is the opposite of the response of the VAR without controls, and the reason is the timing of the events: EB switches on as unemployment in a state starts to rise, and the rural states that never triggered it went through a much milder recession, so the treated states keep diverging from them after the event. The test of the design reaches the same conclusion from the trends, since it finds that these states do not share the stochastic trends of the treated states, which is the condition that parallel trends would require. With clean controls, which include the treated states in the years in which they had no extended benefits and are therefore closer to them, the estimate is close to zero at every horizon and agrees with the CVAR. Its bands are several times wider than those of the CVAR, because each event contributes a single comparison while the CVAR uses every month of the policy.

\safeinput{tab_eb_lpdid.tex}

\safeinput{tab_eb_states.tex}

\section{The Low Carbon Fuel Standard\label{sec:lcfs}}

This section reports in full the application that Section \ref{sec:lcfssum} of the paper summarises. Unlike the dividend and extended benefits, which react to the business cycle, the policy here is a cost that changes every month. California's Low Carbon Fuel Standard (LCFS) has been in force since 2011, with the stated goal of reducing the greenhouse gas emissions of transportation. It sets an annual benchmark for the average carbon intensity of the fuels sold in the state, which declines over time; the original target was a reduction of ten percent by 2020 relative to 2010. Fuels below the benchmark generate credits, and refiners and importers whose gasoline exceeds it incur deficits, which they cover by buying credits. The standard thereby creates a market for carbon intensity, and the cost of compliance per gallon of gasoline is the price of a credit multiplied by the deficit that a gallon generates, which grows as the benchmark tightens.

Whether this cost is passed through to consumers as higher gasoline prices is disputed. The California Air Resources Board has argued that credit prices and retail gasoline prices show no relationship \citep{CARB2024}, while the cost per gallon that refiners report to the California Energy Commission follows the credit price closely \citep{KleinmanCenter2024}. The closest evidence comes from the federal Renewable Fuel Standard, for which \citet{KnittelMeiselmanStock2017} find that the price of compliance credits passes fully to wholesale fuel prices within days. They estimate a VAR in the price of the credits and in the spread between a fuel that must comply and a comparable fuel that need not, which is a CVAR with one control and a weight fixed at one. Their result is a useful benchmark, but it concerns a different programme and a different stage of the supply chain, and the pass-through of fuel-standard credits to retail prices has received little attention.

Incidence theory predicts that a cost per gallon borne by suppliers passes almost fully to consumers, because the short-run demand for gasoline is very inelastic. The difficulty is empirical. Retail prices in California move with crude oil, with national refining margins and with the outages of its own refineries, and the credit price moves with the balance of the credit market and with the rules of the programme. A comparison of the two series mixes all of these movements. The CVAR compares California with regions that buy crude oil and refined products in the same markets without the standard, which removes the national movements in the price.

The data for California are monthly, from January 2014 to July 2026. The outcome is the retail price of regular gasoline in California, in cents per gallon, the monthly average of the weekly prices of the Energy Information Administration, with calendar-month effects removed from its changes. The controls are the retail prices of regions without a comparable standard, in three sets: the Gulf Coast, where much of the refining capacity of the country is located; the average of the East Coast, Midwest, Gulf Coast and Rocky Mountain regions; and the average of Texas and Florida. I exclude the West Coast outside California, because Oregon and Washington run clean fuel standards of their own and Arizona and Nevada buy much of their gasoline from California refineries.

The policy is the cost of the standard per gallon, in cents. Until April 2024 I compute it from the monthly average price of credits in the transfer reports of the Air Resources Board and the formula of the programme, with the gasoline benchmark of each year.\footnote{The cost per retail gallon is the credit price times the difference between the carbon intensity of the gasoline blendstock and the benchmark of the year, times the energy density of the blendstock and its share in a gallon of retail gasoline, which contains ten percent ethanol. The benchmarks are those of the regulation in force before the amendments of 2024. Appendix \ref{sec:data} gives the sources.} From May 2024 I use the cost that refiners report to the California Energy Commission. Over the months of 2024 in which both measures exist, the formula tracks the reported cost closely; from January 2025, after the amendments of 2024 tightened the benchmarks, the reported cost exceeds the formula. The cost has a stochastic trend, so the policy enters the VAR as its monthly change, and the response is to a permanent increase of one cent per gallon. Full pass-through is a response of one.

\begin{figure}[htbp]
\centering\safegraphic[width=0.85\textwidth]{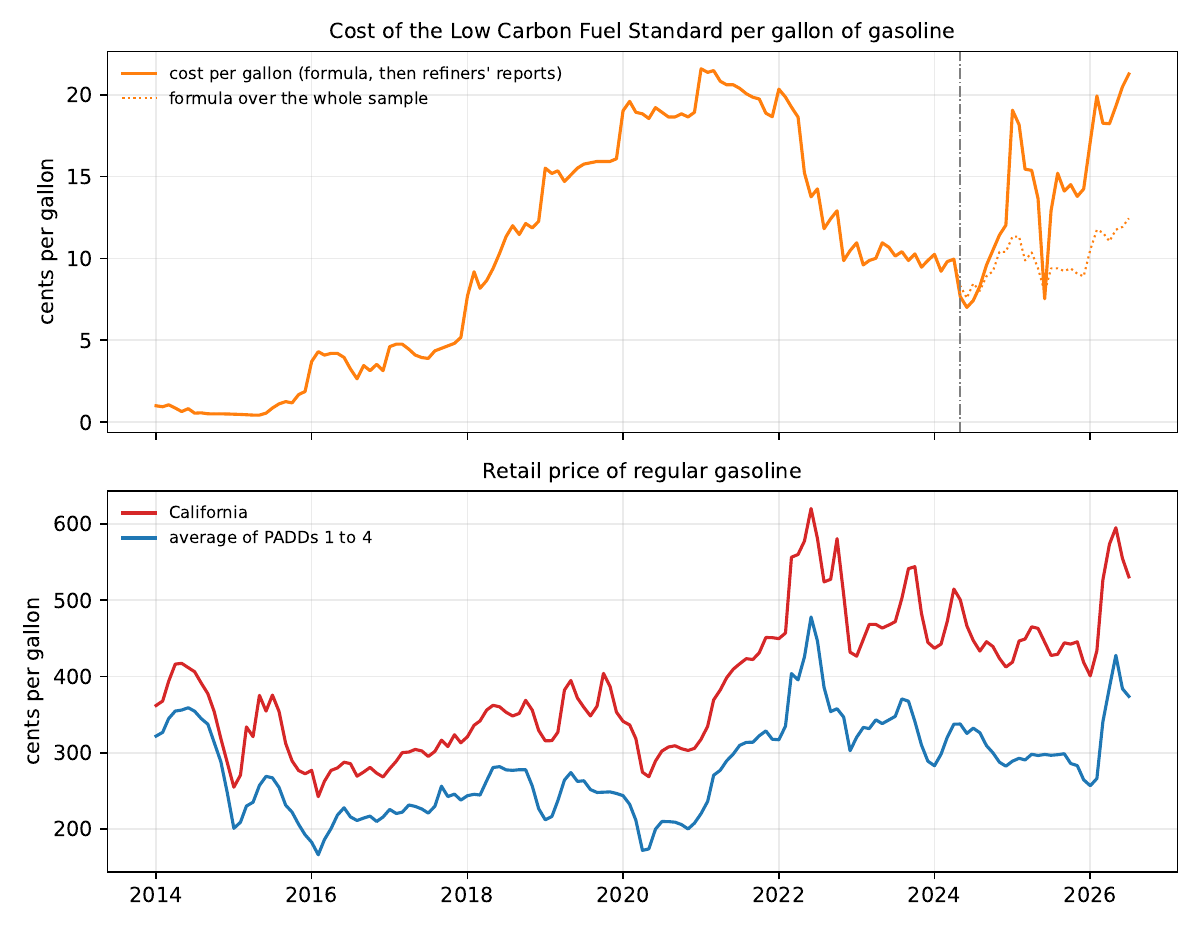}
\caption{Low Carbon Fuel Standard: policy and outcomes. Top: cost of the standard per gallon of retail gasoline, from the formula of the programme until April 2024 and as reported by refiners to the California Energy Commission from May 2024 (vertical line); dotted, the formula over the whole sample. Bottom: retail price of regular gasoline in California and average of the East Coast, Midwest, Gulf Coast and Rocky Mountain regions, monthly averages of weekly prices. Cents per gallon, January 2014 to July 2026.\label{fig:ca-data}}
\end{figure}

Figure \ref{fig:ca-data} shows the cost of the standard and the retail prices. The price in California and the average price of the controls move together with crude oil, and the gap between them changes slowly, which is the movement specific to California that the test of the design leaves in the cycle.

As for extended benefits, the test of Section \ref{sec:test} does not reject a stochastic trend specific to California with any of the three sets of controls, so I estimate the CVAR in changes, with equal weights on the controls, which buy crude oil and refined products in the same national markets as California. The VAR contains the change in the cost and the change in the cycle, with the cost ordered first, and the VAR without controls replaces the cycle with the price in California. The lag order of each model follows the rule of Section \ref{sec:eb}. Assumption \ref{assu:A4d} requires that a change in the cost carries no news about the future path of the component of the price specific to California. The credit price is set in a market that responds to the balance of credits and deficits accumulated over years and to news about the rules of the programme, and in the data the past changes of the cycle do not predict it: the Wald test that the cycle does not Granger-cause the cost in the VAR of the CVAR does not reject with any set of controls, measure of the cost, sample or lag order.

Figure \ref{fig:california_main} and Table \ref{tab:california_main} report the results. With every set of controls, the CVAR finds that the price of gasoline in California rises by about the increase in the cost within a month and stays near that level. The 68 percent bands exclude zero one month after the increase with every set of controls, and the 90 percent bands contain full pass-through at every horizon. The VAR without controls finds a response that keeps rising well above full pass-through, with bands up to about twice as wide as those of the CVAR that contain both zero and full pass-through. The CVAR delivers the answer that incidence theory predicts and that \citet{KnittelMeiselmanStock2017} find for federal credits at wholesale, with bands narrow enough to exclude the large responses that the VAR without controls cannot rule out.

\safeinput{fig_california_main_float.tex}

\safeinput{tab_california_main.tex}

The result holds across a wide range of specifications, which Tables \ref{tab:california_robust}, \ref{tab:california_appendix}, \ref{tab:california_measures} and \ref{tab:california_lp} report. In the samples that start in 2017 and in 2020, when the cost per gallon is largest, the response at impact is larger, and in most of these specifications its 90 percent band excludes zero. Dropping any one region from the average of the controls leaves the response close to that of the main specification, and so does removing the years from 2020 to 2022 with a dummy for every month, which lowers the response at impact and leaves it unchanged one month later. The response at impact and one month later is stable across lag orders from one to twelve. With the cost computed from the formula over the whole sample, the responses are larger, as expected from a measure that understates the cost from 2025. Between 2014 and 2019 the cost per gallon was small and the policy varied little, and the responses are imprecise and of uncertain sign. Local projections reproduce the response of the CVAR and find no difference between increases and decreases of the cost, or between periods of high and low credit prices, that their bands can detect.

\safeinput{tab_california_robust.tex}

\safeinput{tab_california_appendix.tex}

\safeinput{tab_california_measures.tex}

\safeinput{tab_california_lp.tex}

\bibliographystyle{apalike}
\bibliography{bibliography}